%% file: main.tex
\documentclass{article} %
\usepackage{llm2cert-preprint,times}
\usepackage{graphicx}

\usepackage{hyperref}
\hypersetup{hidelinks}
\usepackage{url}

\usepackage{algorithm}
\usepackage{algorithmic}
\usepackage{soul}
\usepackage{tabularx}
\usepackage{newfloat}
\usepackage{listings}
\usepackage{booktabs}
\usepackage{subcaption}
\usepackage{amsmath,amsthm,amssymb}
\newtheorem{corollary}{Corollary}
\usepackage{mathtools}
\usepackage{enumitem}
\usepackage[utf8]{inputenc}
\usepackage{xcolor}
\usepackage[T1]{fontenc}
\usepackage{helvet}
\usepackage{comment}
\usepackage{natbib} 
\usepackage{courier}
\usepackage{multirow}
\usepackage{bm}
\usepackage{float}
\newtheorem{theorem}{Theorem}
\newtheorem{definition}{Definition}

\newtheorem{remark}{Remark}

\newtheorem{problem}{Problem}

\newcounter{systeminstance}
\renewcommand{\thesysteminstance}{\arabic{systeminstance}}

\newenvironment{systeminstance}[1]{%
  \refstepcounter{systeminstance}%
  \noindent\textit{System \thesysteminstance\ (#1).}\ %
}{%
  \par\medskip
}

\usepackage{array}
\newcolumntype{L}[1]{>{\raggedright\arraybackslash}p{#1}}

\lstdefinestyle{jsonstyle}{
    basicstyle=\ttfamily\small,
    numbers=left,
    numberstyle=\tiny\color{gray},
    stepnumber=1,
    numbersep=5pt,
    backgroundcolor=\color{white},
    showspaces=false,
    showstringspaces=false,
    showtabs=false,
    frame=single,
    rulecolor=\color{black},
    tabsize=2,
    captionpos=b,
    breaklines=true,
    breakatwhitespace=false,
    stringstyle=\color{blue},
    keywordstyle=\color{magenta},
    commentstyle=\color{gray},
    morecomment=[l]{//},
    morecomment=[s]{/*}{*/}
}

\usepackage{placeins}

\newcommand{\apptablestyle}{%
    \small
    \renewcommand{\arraystretch}{1.12}%
    \setlength{\tabcolsep}{4pt}%
}

\newlength{\apptablewidth}
\newcommand{\apptabnote}[1]{%
    \par\vspace{3pt}%
    \begin{minipage}{\apptablewidth}\footnotesize\raggedright
    \textit{Note.} #1
    \end{minipage}%
}


\newcolumntype{Y}{>{\raggedleft\arraybackslash}X}

\usepackage[most]{tcolorbox}

\definecolor{PromptProposal}{HTML}{456B82}
\definecolor{PromptLean}{HTML}{4F7065}
\definecolor{PromptBackground}{HTML}{F7F8F8}
\definecolor{PromptRule}{HTML}{D7DCDF}
\definecolor{PromptMeta}{HTML}{626A70}

\usepackage{needspace}

\newtcolorbox{promptblock}{
    enhanced,
    breakable,
    frame hidden,
    colback=white,
    boxrule=0pt,
    borderline north={0.45pt}{0pt}{black},
    borderline south={0.45pt}{0pt}{black},
    left=8pt,
    right=8pt,
    top=7pt,
    bottom=7pt,
    before skip=6pt,
    after skip=10pt,
    parbox=false,
    fontupper=\normalfont\footnotesize
}

\newenvironment{promptbody}
{
    \normalfont
    \footnotesize
    \setlength{\parindent}{0pt}
    \setlength{\parskip}{0.55em}
}
{
    \par
}

\newcommand{\promptrole}[1]{\textbf{#1:}}
\newcommand{\promptplaceholder}[1]{\textit{#1}}

\usepackage{fix-cm}
\usepackage{tcolorbox}
\tcbuselibrary{skins,breakable}
\definecolor{CertiExampleBackground}{HTML}{F0F5F8}
\definecolor{CertiExampleRule}{HTML}{7C9CAF}
\definecolor{CertiExampleTitle}{HTML}{34586D}
\definecolor{CertiExampleText}{HTML}{22323D}

\newtcolorbox{certileanexample}[1]{%
  enhanced,
  breakable,
  colback=CertiExampleBackground,
  colbacktitle=CertiExampleBackground,
  coltitle=CertiExampleTitle,
  coltext=CertiExampleText,
  frame hidden,
  boxrule=0pt,
  borderline west={1.7pt}{0pt}{CertiExampleRule},
  arc=1pt,
  outer arc=1pt,
  left=8pt,
  right=8pt,
  top=2pt,
  bottom=3pt,
  toptitle=1.5pt,
  bottomtitle=1pt,
  fontupper=\fontsize{9.5}{11.5}\selectfont,
  fonttitle=\fontsize{9.5}{11.5}\selectfont\bfseries,
  title={Running example: #1},
  title after break={Running example: #1 (continued)},
  before skip=4pt plus 1pt minus 1pt,
  after skip=3pt plus 1pt minus 1pt,
  before={\par\Needspace{4\baselineskip}\noindent},
  after={\par},
}

\newcommand{\methodname}{\textsc{LLM2Cert}}
\newcommand{\benchmarkname}{\textsc{DynaCertBench}}

\title{\methodname: LLM Synthesis of Certificates with SMT/Lean Verification}

\author{%
\textbf{Zhipeng Cao$^{1,2,*}$, Peiyang Song$^{3,4,*}$,
Nikolaus Vertovec$^{5,6}$, Weiru Cui$^{1,8}$,
Zhihao Yang$^{1,2}$,} \\
\textbf{Mingkang Liu$^{7,8}$, Luxia Lin$^{1,8}$,
Bai Xue$^{1,8,\dagger}$, Alessandro Abate$^{5\dagger}$} \\[4pt]
$^{1}$Institute of Software, Chinese Academy of Sciences (CAS) \\
$^{2}$School of Advanced Interdisciplinary Sciences,
University of Chinese Academy of Sciences \\
$^{3}$Language Technologies Institute, School of Computer Science, Carnegie Mellon University \\
$^{4}$Computing + Mathematical Sciences Department, California Institute of Technology \\
$^{5}$Department of Computer Science, University of Oxford \\
$^{6}$Gigaton Technologies Ltd \\
$^{7}$Institute of Geographic Sciences and Natural Resources Research, CAS\\
$^{8}$University of Chinese Academy of Sciences \\[4pt]
{\small $^{*}$Equal contribution. \quad
$^{\dagger}$Equal advising.}
}

\begin{document}

\maketitle

\begin{abstract}
Lyapunov functions and barrier certificates give state-space guarantees of stability and safety, but synthesizing and verifying them---especially for nonlinear, high-dimensional systems---remains an open challenge. Existing pipelines verify candidates with numerical solvers (sum-of-squares programs), interval methods, or satisfiability-modulo-theories (SMT) solvers, which often require relaxations, are conservative, or scale poorly with model dimension and nonlinearity. These limitations have also shaped benchmark design: standard test suites are dominated by low-dimensional polynomial systems chosen to suit what solvers can handle rather than what engineering applications require. We address both sides of this challenge. First, we introduce \benchmarkname, a benchmark of 105 instances: 80 non-redundant systems curated from the literature, extended with high-dimensional and control-constrained variants and new nonlinearly coupled systems, comprehensively spanning stability, region-of-attraction, forward-invariance, and safety tasks. Second, we present \methodname, an LLM-driven framework that synthesizes candidate certificates and verifies them against their proof obligations: each obligation is routed to an SMT solver or directly to the Lean Proof-Repair agent. On \benchmarkname, \methodname{} verifies 81 of 105 instances ($77.1\%$), a $17.1$ percentage point improvement over one-shot proposal without feedback.  On the 81 fixed certificates, \methodname{} verifies 22 that All-SMT leaves unresolved. Code and data are available at \url{https://github.com/FormalCert-AI/llm2cert}.
\end{abstract}

\input{sections/introduction}
\input{sections/problem_setting_and_certificates}
\input{sections/contract_preserving_agentic_synthesis_and_verification}
\input{sections/benchmark_construction}
\input{sections/experiments}
\input{sections/conclusion}

\clearpage
\bibliography{references}
\bibliographystyle{references}

\newpage
\appendix

\input{sections/limitations_and_future_work}
\input{sections/related_work}
\input{sections/dynamical_models_and_notation}
\input{sections/prompt}
\input{sections/experiments_details}
\input{sections/benchmark_details}

\end{document}

%% file: sections/introduction.tex
\section{Introduction}

Safety-critical autonomous systems, such as robotic platforms, autonomous vehicles, and aerospace systems, require formal guarantees that go beyond finite simulation and testing. For dynamical systems, Lyapunov functions and barrier certificates are the standard mathematical witnesses of stability and safety~\citep{khalil2002nonlinear,prajna2007framework}. A Lyapunov function certifies stability by decreasing along system trajectories, while a barrier certificate certifies safety by defining a forward-invariant boundary that separates trajectories from unsafe regions. Because these certificates reduce trajectory-level properties to inequalities over the state space, they avoid explicit trajectory computation and are indispensable for analyzing complex nonlinear systems.

Verification methods for Lyapunov and barrier functions include optimization-based techniques such as sum-of-squares (SOS) programming, interval-based methods such as interval bound propagation (IBP), and formal methods based on Satisfiability Modulo Theories (SMT) solvers. Each comes with caveats. Numerical approaches such as SOS are affected by floating-point errors and numerical instability. SMT-based methods, in contrast, can provide satisfiability results under $\delta$-\textsc{sat}, but such results do not necessarily establish the exact mathematical validity of the underlying proof obligations. In practice, both families also rely on ad-hoc relaxations, such as excluding a neighborhood of the origin or replacing transcendental terms by Taylor approximations \cite{chang2019neural,zhou2022neural,taheri2025barrierbench}, and both scale poorly to high-dimensional state spaces and strongly nonlinear dynamics. These limitations motivate a complementary approach based on proof assistants such as Lean~\citep{moura2021lean}, in which the verified statement is an exact mathematical theorem checked by a small trusted kernel, rather than a numerically tolerated or approximately satisfiable surrogate.

These methodological bottlenecks have not only limited verification itself but have also skewed how verification algorithms are evaluated. Standard benchmarks remain largely homogeneous: most instances are low-dimensional, parameter-free systems on tightly bounded domains. This lack of diversity does not reflect engineering requirements; rather, existing datasets are implicitly designed around the limits of the solvers used to verify them. High-dimensional benchmarks are scarce, and nearly all of those that exist are polynomial. Moving beyond these constraints requires both a rigorous verification paradigm and benchmark tasks that reflect richer, more realistic dynamics.

We report the following main contributions:

\begin{enumerate}
    \item \textbf{\benchmarkname: A Comprehensive Benchmark for Verifying Dynamical Systems.} We curate and extend a benchmark that covers stability, region-of-attraction, forward-invariance, and safety verification for continuous- and discrete-time systems. It deliberately includes realistic scenarios that existing suites avoid: high-dimensional state spaces, control-input constraints, strongly nonlinear coupling, and parameterized systems. Every instance is posed with its exact, unrelaxed certificate conditions.
    
    \item \textbf{A Formal Verification Paradigm Combining SMT and Lean.} To the best of our knowledge, this is the first work to integrate the Lean theorem prover into the end-to-end synthesis and verification of Lyapunov and barrier certificates for both continuous- and discrete-time dynamical systems. Our method delegates proof obligations that SMT solvers cannot discharge to Lean, enabling the construction of rigorous mathematical proofs. This design eliminates the need for conservative assumptions commonly introduced by SMT-based approaches, including Taylor approximations, $\delta$-satisfiability, and the exclusion of a neighborhood around the origin. Our approach thereby provides machine-checkable formal guarantees of safety and stability for the system under the explicitly stated assumptions.
    
    \item \textbf{\methodname: An Agentic Framework for Efficient Formal Verification.} We propose an LLM-driven framework that balances computational efficiency with mathematical soundness. To address the computational bottleneck of Lean, our framework adopts a hybrid verification strategy. Specifically, it utilizes SMT solvers for faster verification and counterexample generation, while seamlessly integrating Lean to finalize formal proofs. The two backends are complementary: SMT efficiently discharges obligations, including some for which Lean proof search does not finish within the budget, while Lean resolves others left unresolved by SMT.
\end{enumerate}

%% file: sections/problem_setting_and_certificates.tex
\section{Problem Setting and Certificates}
\label{sec:preliminaries}

\subsection{Dynamical Systems and Verification Tasks}
\label{sec:dynamical_systems}

We consider continuous-time and discrete-time dynamical systems:
\begin{equation}
    \dot{x}(t) = f(x(t),u(t)) \quad \text{or} \quad x_{k+1} = F(x_k,u_k), \label{eq:systems}
\end{equation}
where $u \in \mathcal{U}\subseteq\mathbb{R}^m$ is the control input and $f, F: \mathbb{R}^n \times \mathcal{U} \to \mathbb{R}^n$. Given a state-feedback policy $\pi:\mathbb{R}^n\to\mathcal{U}$, we write the closed-loop dynamics as $f_\pi(x):=f(x,\pi(x))$ and $F_\pi(x):=F(x,\pi(x))$, and let $\phi_\pi^{x_0}(t)$ denote the unique closed-loop trajectory from $x_0$. Let $\mathcal{X}\subseteq\mathbb{R}^n$ be the domain of interest, and let $\mathcal{X}_I\subseteq\mathcal{X}$ and $\mathcal{X}_U$ denote the initial and unsafe sets, where either $\mathcal{X}_U\subseteq\mathcal{X}$ or $\mathcal{X}_{U} = \mathbb{R}^n\setminus \mathcal{X}$. Let $\mathcal{N}$ be a neighborhood of the origin, and let $\Omega_c:=\{x:V(x)\le c\}$ be the sublevel set of a Lyapunov candidate $V$ for $c>0$.

As summarized in Table~\ref{tab:task_certificate_mapping}, we consider four verification tasks concerning stability and safety: \textbf{(i) stability certification} of the equilibrium; \textbf{(ii) region-of-attraction (ROA) verification}; \textbf{(iii) forward-invariant set synthesis} avoiding an unsafe set $\mathcal{X}_U$; and \textbf{(iv) safety verification}, ensuring that trajectories from $\mathcal{X}_I$ never reach $\mathcal{X}_U$. Each task is established by a scalar certificate $W: \mathbb{R}^n \to \mathbb{R}$, which instantiates either a Lyapunov function $V$ or a barrier function $B$. The change of $W$ along the dynamics is the Lie derivative $\mathcal{D}_\pi W(x) := \nabla W(x)^\top f_\pi(x)$ in continuous time and the one-step difference $\mathcal{D}_\pi W(x) := W(F_\pi(x)) - W(x)$ in discrete time. Formal definitions and certificate conditions are given in Appendix~\ref{sec:dynamical_models}.

\begin{figure}[t]
    \centering
    \includegraphics[width=1\linewidth]{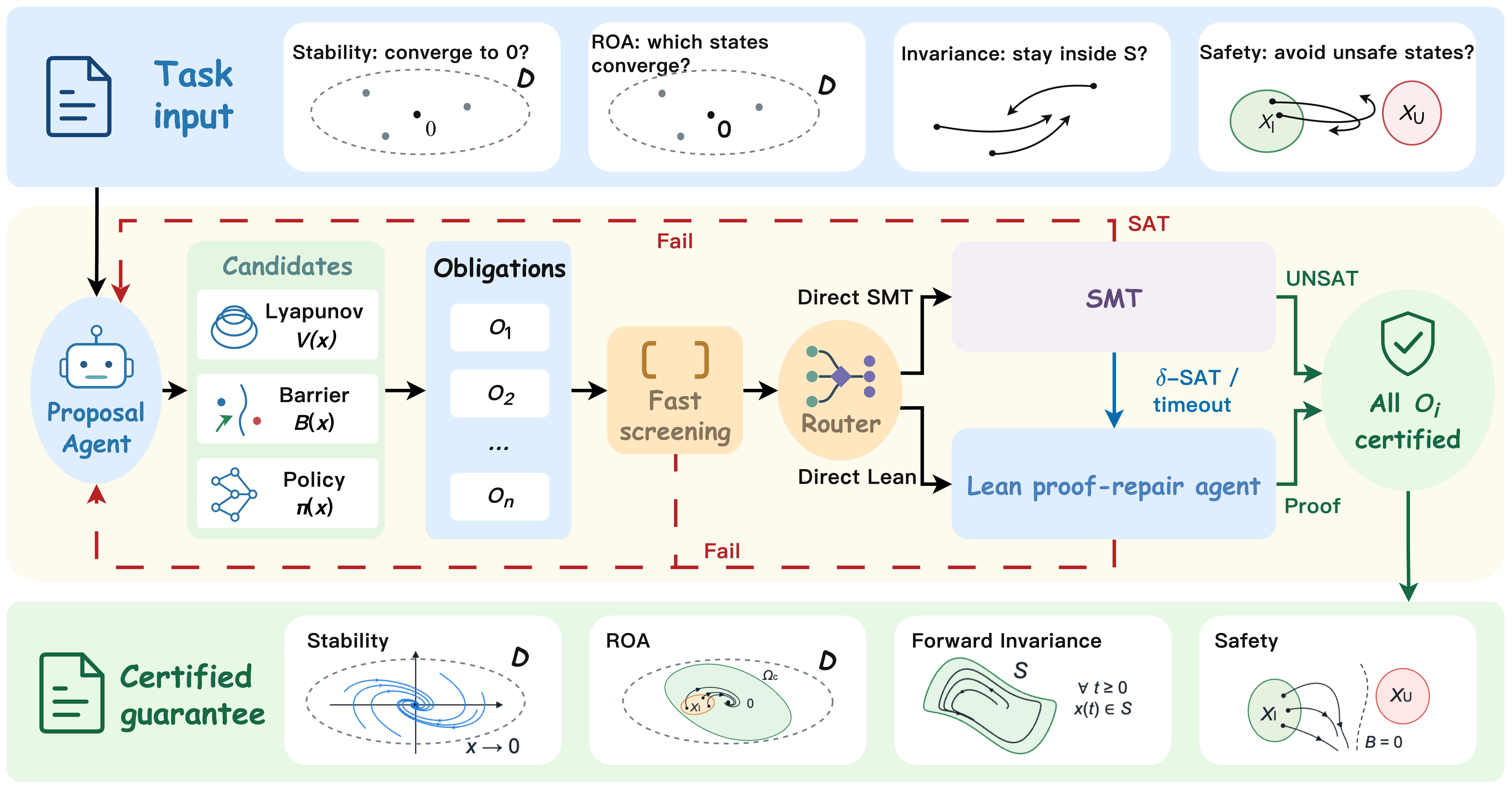}
    \caption{\methodname{} framework. The Proposal Agent proposes a candidate certificate (and policy), which induces a list of proof obligations. Obligations without a compatible SMT encoding go directly to Lean. For the remaining obligations, the router selects either direct SMT or direct Lean. Obligations left unresolved by SMT are escalated to the Lean Proof-Repair Agent. Counterexamples and unresolved obligations are fed back to the Proposal Agent for the next candidate.}
    \label{fig:framework_overview}
\end{figure}

\begin{table}[t]
\caption{Verification tasks, the corresponding certificate conditions, and the theorems (Appendix~\ref{sec:dynamical_models}) that connect them. Sets $\mathcal{X}$, $\mathcal{X}_I$, $\mathcal{X}_U$, $\mathcal{N}$, and $\Omega_c$ are defined in Section~\ref{sec:dynamical_systems}.}
\label{tab:task_certificate_mapping}
\centering
\small
\renewcommand{\arraystretch}{1}
\begin{tabularx}{\columnwidth}{@{}l >{\raggedright\arraybackslash}X >{\centering\arraybackslash}p{2.2cm}@{}}
\toprule
\textbf{Task} & \textbf{Certificate conditions} & \textbf{Theorems} \\
\midrule
Stability & $V(0)=0$, $V(x)>0$ for $x\neq 0$, and $\mathcal{D}_\pi V\le 0$ (or $<0$ for $x\neq 0$) & Thm.~\ref{thm:lyapunov} \\
ROA & Lyapunov $V$ and corr. sublevel set $\Omega_c$ with $\mathcal{X}_I \subseteq \Omega_c \subseteq \mathcal{N}$ & Cor.~\ref{cor:roa} \\
Invariant Set & Barrier $B$ with $B < 0$ on $\mathcal{X}_U$, boundary/invariance condition & Thm.~\ref{thm:barrier_safety}, Cor.~\ref{cor:barrier_set_synthesis} \\
Safety & Barrier $B$ with $B \ge 0$ on $\mathcal{X}_I$, $B < 0$ on $\mathcal{X}_U$, barrier condition & Cor.~\ref{cor:barrier_safety} \\
\bottomrule
\end{tabularx}
\end{table}

\begin{remark}
Two aspects of our setting differ from prior work. \emph{\textbf{(i) Verification vs.\ optimization:}} ROAs and invariant sets are generally not unique, and prior methods often optimize their size. We instead verify the set induced by a proposed candidate $W$. \emph{\textbf{(ii) Control input:}} \citet{vertovec2025scalable} verify that an admissible control exists at each state ($\forall x\,\exists u$), whereas we verify $W$ against a specific feedback policy $u=\pi(x)$, i.e., on the closed-loop system.
\end{remark}

\subsection{Soundness of Certificate Conditions}
\label{sec:trust_model}

Soundness has two layers. First, the certificate conditions for each task in Table~\ref{tab:task_certificate_mapping} must imply the corresponding property: if a candidate satisfies the conditions, the theorem holds. We prove every such condition-to-theorem implication in Lean, and each proof is checked by the Lean kernel. Second, given a candidate, we establish that it satisfies the certificate conditions. This layer is provided by \methodname{} (Section~\ref{sec:termination}), which accepts a candidate only when every condition has been discharged by an SMT \texttt{unsat} verdict or a kernel-checked Lean proof. Together, the two layers give the end-to-end guarantee
\[
\text{Proposed candidate}
\;\xrightarrow{\text{\methodname}}\;
\text{Verified candidate}
\;\xrightarrow{\text{Lean}}\;
\text{Theorem holds}.
\]

%% file: sections/contract_preserving_agentic_synthesis_and_verification.tex
\section{\methodname: LLM-based Certificate Synthesis with SMT and Lean Verification}
\label{sec:LLM2Cert}

\methodname{} combines LLM-based candidate synthesis with verification by SMT solvers and Lean (Figure~\ref{fig:framework_overview}). Throughout the verification of a candidate, the candidate and its obligations remain fixed: only proofs are searched for, and the target statements are never relaxed. 
The shaded boxes follow a single running example (System~\ref{sys:bb-s011-ext}) throughout the section. 

\subsection{Candidate Search, Obligation List, and Fast Screening}
\label{sec:candidate_search}
\label{sec:screening}
\label{sec:running_example}

Given a typed specification $R=(f/F,\mathcal X,\mathcal N,\mathcal X_I,\mathcal X_U,\mathcal U,\mathrm{Task})$, the Proposal Agent proposes, at each iteration $k$, a symbolic certificate and, for controlled systems, a feedback policy. A format check parses the response into a candidate $C^{(k)}=(W^{(k)},\pi^{(k)})$; if parsing fails, the error message is returned to the Proposal Agent for correction. A parsed candidate is compiled into a typed expression tree $A^{(k)}$, and the certificate theorem of the task (Table~\ref{tab:task_certificate_mapping}) instantiated at $A^{(k)}$ yields the obligation list $\mathcal M(A^{(k)})=\{O_1,\ldots,O_n\}$, with one obligation per theorem premise. The obligations retain the original domain, boundary, and parameter constraints of $R$ without relaxation.

Before formal verification, fast screening evaluates each obligation on finite sampled states. A detected violation rejects the candidate, and the violating samples are returned to the Proposal Agent as counterexample feedback for the next candidate $C^{(k+1)}$. Because screening samples a continuous domain, passing it provides no guarantee: all obligations in $\mathcal{M}(A^{(k)})$ remain to be formally verified.  

\begin{certileanexample}{Candidate and obligations}
We use the two-dimensional continuous-time safety instance (\texttt{Extend-S011})
\citep{taheri2025barrierbench}:
\[
\dot{x}_1=x_2+u_0,\qquad
\dot{x}_2=-\frac{3}{10}x_1-\frac15(1+\cos x_1)x_2-\frac1{10}x_1^3+u_1.
\]
The specification is $\mathcal X=\mathbb R^2$,
$\mathcal X_I=\{x:\|x\|_2\leq3/5\}$,
$\mathcal X_U=\{x:\|x-(2,2)^{\mathsf T}\|_2\leq1\}$, and
$\mathcal U=[-1,1]^2$. At iteration $k=1$, the Proposal Agent returns the candidate $C^{(1)}=(W^{(1)},\pi^{(1)})$ with barrier function $W^{(1)}$ and policy $\pi^{(1)}(x)=(u_0(x),u_1(x))^{\mathsf T}$ where 
\[
W^{(1)}(x)=1-x_1^2-x_2^2,\qquad
u_0(x)=-\frac{x_1}{1+x_1^2},\qquad
u_1(x)=-\frac{x_2}{1+x_2^2}.
\]
Writing $L_{f_{\pi^{(1)}}}W^{(1)}$ for the Lie derivative of $W^{(1)}$ along the closed-loop dynamics $f_{\pi^{(1)}}$, the safety task (Cor.~\ref{cor:barrier_safety}) induces four obligations:
\[
\begin{aligned}
O_1 &: \pi^{(1)}(x)\in\mathcal U\quad(x\in\mathcal X),&
\qquad O_2 &: W^{(1)}(x)\geq0\quad(x\in\mathcal X_I),\\
O_3 &: W^{(1)}(x)<0\quad(x\in\mathcal X_U),&
\qquad O_4 &: L_{f_{\pi^{(1)}}}W^{(1)}(x)>0\quad(x\in\mathcal X,\ W^{(1)}(x)=0).
\end{aligned}
\]
Fast screening finds no violation, so the candidate is passed to the Verification Router (Fig.~\ref{fig:framework_overview}), which dispatches each obligation to a backend: obligations without a compatible SMT encoding go to Lean. For the others, the router selects direct SMT or direct Lean, and obligations sent to SMT are routed to Lean if the solvers return $\delta$-\texttt{sat} or time out. In this example, all four obligations are routed to  direct SMT.
\end{certileanexample}

\subsection{Verification Router} 
\label{sec:routing}

The Verification Router selects a verification path(direct SMT or direct Lean) for each obligation according to its type and solver compatibility. Obligations that have no compatible SMT encoding, such as radial unboundedness, are sent directly to Lean. For verification tasks sent directly to SMT solvers (Z3, cvc5, Yices, and dReal), we set a timeout of $30$\,s for each obligation. An obligation has the form $O_i:\forall z,\ P_i(z;R,C^{(k)})\Rightarrow Q_i(z;R,C^{(k)})$, where $z$ collects all quantified state and parameter variables and $P_i$ retains the original premises. The solvers are given the counterexample query $\exists z,\ P_i(z;R,C^{(k)})\land\neg Q_i(z;R,C^{(k)})$.

An SMT solver returns one of four outcomes: \texttt{unsat}, \texttt{sat}, $\delta$-\texttt{sat}, or timeout. An \texttt{unsat} result verifies the obligation. A \texttt{sat} result means the candidate violates the obligation, and the solver's model is returned to the Proposal Agent as a counterexample for the next candidate $C^{(k+1)}$. 
A dReal $\delta$-\texttt{sat} result is inconclusive: it establishes satisfiability only of a $\delta$-weakened version of the query, not of the query itself, so it neither refutes nor verifies the obligation. Obligations that end in $\delta$-\texttt{sat} or timeout therefore remain unresolved by SMT and are escalated to the Lean Proof-Repair Agent, which attempts to construct a kernel-checked proof of the unchanged statement in Lean.

\begin{certileanexample}{SMT verification and escalation to Lean}
The Verification Router sends $O_1,O_2,O_3,O_4$ to SMT. cvc5 verifies $O_1$, and Z3 verifies $O_2$ and $O_3$. The remaining obligation $O_4$ contains $\cos x_1$, which Z3 and Yices do not support. dReal supports $\cos x_1$ but requires finite variable bounds, which are absent here because $\mathcal X=\mathbb R^2$. cvc5 is therefore the only applicable backend; it checks the counterexample query $\exists x\in\mathcal X:W^{(1)}(x)=0\land L_{f_{\pi^{(1)}}}W^{(1)}(x)\leq0$ but reaches the $30$\,s timeout. $O_4$ remains unresolved and is passed to Lean.
\end{certileanexample}

\subsection{Lean Proof-Repair agent}
\label{sec:lean-repair}

For each obligation routed to Lean, the framework constructs the Lean context and theorem statement from the fixed candidate and obligation. The Lean Proof-Repair Agent may generate only the proof body, while the definitions, hypotheses, and theorem statement remain fixed. Each proof attempt is elaborated and checked in the pinned Lean environment.

A failed attempt returns \emph{structured repair feedback}: the classified failure (an unknown identifier, an application type mismatch, or an unsolved goal), its source location, and the remaining goal. For an unknown identifier, the feedback also includes the declaration lines of matching lemmas found by a local search in Mathlib. The agent revises the failing part of its previous proof using this feedback. 

A proof is accepted only if it type-checks against the fixed theorem and passes an audit that rejects proof placeholders (the \texttt{sorry} tactic in Lean) and any axiom beyond Lean's standard ones. 
The repair loop is bounded by a proof-time budget and a maximum number of rounds. If no proof is accepted within these limits, the obligation remains unresolved, and its last proof attempt and diagnostics are returned to the Proposal Agent to inform the synthesis of the next candidate $C^{(k+1)}$.

\begin{certileanexample}{Repairing the proof of $O_4$ in Lean}
The first two proof attempts fail on a missing lemma name and an implicit-argument mismatch; the compiler feedback guides a third attempt, which is accepted. On $W^{(1)}=0$, the proof derives $x_1^2+x_2^2=1$ and $x_i^2\leq1$. Writing $q=-7/5+x_1^2/5$, it establishes $q x_1x_2\geq-7/10$, $2x_i^2/(1+x_i^2)\geq x_i^2$, and $1+\cos x_1\geq0$. The tactic \texttt{ring} exposes the identity
\[
L_{f_{\pi^{(1)}}}W^{(1)}=\frac{2x_1^2}{1+x_1^2}+\frac{2x_2^2}{1+x_2^2}+q x_1x_2+\frac25(1+\cos x_1)x_2^2,
\]
and \texttt{nlinarith} combines these bounds to conclude $L_{f_{\pi^{(1)}}}W^{(1)}\geq1-7/10=3/10>0$.
\end{certileanexample}

\subsection{Soundness and Termination}
\label{sec:termination}

The outcome for a candidate is determined at the obligation level. A candidate $C^{(k)}$ is \textbf{verified} when every obligation in $\mathcal M(A^{(k)})$ has proof evidence, i.e., an SMT \texttt{unsat} result or an accepted kernel-checked Lean proof; \textbf{rejected} when fast screening detects a violation or SMT returns \texttt{sat}; and \textbf{unresolved} when at least one obligation is  open after the Lean budget is exhausted. A rejected or unresolved candidate returns its counterexamples or failure diagnostics to the Proposal Agent, which generates $C^{(k+1)}$; the loop continues until a candidate is verified or the time budget is exhausted. Exhausting the candidate budget without a verified certificate is reported as \textbf{search exhaustion}. 

\begin{certileanexample}{Aggregating evidence}
The recorded final label is \texttt{hybrid\_verified}: $O_1$--$O_3$ are discharged by SMT with $\texttt{unsat}$, and $O_4$ by a Lean proof. This example shows how Lean discharges an obligation that no SMT solver could verify within the allotted time, and consequently how \methodname{} establishes the exact safety conditions of Cor.~\ref{cor:barrier_safety} for the synthesized candidate.
\end{certileanexample}

%% file: sections/benchmark_construction.tex
\section{{\benchmarkname}: Benchmark Construction}
\label{dynacertbench}

\subsection{Motivation and Design Goals}

Existing certificate-synthesis benchmarks routinely compromise the verification target to accommodate solver limitations: they exclude a neighborhood of the origin, replace transcendental terms by Taylor approximations, or accept $\delta$-\texttt{sat} verdicts. They have also remained largely homogeneous over the past decade, consisting mostly of low-dimensional polynomial systems. This implicit adaptation to solver limitations distorts the evaluation of algorithms.
\benchmarkname{} addresses this gap in two ways. First, every instance is posed with its exact verification target: as detailed in Appendix~\ref{benchmark}, conditions that source papers verified \textit{only approximately} are restated as the \textit{strict conditions} of Table~\ref{tab:task_certificate_mapping}. Second, it systematically adds high-dimensional, control-constrained systems.

\subsection{Data Collection and Curation}

To construct a non-redundant benchmark, we collected 233 candidate entries from 15 representative papers and BarrierBench~\citep{taheri2025barrierbench}. We then applied five curation rules: merging exact duplicates, consolidating multiple configurations of the same dynamics, keeping one representative per system family, removing structurally redundant systems, and excluding out-of-scope cases (Appendix~\ref{benchmark}, Table~\ref{tab:curation-reasons})---which left 80 distinct core systems (Table~\ref{tab:source-system-distribution}). Notably, only 7 of the 29 nominally high-dimensional entries ($n \geq 5$) survive curation (3 polynomial and 4 non-polynomial). To address this scarcity, we add two tiers beyond the \textbf{Literature-derived} core: \textbf{Extended} instances, which scale the state dimension, enlarge the domain, or add control constraints or perturbations to retained systems, and \textbf{New} instances, which are constructed high-dimensional systems. Unlike BarrierBench, which is limited to safety verification, \benchmarkname{} spans all four tasks in Table~\ref{tab:task_certificate_mapping}.

\subsection{Dataset Statistics and Analysis}

\begin{figure}[t]
    \centering
    \begin{minipage}{\linewidth}
        \centering
        \includegraphics[width=1\linewidth]{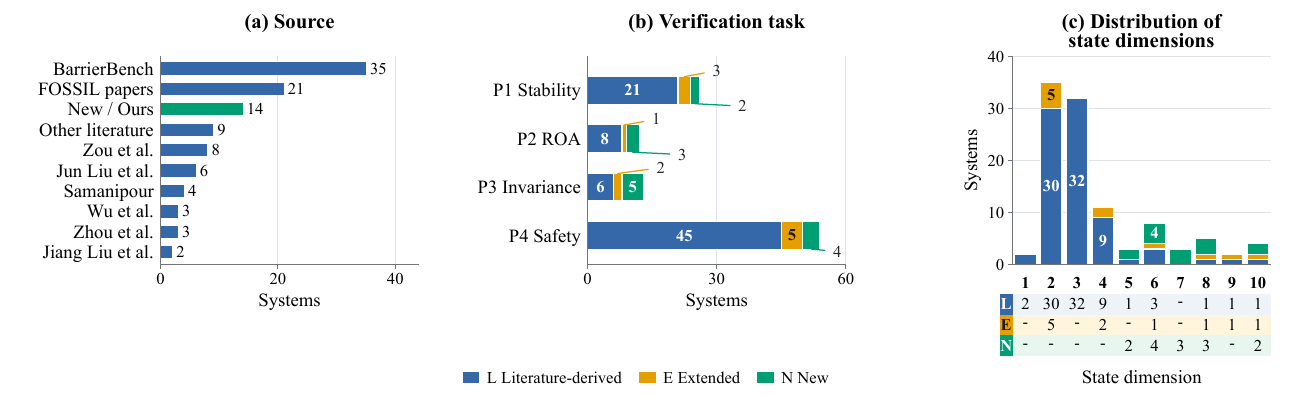}
        \caption{\benchmarkname{} overview by (a) source, (b) verification task, and (c) state dimension, split into the Literature-derived (L), Extended (E), and New (N) tiers.} 
        \label{fig:benchmark}
    \end{minipage}
\end{figure}

Figure~\ref{fig:benchmark} summarizes the full benchmark by source, task, and state dimension. The Extended and New tiers concentrate the high-dimensional instances: they raise the number of instances with $n\ge5$ from 7 to 25. In Section~\ref{experiments} we report results per tier, so that the added high-dimensional instances are not conflated with the literature-derived core. 

%% file: sections/experiments.tex
\section{Experiments}
\label{experiments}

\subsection{Experimental Setup}

\paragraph{Benchmarks and evaluation protocol.}

We evaluate \methodname{} on \benchmarkname{} (Section~\ref{dynacertbench}), reporting results separately for the Literature-derived, Extended, and New tiers. The benchmark covers stability, region-of-attraction, forward-invariance, and safety tasks for both continuous- and discrete-time systems. All methods are evaluated against the same original verification targets. A candidate counts as verified only if every obligation is discharged by a definitive SMT \texttt{unsat} result or a kernel-checked Lean proof.

\textbf{Baselines.} We compare \methodname{} with five pipelines that pair different synthesizers and verifiers (Table~\ref{tab:main_experiments}(a)). Method~1 synthesizes polynomial certificates by SMT-based CEGIS with Z3 and supports only polynomial systems~\citep{ahmed2020automated}. Method~2, FOSSIL~2.0, trains neural candidates by CEGIS and verifies them with its native SMT verifier~\citep{edwards2024fossil}. Method~3, the neural-CBF method of \citet{vertovec2025scalable}, learns neural candidates and certifies them with linear bound propagation (LBP). Method~4, the RL-based method of \citet{zou2025analytical}, generates analytic Lyapunov expressions and verifies them with its native falsification and SMT pipeline. Method~5, SynNLF, learns a polynomial neural Lyapunov candidate and checks it through SOS constraints encoded as LMI feasibility problems~\citep{zhao2025learning}. For each baseline we report \emph{Native}, the number of systems it verifies under its own conditions (e.g., with an excluded origin neighborhood or a $\delta$-\texttt{sat} tolerance), and \emph{Strict}, the number of these native certificates that remain accepted under unrelaxed conditions using the corresponding SMT, LBP, or SOS certification (Appendix~\ref{app:implementation_details}).

\paragraph{Comparisons.}

\textbf{(1) End-to-end comparisons.} We compare four configurations using the same implementation: \emph{One-shot Proposal}, which generates one candidate without any feedback; \methodname{} (SMT only); \methodname{} (Lean only); and the full \methodname{} (SMT with Lean). External tools are evaluated on their compatible benchmark subsets.  \textbf{(2) Fixed candidate for backend analysis.} To isolate the effect of the verification backend, RQ1--RQ3 use a fixed candidate: we freeze the candidate certificates and the exact verification obligations produced by the full \methodname{} run. Thus, all strategies are compared on identical proof obligations. This is necessary because different obligations would confound the comparison of verification backend. Overall, the former includes candidate search and feedback, whereas the latter holds the candidate certificates and verification obligations fixed to distinguish backend behavior from changes in candidate search. 

\paragraph{Metrics and implementation.}

We report the number of verified systems, wall-clock time. The Proposal Agent proposes at most $K_{\text{cand}}=5$ candidates, one per round. 
For each obligation, the Lean Proof-Repair Agent uses at most $K_{\text{lean}}=6$ rounds within $900\,\mathrm{s}$. All methods are limited to a total wall-clock time of $T_{\text{run}}=2400\,\mathrm{s}$ per system.
The full \methodname{} run took $2.5\,\mathrm{h}$.
On the same fixed certificates (RQ1/RQ2), All-SMT took $0.25\,\mathrm{h}$, All-Lean took $1.4\,\mathrm{h}$ and the RQ3 repair sweep took $2.1\,\mathrm{h}$.
These runs took about $6.2\,\mathrm{h}$ of wall-clock time in total. The reported experiments record $7.83$M model tokens in total. Further implementation details are given in Appendix~\ref{app:implementation_details}.

\subsection{Main experiment: end-to-end synthesis and verification}

\begin{table*}[t]
\centering
\small
\setlength{\tabcolsep}{6pt}
\caption{End-to-end synthesis and verification results. (a) Comparison with existing pipelines under the same time limit; each baseline runs on the subset of systems it supports, and the bottom row aggregates over the union of these subsets. (b) Ablations of \textsc{LLM2Cert} on the full 105-system benchmark under strict conditions.}
\label{tab:main_experiments}
\begin{tabular*}{\textwidth}{@{\extracolsep{\fill}}lllrrrr@{}}
\multicolumn{7}{@{}l}{\textit{(a) Comparison with existing end-to-end methods}}\\
\toprule
Method & Synthesis & Verification & Systems
& Native & Strict & \methodname\\
\midrule
1 & SMT(Z3) & SMT(Z3) & 8 & 0/8 & -- & 8/8 \\
2 & NN CEGIS & SMT & 49 & 8/49 & 5/8 & 38/49 \\
3 & NN CEGIS & LBP & 8 & 4/8 & 4/4 & 7/8 \\
4 & RL & SMT & 7 & 1/7 & 0/1 & 5/7 \\
5 & Poly.\ NN & SOS/LMI & 8 & 7/8 & 3/7 & 8/8 \\
\midrule
\multicolumn{3}{@{}l}{Verified by at least one baseline}
& 51 & 19/51 & 11/19 & 40/51 \\
\bottomrule
\end{tabular*}
\begin{tabular*}{\textwidth}{@{\extracolsep{\fill}}lcccc@{}}
\toprule
\multicolumn{5}{@{}l}{\textit{(b) End-to-end ablations for \methodname }}\\
Method & Literature & Extended
& New & Overall \\
\midrule
One-shot Proposal\textsuperscript{\P} & 54/80 & 4/11 & 5/14 & 63/105\\
\methodname{} (SMT only) & 53/80 & 3/11 & 1/14 & 57/105\\
\methodname{} (Lean only) & 57/80 & 2/11 & 1/14 & 60/105\\
\methodname{} (Full) & \textbf{66/80} & \textbf{7/11} & \textbf{8/14} & \textbf{81/105}\\
\bottomrule
\end{tabular*}
\begin{minipage}{\textwidth}
\footnotesize
\textit{Note.}
\textsuperscript{\P}This row is reconstructed from the successful runs that did not require feedback.
\end{minipage}
\end{table*}

\begin{figure}[t]
    \centering
    \includegraphics[width=1\linewidth]{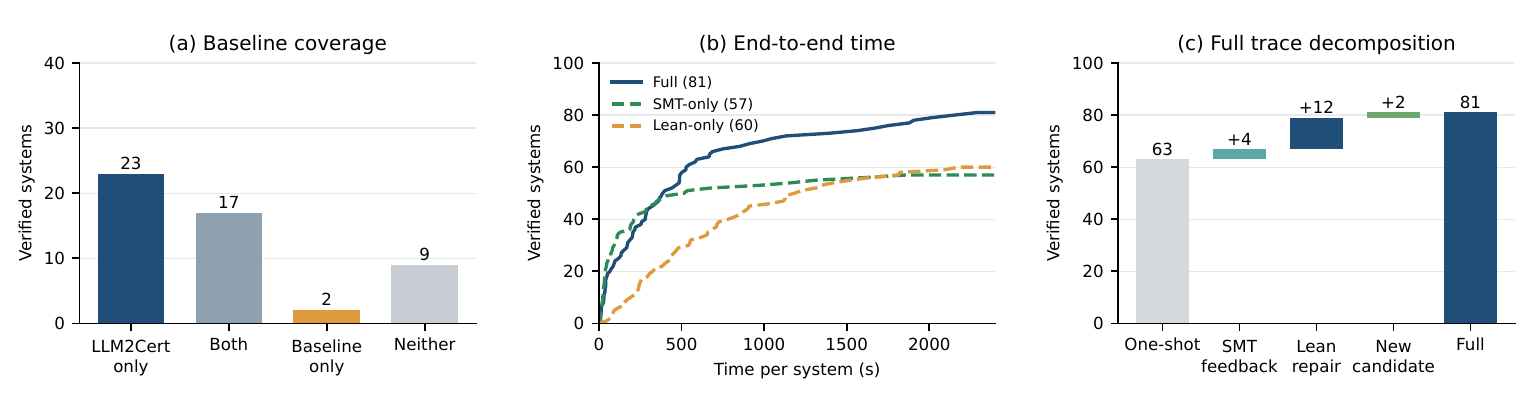}
    \caption{ \textbf{End-to-end results.} (a) Systems verified by \methodname{} and by the union of baselines on the 51 baseline-supported systems. (b) Verified systems vs.\ per-system wall-clock time for the three \methodname{} configurations. (c) Gain contributed by each component over one-shot proposal.}
    \label{fig:rq1_runtime_and_iterations}
\end{figure}

\textbf{How does \methodname{} compare with existing end-to-end synthesis methods?}
Table~\ref{tab:main_experiments}(a) compares complete synthesis and verification pipelines on their supported benchmark subsets.
\emph{Native} reports each baseline under the conditions stated in its paper, and \emph{Strict} does not permit any change to the obligations.
The baselines verify 19/51 instances natively, and 11 of these native successes remain valid under strict conditions (4 certificates are $\delta$-\texttt{sat}, and 4 SynNLF certificates fail the strict recheck), whereas \methodname{} verifies 40/51 instances.

Among these 51 systems, \methodname{} verifies 23 that no baseline verifies natively, and 17 that a baseline also verifies. Conversely, the baselines verify two systems that \methodname{} leaves unresolved. This suggests that combining neural certificate synthesis with \methodname{} is a promising direction for future research. Overall, these results demonstrate broader end-to-end coverage of \methodname{} under unchanged obligations.

\subsection{End-to-end Ablations}

\textbf{What do the feedback loop and Lean contribute to \methodname?}  LLM proposal alone is not sufficient. Table~\ref{tab:main_experiments}(b) shows that One-shot Proposal verifies only 63/105 instances, whereas the full \methodname{} verifies 81/105, an improvement of 15.0, 27.3, and 21.4 percentage points on the Literature-derived, Extended, and New tiers, respectively. The SMT-only configuration corresponds to the setup of BarrierBench~\citep{taheri2025barrierbench}, which also feeds SMT counterexamples back to the LLM. SMT-only and Lean-only verify 57/105 and 60/105 systems, respectively, whereas the full framework reaches 81/105, showing that neither backend alone recovers the coverage of their integration. \methodname{} also verifies 8 of the 10 systems that \citet{taheri2025barrierbench} report as unsolved. Fig.~\ref{fig:rq1_runtime_and_iterations}(b) shows verified systems over time for the three configurations, and Fig.~\ref{fig:rq1_runtime_and_iterations}(c) attributes the gain over one-shot proposal to SMT feedback, Lean proof repair, and new candidates. These results show that feedback and proof repair are most effective when combined within the full framework: counterexamples and unresolved obligations guide subsequent candidates, while routing and structured proof repair discharge the obligations that one-shot synthesis leaves open. The results also demonstrate that Lean and SMT play complementary roles in \methodname{}.

\subsection{Verification Framework Analysis}

To separate the verification backend from candidate search, RQ1--RQ3 use fixed candidates. \textbf{RQ1} asks when Lean adds value over SMT-only verification; \textbf{RQ2} asks whether per-obligation routing improves success and cost; \textbf{RQ3} asks whether structured, compiler-guided feedback improves Lean proof repair. RQ1 and RQ2 fix the 81 certificates verified by the full framework, together with their 296 obligations, and RQ3 fixes the failed first Lean proof attempts recorded in the full run. An analysis of the 24 unresolved systems is given in Appendix~\ref{app:unresolved}.

\textbf{RQ1: When does Lean add value beyond SMT-only verification?} We compare strategies on the same obligations: \textsc{All-SMT}, \textsc{All-Lean}, and \textsc{Routed}. \textsc{All-SMT} and \textsc{All-Lean} allow $900\,\mathrm{s}$ per obligation. Table~\ref{tab:fixed_verifier} shows that Lean's additional coverage is on the non-polynomial systems. 
\label{sec:rq1} 

All-SMT verifies 59/81 certificates and All-Lean verifies 64/81. The two backends cover complementary coverage. \textsc{All-Lean} verifies 21 certificates that \textsc{All-SMT} leaves unresolved, all of them non-polynomial. \textsc{All-SMT} resolves additional certificates for which Lean proof search does not finish within the budget. On polynomial systems, SMT covers all certificates verified by Lean, whereas Lean contributes exclusive successes on the non-polynomial subset: Lean verifies 34/40, compared with 18/40 for SMT. This complementary behavior motivates obligation routing: SMT handles obligations in its effective fragments efficiently, while Lean extends coverage beyond them. 

\textbf{RQ2: Does per-obligation routing improve success and time cost?} As shown in the Table~\ref{tab:fixed_verifier}, the two backends have different coverage time profiles. \textsc{All-SMT} obtains its coverage early: 58 of the 81 fixed certificates are verified within $30\,\mathrm{s}$ per obligation, and this number rises only to 59 at $900\,\mathrm{s}$. \textsc{All-Lean} verifies 64/81. On the systems it verifies, its mean time is $511.99\,\mathrm{s}$ for polynomial systems and $638.44\,\mathrm{s}$ for non-polynomial systems, against $1.65\,\mathrm{s}$ and $8.30\,\mathrm{s}$ for \textsc{All-SMT}. 17 certificates remain unresolved by \textsc{All-Lean}. SMT is thus fast where it succeeds. This complementarity motivates per-obligation routing: SMT handles compatible checks, while Lean handles directly assigned obligations and those left unresolved by SMT.
\label{sec:rq2}

\textsc{Routed} implements this strategy and verifies all 81 certificates in the end-to-end experiment. Of the 296 obligations, SMT verifies 253 (85.5\%) by \texttt{unsat}, and 43 reach Lean (14.5\%): 21 after an SMT result other than \texttt{unsat} (8 $\delta$-\texttt{sat}, 13 timeout), 19 by a direct router decision, and 3 with no SMT encoding (radial unboundedness). Selective use of Lean thus resolves the remaining hard obligations and covers more than either backend alone at a fraction of the all-Lean cost (Table~\ref{tab:fixed_verifier}).

\begin{table}[t]
\centering
\small
\setlength{\tabcolsep}{2.3pt}
\caption{Fixed-candidate verification on the 81 certificates verified by the full framework, split by polynomial and non-polynomial systems.}
\label{tab:fixed_verifier}
\begin{tabular*}{\linewidth}{@{\extracolsep{\fill}}l*{6}{c}}
\toprule
& \multicolumn{3}{c}{Polynomial ($N_{\mathrm{sys}}=41$)}
& \multicolumn{3}{c}{Non-polynomial ($N_{\mathrm{sys}}=40$)} \\
\cmidrule(lr){2-4}\cmidrule(lr){5-7}
Strategy & Verified & Time/sys.  & Lean/sys. & Verified & Time/sys.  & Lean/sys.  \\
\midrule
\textsc{All-SMT} & \textbf{41/41} & \textbf{1.65} & 0.0\% & 18/40 & \textbf{8.30} & 0.0\% \\
\textsc{All-Lean} & 30/41 & 511.99 & 100.0\% & 34/40 & 638.44 & 100.0\% \\
\textsc{Routed} & \textbf{41/41} & 2.49 & 0.0\% & \textbf{40/40} & 239.05 & 72.5\% \\
\bottomrule
\end{tabular*}
\begin{minipage}{\linewidth}
\textit{Note.} Time/sys.: mean verification time (s) over each strategy's verified systems. Lean/sys.: fraction of systems with at least one obligation routed to Lean.
\end{minipage}
\end{table}

\textbf{RQ3: Does structured repair feedback improve Lean proof completion?} Lean is the expensive backend, so the cost of \methodname{} depends on how quickly the Lean Proof-Repair Agent converges. The agent used in RQ1--RQ2 already receives our structured repair feedback (Section~\ref{sec:lean-repair}); RQ3 isolates the contribution of this feedback by replacing it with less informative alternatives. 
\label{sec:rq3}

On the 19 systems where the full run repaired at least one Lean obligation, 26 obligations failed their first proof attempt: 18 after SMT and 8 through direct routing. Of these, 21 were eventually repaired and 5 remained unresolved. We compare \textsc{Structured Repair} with \textsc{Generic Retry} and \textsc{Raw Diagnostics}. \textsc{Generic Retry} reports only failure, and \textsc{Raw Diagnostics} returns all the unprocessed Lean compiler output. We set a same $900,\mathrm{s}$ budget for every obligation.

\begin{table}[t]
\centering
\small
\setlength{\tabcolsep}{4pt}
\caption{Lean proof repair under three feedback strategies ($N_{\mathrm{obl}}=26$).}
\label{tab:repair}
\begin{tabular*}{\linewidth}{@{\extracolsep{\fill}}lccc@{}}
\toprule
Strategy & Repaired obl. & Rounds/obl. & Time/obl. \\
\midrule
\textsc{Generic Retry} & 1/26 & 3.58 & 873.66 \\
\textsc{Raw Diagnostics} & 17/26 & 2.35 & 491.02 \\
\textsc{Structured Repair}
& \textbf{21/26} & \textbf{2.19} & \textbf{407.42} \\
\bottomrule
\end{tabular*}
\begin{minipage}{\linewidth}
\textit{Note.}
Times in seconds. Unrepaired obligations are charged the full 900\,s budget. All three strategies are evaluated from the same frozen failed first attempts.
\end{minipage}
\end{table}

Table~\ref{tab:repair} shows that feedback quality substantially affects proof repair. \textsc{Generic Retry} repairs 1/26 obligations and no complete system, whereas \textsc{Raw Diagnostics} repairs 17/26 obligations and 13/19 systems. \textsc{Structured Repair} achieves the highest completion rate, repairing 21/26 obligations and 15/19 systems, while requiring fewer rounds and reducing mean repair time to $407.42\,\mathrm{s}$ per obligation. Structured feedback improves both proof completion and verification efficiency.

%% file: sections/conclusion.tex
\section{Conclusion}

We studied certificate synthesis and verification for dynamical systems end-to-end. Solver limitations have long pushed both certificate verification and benchmark design toward conservative relaxations of the true stability and safety conditions. We instead keep the original, unrelaxed certificate conditions as a fixed target and introduce \methodname, an agentic framework that synthesizes candidate certificates and verifies every obligation either by an SMT solver or by a kernel-checked Lean proof, routing per obligation and falling back to Lean when solvers time out or return $\delta$-\texttt{sat}. On \benchmarkname, a curated benchmark that restores the high-dimensional, constrained, and nonlinearly coupled systems that solver-oriented datasets have excluded, \methodname{} verifies 77.1\% of instances end to end and, where solvers alone fail, obtains machine-checked proofs of the exact conditions rather than of a relaxed surrogate. We see this as a foundational first step toward kernel-checked proof as a practical verifier for synthesized dynamical-system certificates.

%% file: sections/limitations_and_future_work.tex
\section{Limitations and Future Work}
\label{app:limitations}

Our current framework has several limitations that suggest natural directions for future work. First, \benchmarkname, while broader than prior benchmarks along the dimensionality and coupling axes, remains a curated collection of 105 instances rather than an exhaustive sample of certificate-synthesis problems. Extending the benchmark to a wider range of high-dimensional, hybrid, and uncertain dynamics would provide a more comprehensive evaluation of formal certificate synthesis. Second, the present formulation focuses primarily on deterministic dynamical systems; extending the framework to stochastic systems with random disturbances and probabilistic guarantees is an important next step. Third, the current synthesis loop targets certificates for prescribed verification domains rather than explicitly optimizing the certified set itself. Incorporating region-size objectives into the proposal and verification loop could enable the formal synthesis of larger regions of attraction and invariant sets.

%% file: sections/related_work.tex
\section{Related work}
\label{sec:related_work}

\paragraph{Certificate synthesis and learner--verifier loops.}

Functional certificates turn trajectory-level properties into inequalities over the state space, but the certificate itself must still be discovered.  Certificate synthesis typically restricts the witness to a parameterized family, using SOS or semidefinite relaxations for polynomial templates \citep{papachristodoulou2002construction,prajna2007framework} and bilinear or difference-of-convex formulations when certificates interact with multipliers or controllers \citep{xue2019robust,wang2022encoding,xue2023reach}.
Learner--verifier loops instead alternate candidate search with global counterexample generation.  This pattern spans polynomial and SMT-guided synthesis \citep{ravanbakhsh2019learning,ahmed2020automated} as well as neural Lyapunov functions, barrier certificates, and controller co-synthesis \citep{chang2019neural,abate2020formal,peruffo2021automated,dai2021lyapunov,abate2021fossil,edwards2024fossil}.
The candidate representation determines the verifier: analytic expressions favor nonlinear SMT, piecewise-linear networks admit mixed-integer encodings, and neural certificates motivate activation-region or bound-propagation methods \citep{gao2013dreal,wu2023neural,zhang2023exact,zhang2024seev,vertovec2025scalable,li2025twostage}.
Language models have recently proposed and repaired symbolic barriers using numerical, SOS, or SMT feedback \citep{taheri2025barrierbench,ren2026generative}.
Across these approaches, soundness rests on the final verifier, while template choices, strengthenings, excluded neighborhoods, and numerical approximations can change the proposition actually certified.

\paragraph{Formal Verification of Dynamical Systems.}

For a fixed certificate, SMT checks the existential negation of universal inequalities, including nonlinear transcendental formulas through delta-complete procedures \citep{gao2013dreal}.
Algebraic methods seek checkable nonnegativity decompositions, whereas deductive hybrid-system provers reason about differential and discrete transitions in a specialized calculus \citep{papachristodoulou2002construction,prajna2007framework,fulton2015keymaerax}.
These artifacts provide different assurance: a delta-qualified verdict, an exact algebraic certificate, and a deductive proof are not interchangeable.
Proof assistants further reduce the trusted base by checking explicit proof terms.  Control-theoretic developments include a Rocq stability proof for an inverted-pendulum controller and constructive checking of exact SOS Lyapunov certificates in Minlog \citep{rouhling2018formal,devadze2025computer}.
Recent Lean work formalizes general stability theory, neural-network semantics, and checked bounds for learned controllers \citep{doll2026foundations,george2026torchlean}.
Related verification pipelines also check learned controllers against Lyapunov or barrier conditions \citep{mandal2024formally}.
Our setting connects these layers: it freezes each synthesized candidate's original contract, decomposes it into obligations, and seeks a replayable Lean proof rather than only a solver verdict.

\paragraph{LLMs for Formal and Agentic Reasoning.}

LLMs support autoformalization \citep{wu2022autoformalization} and proof generation through retrieved premises, tactic suggestions, synthetic data, and reinforcement learning \citep{yang2023leandojo,song2024lean,xin2024deepseek,hubert2026olympiad,george2025leanprogress}.
Agentic systems additionally exploit proof-state history and backtracking \citep{thakur2024copra,kumarappan2025leanagent} or repair failed subproofs from compiler diagnostics \citep{ospanov2025apollo,wang2026learning}.
Theorem-proving environments have also been used for LLM-assisted program verification \citep{lin2024fvel,zhao2026algoveri,ye2026verina,thakur2026clever,ye2026vero,li2026p}.
Most such systems begin with a fixed formal theorem \citep{collins2026characterizing,yang2026formal}.  Certificate synthesis adds an earlier semantic choice: the agent must select a witness whose exact stability or safety obligations are both true and provable.  We therefore separate candidate synthesis, route selection, and Lean proof construction, returning structured failures without allowing the loop to change the frozen theorem.

%% file: sections/dynamical_models_and_notation.tex
\section{Dynamical Models and Notation}
\label{sec:dynamical_models}

Let $\mathbb{R}$ and $\mathbb{R}^n$ denote the sets of real numbers and $n$-dimensional real vectors, respectively. We consider continuous-time and discrete-time dynamical systems of the form
\begin{align}
    \dot{x}(t) &= f(x(t),u(t)),                                  \label{eq:ct_system}\\
    x_{k+1} &= F(x_k,u_k),                                      \label{eq:dt_system}
\end{align}
where $ u \in \mathcal{U}\subseteq\mathbb{R}^m$ denotes the control input, $f:\mathbb{R}^n\times\mathcal{U}\to\mathbb{R}^n$ and $F:\mathbb{R}^n\times\mathcal{U}\to\mathbb{R}^n$. Given a state-feedback policy $\pi:\mathbb{R}^n\to\mathcal{U}$, define the closed-loop dynamics as
\[
    f_\pi(x):=f(x,\pi(x)),
    \qquad
    F_\pi(x):=F(x,\pi(x)).
\]

\subsection{System Properties}
\label{sec:system_properties}

Throughout this paper, $t\geq 0$ denotes $t\in[0,\infty)$ for continuous-time systems and $t\in\mathbb{N}_0$ for discrete-time systems. 
Given a policy $\pi$ and an initial state $x_0\in\mathbb{R}^n$, let $\phi_\pi^{x_0}$ denote the corresponding closed-loop trajectory, which satisfies $\phi_\pi^{x_0}(0)=x_0$ and
\[
    \dot{\phi}_\pi^{x_0}(t)=f_\pi\bigl(\phi_\pi^{x_0}(t)\bigr)
\]
for continuous-time systems, and
\[
    \phi_\pi^{x_0}(t+1)=F_\pi\bigl(\phi_\pi^{x_0}(t)\bigr)
\]
for discrete-time systems. We assume that the closed-loop trajectory is uniquely defined for all future time.

\begin{definition}[Equilibrium point]
A state $x^\ast\in\mathbb{R}^n$ is an equilibrium point of the closed-loop system if $f_\pi(x^\ast)=0$ in continuous time or $F_\pi(x^\ast)=x^\ast$ in discrete time. Without loss of generality, we take the equilibrium point of interest to be the origin.
\end{definition}

\begin{problem}[Stability certification]
\label{prob:stability}
Given the equilibrium point $x^\ast=0$, determine whether it is stable, asymptotically stable, or exponentially stable under the closed-loop dynamics.
\end{problem}

\begin{problem}[Region-of-attraction verification]
\label{prob:roa}
Given a set of initial states $\mathcal{X}_I\subseteq\mathbb{R}^n$, determine whether
\[
    \lim_{t\to\infty} \phi_\pi^{x_0}(t) = 0,
    \qquad
    \forall x_0\in\mathcal{X}_I.
\]
If the property holds, we write $\mathcal{X}_I\subseteq\mathcal{R}_A(0)$.
\end{problem}

\begin{problem}[Forward-invariant set synthesis]
\label{prob:set_synthesis}
Given a domain of interest $\mathcal{X}\subseteq\mathbb{R}^n$ and an unsafe set $\mathcal{X}_U\subseteq\mathcal{X}(\text{or} \;\mathcal{X}_U=\mathbb{R}^n\setminus\mathcal{X})$, find a nonempty set $\mathcal{S}\subseteq\mathcal{X}$ such that $\mathcal{S}$ is disjoint from $\mathcal{X}_U$ and is forward invariant under the dynamics, i.e., every trajectory starting from $\mathcal{S}$ remains in $\mathcal{S}$ for all future time. Formally:
\[
\mathcal{S}\cap\mathcal{X}_U=\varnothing
\quad\text{and}\quad
\phi_\pi^{x_0}(t)\in\mathcal{S},\quad
\forall x_0\in\mathcal{S},\ \forall t\ge0.
\]
\end{problem}

\begin{problem}[Safety verification, \cite{prajna2007framework}]
\label{prob:safety}
Given a domain of interest $\mathcal{X}\subseteq\mathbb{R}^n$, an initial set $\mathcal{X}_I\subseteq\mathcal{X}$, and an unsafe set $\mathcal{X}_U\subseteq\mathcal{X}$ with $\mathcal{X}_I\cap\mathcal{X}_U=\varnothing$, determine that there does not exist any time instant $T\ge 0$ and initial state $x_0\in\mathcal{X}_I$ such that the trajectory $\phi_\pi^{x_0}(t)$ remains in $\mathcal{X}$ for all $t\in[0,T]$ but reaches $\mathcal{X}_U$ at time $T$, i.e., 
\[
\forall x_0\in \mathcal{X}_I, \forall T \in (0,\infty), \left[\forall t\in[0,T], \phi_\pi^{x_0}(t)\in\mathcal{X}\right] \implies \left[\forall t\in[0,T],\phi_\pi^{x_0}(t)\in\mathcal{X}\setminus \mathcal{X}_U\right]
\]

\end{problem}

\begin{remark}
\label{rem:unsafe_set_choice}
We consider two unsafe-set formulations. (1) The unsafe set is a subset of the domain, $\mathcal{X}_U\subseteq\mathcal{X}$. This is a common way to formulate safety specifications: unsafe states are given explicitly and trajectories must avoid them while they remain in $\mathcal{X}$ \cite{abate2021fossil,vertovec2025scalable}. (2) A safe set is prescribed and trajectories must remain within it \cite{mathiesen2022safety,xue2026sufficient}. This formulation is recovered by taking $\mathcal{X}$ as the safe set and setting $\mathcal{X}_U=\mathbb{R}^n\setminus\mathcal{X}$. More generally, some formulations include both a safe set and an unsafe set, which need not be complementary. 
\end{remark}

The Lyapunov conditions in Theorem~\ref{thm:lyapunov} and Corollary~\ref{cor:roa} address Problems~\ref{prob:stability} and~\ref{prob:roa}. For the two safety-related problems, Theorem~\ref{thm:barrier_safety} establishes the basic barrier conditions, and Corollaries~\ref{cor:barrier_set_synthesis} and~\ref{cor:barrier_safety} give sufficient conditions for Problems~\ref{prob:set_synthesis} and~\ref{prob:safety}, respectively.

\subsection{Certificate Conditions}
\label{sec:certificate_conditions}

For a scalar certificate function $W:\mathbb{R}^n\to\mathbb{R}$, continuously differentiable in the continuous-time case, define
\begin{equation}
    \mathcal{D}_\pi W(x)
    :=
    \begin{cases}
        \nabla W(x)^\top f_\pi(x),
            & \text{continuous time},\\[1mm]
        W(F_\pi(x))-W(x),
            & \text{discrete time}.
    \end{cases}
    \label{eq:certificate_change}
\end{equation}

\paragraph{Lyapunov certificates.}

\begin{theorem}[\cite{khalil2002nonlinear}]
\label{thm:lyapunov}
\normalfont
Let $\mathcal{N}\subseteq\mathbb{R}^n$ be a neighborhood of the origin, and let $V:\mathbb{R}^n\to\mathbb{R}$ be continuous, and continuously differentiable in the continuous-time case. Suppose that
\[
    V(0)=0,
    \qquad
    V(x)>0
    \quad
    \text{for all }x\in\mathcal{N}\setminus\{0\}.
\]
Then:
\begin{enumerate}
    \setlength{\itemsep}{1pt}
    \setlength{\parskip}{0pt}
    \setlength{\parsep}{0pt}

    \item If $\mathcal{D}_\pi V(x)\leq0$ for all
    $x\in\mathcal{N}$, then the origin is stable in the sense of Lyapunov on $\mathcal{N}$.

    \item If $\mathcal{D}_\pi V(x)<0$ for all $x\in\mathcal{N}\setminus\{0\}$, then the origin is asymptotically stable on $\mathcal{N}$.

    \item If $V$ is positive definite and radially unbounded on $\mathbb{R}^n$, and $\mathcal{D}_\pi V(x)<0$ for all $x\neq0$, then the origin is globally asymptotically stable.

    \item Local exponential stability:
    If there exist $c_1,c_2>0$ and $c_3>0$ such that
    \[
        c_1\|x\|^2 \leq V(x) \leq c_2\|x\|^2,
        \qquad
        \mathcal{D}_\pi V(x) \leq -c_3 V(x),
        \qquad
        \forall x\in\mathcal{N},
    \]
    then the origin is exponentially stable.
    In discrete time, the condition $\mathcal{D}_\pi V(x) \leq -c_3 V(x)$
    is replaced by $V(F_\pi(x)) \le \rho\,V(x)$ for some $\rho\in(0,1)$,
    which equivalently means $c_3\in(0,1)$ in the above form.
    Global exponential stability holds if these inequalities are
    satisfied on $\mathbb{R}^n$.
\end{enumerate}
\end{theorem}

\begin{corollary}[Region-of-attraction certificate]
\label{cor:roa}
\normalfont
Assume that the origin is an equilibrium of the closed-loop system.
Let $\mathcal{N}\subseteq\mathbb{R}^n$ be a neighborhood of the origin,
let $V:\mathbb{R}^n\to\mathbb{R}$ be continuous and, in the
continuous-time case, continuously differentiable, and let $c>0$.
Define the sublevel set $\Omega_c := \{x\in\mathbb{R}^n : V(x)\leq c\}$.
Suppose that:
\begin{enumerate}
\setlength{\itemsep}{1pt}
\setlength{\parskip}{0pt}
\setlength{\parsep}{0pt}

\item $V(0)=0$.

\item $V(x)>0$ for all
$x\in\Omega_c\setminus\{0\}$.

\item $\mathcal{D}_\pi V(x)<0$ for all
$x\in\Omega_c\setminus\{0\}$.

\item $\Omega_c$ is compact and satisfies $\mathcal{X}_I\subseteq\Omega_c\subseteq\mathcal{N}$.

\end{enumerate}
Then $\mathcal{X}_I\subseteq\mathcal{R}_A(0)$, i.e., $\Omega_c$ is a region of attraction.
\end{corollary}

\paragraph{Barrier certificates.}

We consider the following barrier certificate conditions for continuous-time and discrete-time systems \cite{prajna2007framework,agrawal2017discrete,ames2019control}.

\begin{theorem}
\label{thm:barrier_safety}
\normalfont
Let $\mathcal{X}\subseteq\mathbb{R}^n$ and $\mathcal{X}_U\subseteq\mathcal{X}(\text{or} \;\mathcal{X}_U=\mathbb{R}^n\setminus\mathcal{X})$. Let
$B:\mathbb{R}^n\to\mathbb{R}$ be continuous and, in the
continuous-time case, continuously differentiable. Suppose that:
\begin{enumerate}[nosep, leftmargin=*]
    \item $B(x)<0$ for all $x\in\mathcal{X}_U$;
    \item \textbf{Continuous time:} for all $x\in\mathcal{X}$ with
    $B(x)=0$, $\nabla B(x)\neq0$ and
    $\mathcal{D}_\pi B(x)>0$;
    \item \textbf{Discrete time:} for all $x\in\mathcal{X}$ with
    $B(x)\geq0$, $B(F_\pi(x))\geq0$.
\end{enumerate}
Then there do not exist $x_0\in\mathcal{X}$ with $B(x_0)\geq0$ and
$T\geq0$ such that
\[
    \phi_\pi^{x_0}(T)\in\mathcal{X}_U
    \quad\text{and}\quad
    \phi_\pi^{x_0}(t)\in\mathcal{X},
    \qquad
    \forall t\in[0,T).
\]
\end{theorem}
Define the superlevel set $\mathcal{C}_B:=\{x\in\mathbb{R}^n:B(x)\geq0\}$.
\begin{corollary}[Forward-invariant set synthesis]
\label{cor:barrier_set_synthesis}
\normalfont
If $B$ satisfies the conditions of Theorem~\ref{thm:barrier_safety}
and $\emptyset\neq\mathcal{C}_B\subseteq\mathcal{X}$, then
$\mathcal{C}_B$ is forward invariant and
$\mathcal{C}_B\cap\mathcal{X}_U=\varnothing$. Hence,
$\mathcal{S}=\mathcal{C}_B$ is a solution to
Problem~\ref{prob:set_synthesis}.
\end{corollary}

\begin{corollary}[Safety verification]
\label{cor:barrier_safety}
\normalfont
If $B$ satisfies the conditions of Theorem~\ref{thm:barrier_safety}
and $B(x)\geq0$ for all $x\in\mathcal{X}_I$, then Problem~\ref{prob:safety} is verified.
\end{corollary}

\subsection{Verification Principles}

This paper uses two kinds of verification tools. 
\begin{enumerate}
    \item SMT solvers (Z3, cvc5, Yices, and dReal). All of them perform counterexample search: the negation of the universal condition $\forall x.\, P(x)$ is encoded as a satisfiability problem over the reals. If the solver returns \texttt{unsat}, no counterexample exists and the original condition holds. Z3, cvc5, and Yices give exact verdicts on decidable arithmetic fragments, whereas dReal performs a $\delta$-complete search for general nonlinear real arithmetic. A dReal $\delta$-\texttt{sat} result is inconclusive with respect to the original formula: it means that a $\delta$-weakened version of the problem is satisfiable, not that the original formula is.
    \item The proof assistant: Lean, which takes a fundamentally different approach. Rather than searching for counterexamples, Lean constructs a formal proof term for $\forall x.\, P(x)$ in dependent type theory. The condition is established once the proof term passes the kernel's type check. 
\end{enumerate} 
The essential distinction lies in the underlying foundations: SMT solvers appeal to model theory and establish the non-existence of a model over the reals, whereas Lean appeals to proof theory and establishes the existence of a derivation in a formal system. Whichever tool is used, once all certificate conditions have been verified, the corresponding Lyapunov or barrier theorem implies the desired stability or safety property.

%% file: sections/prompt.tex
\section{Prompts Details in \methodname}
\label{app:prompts}

This appendix gives the template-level prompt interfaces used by \methodname. Runtime values are injected by the harness at the marked placeholders. The complete serialized prompts, JSON schemas, and case traces are provided in the supplementary material. Table~\ref{tab:prompt-interface} summarizes the message roles and the evidence flow between the harness, the LLM, and the verification backends. Table~\ref{tab:harness-actions} summarizes how the harness processes LLM outputs.

\paragraph{Model and decoding configuration.}
The model, reasoning-effort setting, output-length limit, and budgets used in the reported runs are listed in Appendix~\ref{app:implementation_details}. The reported runs use \texttt{gpt-5.6-sol} through the OpenRouter chat-completions API, with reasoning effort \texttt{high}. The maximum output length is 16,384 tokens. Recorded model usage is reported in Appendix~\ref{app:experimental_details}.

\begin{table}[!htbp]
\centering
\small
\caption{Prompt interfaces and evidence flow.}
\label{tab:prompt-interface}
\setlength{\tabcolsep}{4pt}
\renewcommand{\arraystretch}{1.12}
\begin{tabularx}{\linewidth}{@{}L{0.18\linewidth}L{0.18\linewidth}*{2}{>{\raggedright\arraybackslash}X}@{}}
\toprule
Agent & Message role & Injected by harness & LLM output \\
\midrule
Proposal Agent & system + user & specification $R$, task profile, locked domain/sets, previous counterexample & JSON candidate $C^{(k)}=(W^{(k)},\pi^{(k)})$ \\
Lean Proof-Repair Agent & system + user & locked theorem, definitions, candidate hash, compiler diagnostics, Mathlib search results & JSON proof body, helper lemmas, or \texttt{cannot\_complete} \\
Local LeanRepair & user & frozen proof, target source span, compiler context, verified declarations & \texttt{LeanLocalPatch} or \texttt{cannot\_patch} \\
\bottomrule
\end{tabularx}
\end{table}

\begin{table}[!htbp]
\centering
\small
\caption{Harness processing of LLM outputs.}
\label{tab:harness-actions}
\begin{tabular}{p{0.22\linewidth}p{0.34\linewidth}p{0.34\linewidth}}
\toprule
LLM output & Harness action & Result \\
\midrule
Candidate JSON & format check, semantic check, compile to $A^{(k)}$ & obligation list $\mathcal{M}(A^{(k)})$ and fast screening \\
Lean proof body & elaboration in pinned Lean/Mathlib & accepted proof, repair feedback, or unresolved trace \\
\texttt{cannot\_complete} & check blocker and optional counterexample & continue proof search or return diagnostics to Proposal Agent \\
\bottomrule
\end{tabular}
\end{table}

\FloatBarrier
\Needspace{8\baselineskip}
The Proposal Agent uses one output protocol across tasks, and it checks whether $W^{(k)}$ is a Lyapunov certificate, a region-of-attraction certificate, or a barrier certificate. The Lean Proof-Repair Agent uses one locked obligation at a time and returns only proof bodies, helper lemmas, or a precise blocker.

\subsection{Proposal Agent}
\label{app:proposal-prompts}

As described in Section~\ref{sec:candidate_search}, the Proposal Agent receives the typed specification $R$ and proposes a candidate $C^{(k)}=(W^{(k)},\pi^{(k)})$ at iteration $k$. The candidate is first checked for format and semantic validity. Once accepted, it is compiled into the candidate representation $A^{(k)}$, which induces the obligation list $\mathcal{M}(A^{(k)})$. The same output protocol is used across tasks, while the task profile specifies whether $W^{(k)}$ is interpreted as a Lyapunov certificate $V$, a region-of-attraction certificate, or a barrier certificate $B$.

\Needspace{8\baselineskip}
\subsubsection{Initial Proposal Prompt}
\label{app:prompt-proposal}

\begin{promptblock}
\begin{promptbody}

\promptrole{Role}

You are a control theory expert specializing in nonlinear dynamical systems, certificate construction, and formal verification.

\promptrole{System information}

\promptplaceholder{The system name, problem identifier, dynamics, time semantics, state variables, control variables, formal domain, screening box, initial set, unsafe set, equilibrium information, fixed parameters, control set, and controller observation pattern are inserted here.}

\promptrole{Verification task}

\promptplaceholder{The task profile is inserted here. For stability and region-of-attraction tasks, the profile specifies the required Lyapunov conditions on $V$ and, when applicable, the level set $\Omega_c=\{x\mid V(x)\leq c\}$. For invariance and safety tasks, the profile specifies the required barrier conditions on $B$, including the initial-set, unsafe-set, boundary, or one-step conditions.}

At iteration $k$, propose a candidate certificate and, when applicable, a feedback controller. The candidate must satisfy the supplied theorem conditions rather than a relaxed numerical surrogate.

Use continuous-time Lie-derivative semantics only for continuous-time systems. For discrete-time systems, use only the one-step closed-loop map $F_{\mathrm{cl}}$ and the corresponding change in the certificate.

If fixed parameters are present, construct one state-only certificate that works uniformly over the full parameter interval. Parameters may appear in the dynamics and in the construction explanation, but not as additional certificate variables.

\promptrole{Candidate construction}

Use only the declared state variables. Every controller assignment must respect the supplied observation pattern and control bounds. Prefer a constant, linear, or affine controller before using a nonlinear controller, and avoid unnecessarily large gains.

Use exact rational constants and the supported expression language. The certificate and controller must be returned as concise expressions rather than recursive AST nodes.

The construction summary is a construction note, not a certificate. It must identify the useful auxiliary quantities, give a factorized form of the Lie expression or one-step change, and state the explicit bounds needed by the downstream SMT and Lean verification stages.

\promptrole{Output format}

Return exactly one JSON object matching the task-specific output contract. Return the certificate, controller assignments when applicable, a construction summary, and the required assumptions or comparison constants.

Do not return Markdown, code fences, recursive AST nodes, unsupported function names, or any text outside the JSON object.

\end{promptbody}
\end{promptblock}

\paragraph{Transition to screening.}
After format and semantic checking, the candidate is compiled into $A^{(k)}$ and induces the finite obligation list $\mathcal{M}(A^{(k)})$. Fast screening checks these obligations on a finite sample-based schedule. A detected violation rejects the candidate and is returned to the Proposal Agent as structured evidence. 
\Needspace{8\baselineskip}
\subsubsection{Proposal Repair Prompt}
\label{app:prompt-proposal-repair}

A rejected candidate is not repaired by changing the verification target. Only a confirmed and independently replayed counterexample authorizes a new candidate-generation turn. Solver timeouts and proof diagnostics are returned as diagnostic evidence.

\begin{promptblock}
\begin{promptbody}

\promptrole{Next turn}

The system message still contains the problem specification, task conditions, and output contract. Do not wait for these sections to be repeated. The following message contains only new evidence from the previous attempt.

Diagnose the failed condition and propose the next candidate. Change the expression structure when coefficient tuning cannot address the evidence. Preserve the continuous-time or discrete-time semantics of the task.

\promptrole{New evidence}

\promptplaceholder{The candidate index, failed candidate, independently verified counterexample, rejected candidate history, and parser or semantic errors are inserted here.}

\promptrole{Task-specific reminders}

For stability, a counterexample in the fixed claimed domain must be addressed by changing the certificate or controller; it may not be removed by introducing an unsupported level restriction.

For region-of-attraction verification, the level $c$ may be changed, but the new candidate must still satisfy the locked containment and Lyapunov obligations.

For invariance and safety, preserve the original working domain, initial set, unsafe set, and barrier semantics.

\promptrole{Output format}

Return exactly one JSON object matching the original task-specific output contract. Return concise certificate and controller expressions, not recursive AST nodes or explanatory text outside the JSON object.

\end{promptbody}
\end{promptblock}

\paragraph{Transition to formal verification.}
When an obligation is selected for Direct Lean, or when SMT leaves the obligation unresolved, the framework constructs a locked Lean theorem from the fixed candidate and obligation. The Lean agent is allowed to generate proof terms, but not to change $R$, $C^{(k)}$, $A^{(k)}$, the theorem statement, or the obligation list.

\subsection{Lean Proof-Repair Agent}
\label{app:lean-prompts}

The Lean Proof-Repair Agent operates on one obligation at a time. The framework compiles each response in the pinned Lean/Mathlib environment. The proof interface is proof-body-only: the theorem header and generated definitions remain fixed.

\Needspace{8\baselineskip}
\subsubsection{Initial Lean Proof Prompt}
\label{app:prompt-lean-initial}

\begin{promptblock}
\begin{promptbody}

\promptrole{Role}

Complete one fixed Lean~4/Mathlib verification condition.

\promptrole{Goal}

Return the complete term inserted after \texttt{:=} in \texttt{proof\_body}. The theorem statement, generated definitions, dynamics, certificate, controller, and candidate hash are locked.

For a Lyapunov obligation, prove exactly the stated equilibrium, positivity, comparison, Lie-derivative, one-step decrease, radial-unboundedness, or exponential-decay condition.

For a barrier or safety obligation, prove exactly the locked boundary, Lie-derivative, initial-set, unsafe-set, or one-step condition. For an existential violation statement, begin from the actual proposition, usually with \texttt{rintro $\langle x,hx\rangle$}.

\promptrole{Fixed inputs}

\promptplaceholder{The candidate hash, condition identifier and kind, locked Lean definitions, locked theorem statement, public framework lemmas, previous proof body, compiler diagnostics, and proof mode are inserted here.}

\promptrole{Division of labour}

The agent does not need a local Lean workspace. The harness compiles the response against the pinned Lean~4 and Mathlib environment and returns exact diagnostics on later rounds.

\promptrole{Constraints}

Do not change the theorem statement, generated definitions, system, controller, certificate, or candidate hash.

Do not use \texttt{sorry}, \texttt{admit}, user-defined axioms, \texttt{unsafe}, imports, external assumptions, Markdown, or a natural language proof sketch. Do not declare a new top-level theorem inside \texttt{proof\_body}.

Helper lemmas are optional. Each helper entry must contain one focused named declaration and is compiled independently. Use local \texttt{have} and \texttt{calc} steps for one-use facts.

Use the generated unfolding and closed-loop simplification lemmas. Use supplied smoothness hypotheses directly. Establish local sign, interval, trigonometric, and algebraic facts before invoking \texttt{linarith} or \texttt{nlinarith}.

Qualified \texttt{Real.*} names are prechecked for existence. This is an early name check, not a restriction to the \texttt{Real} namespace.

\promptrole{Acceptance condition}

The proof must compile, preserve the locked theorem hash, and pass the axiom audit. The accepted dependency set is
\[
\{\texttt{propext},\texttt{Classical.choice},\texttt{Quot.sound}\}.
\]

\promptrole{Output format}

Return exactly one JSON object with the fields \texttt{status}, \texttt{helper\_lemmas}, \texttt{proof\_body}, \texttt{blocking\_reason}, and \texttt{counterexample}.

Return \texttt{candidate\_proof} only with a concrete proof body. Return \texttt{cannot\_complete} only with a precise blocker. A claimed counterexample must contain exact rational coordinates and is replayed independently by the harness.

\end{promptbody}
\end{promptblock}

\Needspace{8\baselineskip}
\subsubsection{Compiler-Grounded Lean Repair}
\label{app:lean-error-repair}

A failed first proof does not trigger an unrestricted retry. The framework classifies the failure and returns the relevant evidence to the same locked proof context.

\begin{description}
\item[Output-format failure.] If the response is not a valid structured object, the framework performs a format-only repair. This repair changes the JSON shape, not the mathematical proof.
\item[Unknown declaration.] Qualified \texttt{Real.*} names are checked before full elaboration. If a name is unknown, local Mathlib search results and Lean-verified replacement declarations are returned.
\item[Syntax, elaboration, or type error.] Lean diagnostics, source locations, and relevant goal states are passed back with the previous proof. The agent repairs the failing helper or main proof while preserving accepted parts.
\item[Tactic failure or unsolved goal.] The repair prompt asks the agent to expose intermediate sign, interval, and algebraic facts before retrying \texttt{linarith}, \texttt{nlinarith}, or other tactics.
\item[Forbidden proof content.] Responses containing \texttt{sorry}, \texttt{admit}, user-defined \texttt{axiom}, \texttt{unsafe}, or \texttt{import} are rejected.
\item[Timeout or proof-budget exhaustion.] The obligation remains unresolved. Its diagnostic trace is returned to the Proposal Agent; the locked theorem is not silently changed.
\item[Axiom-audit failure.] A compiled proof is rejected if \texttt{\#print axioms} contains \texttt{sorryAx} or an axiom outside the accepted dependency set.
\item[Rejected falsity claim.] A \texttt{cannot\_complete} response claiming falsity must include an exact rational counterexample. The harness replays it independently. A rejected counterexample does not trigger candidate regeneration and the agent is instructed to continue proving the theorem.
\end{description}

The implementation bounds these repairs by three format-only repairs, six ordinary Lean proof rounds, and three qualified-name repair rounds. In the optional contextual mode, a small number of line-local repairs may be requested before returning to the full proof-repair path.

\Needspace{8\baselineskip}
\subsubsection{Lean Proof Follow-up Prompt}
\label{app:prompt-lean-followup}

\begin{promptblock}
\begin{promptbody}

\promptrole{Next turn}

The system message still contains the Lean role and output contract. Helper lemmas remain one named declaration per entry and are compiled independently. Keep accepted helpers unchanged and edit only the failing helper or main proof.

\promptrole{New evidence}

\promptplaceholder{The candidate hash, condition identifier and kind, locked definitions, locked theorem statement, previous proof body, compiler diagnostics, and repair mode are inserted here.}

Continue from the previous proof. Fix every reported error while preserving the locked theorem and candidate. Do not return a proof sketch or change the verification target.

\promptrole{Output format}

Return exactly one JSON object: either \texttt{candidate\_proof} with a concrete proof body or \texttt{cannot\_complete} with the precise blocker.

\end{promptbody}
\end{promptblock}

\Needspace{8\baselineskip}
\subsubsection{Local LeanRepair Prompt}
\label{app:prompt-leanrepair}

When contextual local repair is enabled, the framework can freeze the current proof and request a patch for one compiler-selected source span.

\begin{promptblock}
\begin{promptbody}

\promptrole{Local repair turn}

The initial proof is frozen. For this turn only, do not return the full \texttt{helper\_lemmas}/\texttt{proof\_body} document. Repair exactly the framework-selected source line and return a \texttt{LeanLocalPatch}.

\promptrole{Compiler-grounded request}

\promptplaceholder{The base output hash, patch identifier, target kind and index, target line, old source text, source window, compiler context, and Lean-verified declarations are inserted here.}

\promptrole{Constraints}

Return replacement code relative to the existing indentation. Preserve all unrelated proof text. The replacement may be a single Lean line or a local multi-line \texttt{by} proof, but it may not declare a new top-level lemma.

Do not reuse declarations reported as unknown by the compiler. A new qualified declaration may be used only when its exact declaration appears in the Lean-verified declaration list. Otherwise prove the required fact locally from accepted surrounding facts.

Do not use \texttt{sorry}, \texttt{admit}, \texttt{axiom}, \texttt{import}, \texttt{unsafe}, Markdown, or explanations outside the structured output. If the selected line cannot be repaired without changing another source line, return \texttt{cannot\_patch}.

\promptrole{Output format}

Return exactly one JSON object. Echo the supplied patch identifier. Return \texttt{patch} with concrete \texttt{replacement\_code}, or return \texttt{cannot\_patch} with an empty replacement and a precise reason.

\end{promptbody}
\end{promptblock}

\subsection{Verification Router Prompt}
\label{app:prompt-verification-router}

After the deterministic fast check has passed, the framework requests one routing decision for each fixed verification condition. The router selects either direct Lean verification or SMT verification and, when SMT is selected, orders all eligible solver backends.

\begin{promptblock}
\begin{promptbody}

\promptrole{Role}

Choose the verification backend for one fixed condition after the deterministic fast check has passed.

\promptrole{Goal}

Choose exactly \texttt{direct\_lean} or \texttt{smt}. If \texttt{smt} is selected, rank every eligible solver exactly once.

\promptrole{Success criteria}

Use only the supplied deterministic capability summary. Do not attempt the proof and do not claim that the condition is valid.

\promptrole{Fixed inputs}

\promptplaceholder{The fixed condition and deterministic capability summary are inserted here. The condition includes its identifier, kind, candidate hash, description, counterexample query, and Lean theorem name. The capability summary specifies whether the query is polynomial NRA, whether it contains transcendental expressions, whether it is formally bounded, and which solver backends are eligible.}

\promptrole{Constraints}

Do not change the condition or solver eligibility. \texttt{direct\_lean} must return an empty \texttt{solver\_order}. \texttt{smt} must return a complete permutation of the eligible backends.

SMT results are checked independently and may be rescued by Lean4.

When the counterexample query includes a bounded fixed parameter, keep it as an ordinary quantified query variable: \texttt{UNSAT} means no violating pair $(x,w)$ exists and therefore establishes the uniform condition. Do not treat the parameter as a state coordinate or as a certificate variable.

\promptrole{Output format}

Return a \texttt{VerificationRouterDecision} containing exactly the following fields:
\texttt{condition\_id}, \texttt{route}, \texttt{solver\_order}, and \texttt{rationale}.

The \texttt{route} must be \texttt{direct\_lean} or \texttt{smt}. The \texttt{solver\_order} must be a JSON array whose entries are eligible backends drawn from \texttt{z3}, \texttt{cvc5}, \texttt{yices}, and \texttt{dreal}. The \texttt{rationale} must be a nonempty string of at most 1,000 characters.

\promptrole{Stop rules}

Return exactly one JSON object and no Markdown.

\end{promptbody}
\end{promptblock}

\subsection{Example Input--Output Trace}
\label{app:case-study}

To illustrate how the prompt interfaces are instantiated in practice, we
show one trace from the safety instance
 (System~\ref{sys:bb-s011-ext}).  The trace is taken from the recorded local run rather than reconstructed from a manually written example.

The human user does not directly provide a separate message at each stage. Instead, the local harness converts the current specification and verification evidence into the \texttt{system} and \texttt{user} messages sent to the LLM.

\paragraph{1. Proposal input.}

The harness supplies the Proposal Agent with the system specification:
\[
\begin{aligned}
\dot{x}_1 &= x_2+u_0,\\
\dot{x}_2 &=
-\frac{3}{10}x_1
-\frac15(1+\cos x_1)x_2
-\frac1{10}x_1^3+u_1,
\end{aligned}
\]
together with
\[
\mathcal X=\mathbb R^2,\qquad
\mathcal X_I=\{x:\|x\|_2\leq 3/5\},\qquad
\mathcal U=[-1,1]^2,
\]
and the unsafe set
\[
\mathcal X_U=\{x:\|x-(2,2)^{\mathsf T}\|_2\leq1\}.
\]
The task profile identifies the problem as continuous-time safety
verification and requires a state barrier, a bounded controller, initial-set
containment, unsafe-set separation, and a strict boundary Lie-derivative
condition.

\paragraph{2. First Proposal-Agent output.}

The first structured response was mathematically plausible but failed the
local expression parser because the certificate was returned with an
assignment:
\[
\texttt{"certificate": "B = 1 - x1**2 - x2**2"}.
\]
The parser expects the expression itself rather than the assignment
\texttt{B =}.  The harness therefore returned a format/semantic diagnostic
to the Proposal Agent without constructing a candidate AST.

\paragraph{3. Proposal repair input and output.}

The next user message to the LLM contained only the new evidence:
\begin{verbatim}
The system message still has the problem,
the locked task conditions, and the output schema.
Below is only new evidence.

parse_or_semantic_errors:
candidate 0: parse error:
certificate: invalid expression
'B = 1 - x1**2 - x2**2'
\end{verbatim}

The Proposal Agent then returned the corrected candidate
\[
B(x)=1-x_1^2-x_2^2,
\qquad
u_0(x)=-\frac{x_1}{1+x_1^2},
\qquad
u_1(x)=-\frac{x_2}{1+x_2^2}.
\]
Its construction summary stated that the initial disk satisfies
$B\geq16/25$, the unsafe set satisfies $B<0$.

\paragraph{4. Routing output.}

The deterministic capability analysis routed the polynomial obligations to
SMT.  The transcendental strict Lie obligation was first exposed to the
verification router, which returned
\[
\texttt{route = smt},\qquad
\texttt{solver\_order = [cvc5]}.
\]
The cvc5 query timed out, so the unresolved obligation was passed to the
Lean Proof-Repair Agent.

\paragraph{5. Initial Lean-proof input and output.}

The Lean prompt contained the locked theorem, generated definitions,
candidate hash, obligation identifier, public framework lemmas, and the
construction summary.  The first proof response used the intended boundary
identity, but cited \texttt{neg\_one\_le\_cos} without its \texttt{Real}
namespace, so the name did not resolve.  The framework returned the compiler error and a
local Mathlib search result suggesting
\texttt{Real.one\_sub\_sq\_div\_two\_le\_cos}.

The second proof response used the suggested theorem with the wrong
application syntax.  Lean reported that the theorem was already specialized
and should not be applied as an ordinary function.  The next repair replaced
the application with a type-correct form:
\begin{verbatim}
simpa using
  (Real.one_sub_sq_div_two_le_cos (x := a))
\end{verbatim}

The third proof response compiled successfully and produced
\[
\texttt{\#print axioms} =
[\texttt{propext},\texttt{Classical.choice},\texttt{Quot.sound}].
\]
The strict Lie obligation was therefore accepted by the Lean kernel.

\paragraph{6. Final harness result.}

The remaining obligations were discharged as follows:
\begin{center}
\begin{tabular}{lll}
\toprule
Obligation & Backend & Result\\
\midrule
Unsafe separation & Z3 & \textsc{unsat}\\
Initial containment & Z3 & \textsc{unsat}\\
Control admissibility & CVC5 & \textsc{unsat}\\
Barrier condition & audited static lemma & accepted\\
Strict Lie condition & Lean~4 & kernel proof accepted\\
Safety instantiation & audited generic theorem & accepted\\
\bottomrule
\end{tabular}
\end{center}

The candidate was finally marked \textbf{verified}.  This trace shows that the LLM proposes certificates and proof terms, while the local harness performs parsing, candidate freezing, routing, solver checks, Lean compilation, and axiom auditing.

%% file: sections/experiments_details.tex
\section{Experimental Details}
\label{app:experimental_details}

\subsection{Implementation Details}
\label{app:implementation_details}

\paragraph{Hardware and software.}
The Full and fixed-certificate experiments use the Linux x86\_64 runtime. The toolchain includes Lean~4.32.2, Mathlib revision \texttt{905b95818eb3}, Z3~4.15.4, cvc5~1.3.3, Yices~2.7.0, and dReal~4.21.06.1; dReal uses precision $\delta=10^{-3}$ and solver seed~0.

\paragraph{Model and budgets.}
Proposal and proof generation use \texttt{gpt-5.6-sol} with reasoning effort \texttt{high} and at most 16,384 output tokens. The router uses the same model when a model decision is requested; polynomial obligations use the deterministic capability route. End-to-end runs have at most $K_{\mathrm{cand}}=5$ candidate rounds under a shared $T_{\mathrm{run}}=2400\,\mathrm{s}$ budget. Each compatible SMT solver has a $30\,\mathrm{s}$ timeout in Full verification. Each Lean obligation has at most $K_{\mathrm{lean}}=6$ proof rounds within $900\,\mathrm{s}$. Independent fixed-certificate runs give each obligation a fresh $900\,\mathrm{s}$ budget.

\paragraph{Fast screening and search.}
Screening uses 1,500 initial-set samples, 1,500 unsafe-set samples, and 2,000 analysis-region or boundary samples, returning at most five violations. Passing this screen does not discharge a formal obligation. Algorithms~\ref{alg:exp_search} and~\ref{alg:exp_verify} describe search and formal verification. Candidates, hypotheses, and theorem statements remain frozen throughout each verification attempt.

\begin{algorithm}[!htbp]
\small
\caption{End-to-end certificate search under a shared deadline}
\label{alg:exp_search}
\begin{algorithmic}[1]
\REQUIRE Original problem $P$, candidate limit $K_{\mathrm{cand}}$, deadline $T$
\STATE $H\gets\varnothing$
\FOR{$k=1,\ldots,K_{\mathrm{cand}}$}
    \IF{the active-work budget is exhausted}
        \RETURN \textsc{Unresolved}
    \ENDIF
    \STATE $A\gets\textsc{GenerateAndParse}(P,H)$
    \IF{no well-formed candidate is obtained}
        \STATE Add parsing diagnostics to $H$; \textbf{continue}
    \ENDIF
    \STATE Freeze $A$ and its obligation manifest $M(A)$
    \STATE $e\gets\textsc{FastCheck}(P,A)$
    \IF{$e$ is an independently validated violation}
        \STATE Add $e$ to $H$; \textbf{continue}
    \ENDIF
    \STATE $v\gets\textsc{VerifyFrozen}(P,A,M(A),T)$
    \IF{all required obligations are verified}
        \RETURN $(A,\textsc{Verified})$
    \ENDIF
    \STATE Add unresolved diagnostics or validated counterexamples from $v$ to $H$
\ENDFOR
\RETURN \textsc{Unresolved}
\end{algorithmic}
\end{algorithm}

\begin{algorithm}[!htbp]
\small
\caption{Verification of a frozen obligation}
\label{alg:exp_verify}
\begin{algorithmic}[1]
\REQUIRE Frozen candidate $A$, manifest $M$, strategy $s$
\STATE Check candidate and manifest hashes
\FORALL{active obligations $O_i$ in the task schedule}
    \STATE Select direct Lean or a compatible SMT according to $s$
    \IF{SMT is selected}
        \STATE Run the compatible solvers concurrently on the negation of $O_i$
        \IF{a definitive \texttt{unsat} verdict is obtained}
            \STATE Mark $O_i$ verified; \textbf{continue}
        \ELSIF{a reported witness validates a violation of $O_i$}
            \RETURN \textsc{Rejected}
        \ELSIF{$s=\textsc{All-SMT}$}
            \RETURN \textsc{Unresolved}
        \ENDIF
    \ENDIF
    \IF{$s=\textsc{All-SMT}$ and no compatible SMT encoding exists}
        \RETURN \textsc{Unresolved}
    \ENDIF
    \STATE Request, compile, audit, and repair a proof of the unchanged Lean theorem
    \IF{no audited proof is obtained within the obligation budget}
        \RETURN \textsc{Unresolved}
    \ENDIF
    \STATE Mark $O_i$ discharged
\ENDFOR
\STATE Apply the audited task-level closing theorem
\RETURN \textsc{Verified}
\end{algorithmic}
\end{algorithm}

\FloatBarrier
\Needspace{10\baselineskip}
\subsection{Additional Results for the Main Experiment}
\label{app:main_additional}

Table~\ref{tab:task_main} complements Table~\ref{tab:main_experiments} with the task-wise Full results. The framework verifies 81/105 systems across all four task categories. ROA has the largest mean recorded token usage, and forward-invariance has the largest mean end-to-end time. Both cost columns average the successful systems within each task.

\begin{table}[!htbp]
\centering
\apptablestyle
\caption{Full end-to-end coverage and mean cost by task.}
\label{tab:task_main}
\begin{tabular*}{\apptablewidth}{@{\extracolsep{\fill}}lrrrr@{}}
\toprule
Task & Verified & Success & Tokens/sys.\ (k) & Time/sys.\ (s) \\
\midrule
Stability          & 21/26  & 80.8\% & 16.00 & 613.57 \\
ROA                & 8/12   & 66.7\% & 19.63 & 427.48 \\
Forward-invariance & 7/13   & 53.8\% & 16.07 & 704.73 \\
Safety             & 45/54  & 83.3\% & 13.95 & 377.58 \\
\midrule
Overall            & \textbf{81/105} & \textbf{77.1\%} & 15.22 & 471.96 \\
\bottomrule
\end{tabular*}
\apptabnote{Token and time means use the successful systems in each row and include candidate search.}
\end{table}

\paragraph{Model usage.}
Table~\ref{tab:agent_tokens} gives per-call means over the 488 returned usage receipts among 707 recorded calls. Visible output excludes reasoning tokens. The largest per-call usage belongs to the Lean proof agent.

\begin{table}[!htbp]
\centering
\apptablestyle
\caption{Mean model usage per call with a returned usage receipt.}
\label{tab:agent_tokens}
\begin{tabular*}{\apptablewidth}{@{\extracolsep{\fill}}lrrrr@{}}
\toprule
Agent & Usage calls & Input (k) & Reasoning (k) & Visible output (k) \\
\midrule
Proposal synthesis  & 119 & 2.24 & 3.66 & 0.44 \\
Verification router & 170 & 1.90 & 0.08 & 0.08 \\
Lean proof agent    & 199 & 7.35 & 3.17 & 1.39 \\
\bottomrule
\end{tabular*}
\end{table}

\FloatBarrier
\Needspace{10\baselineskip}
\subsection{Additional Analysis for RQ1 and RQ2}
\label{app:rq12_additional}

We examine the same 81 frozen certificates and 296 active obligations. \textsc{Routed (Full)} denotes the observed Full traces. \textsc{SMT-first + Lean} composes the $30\,\mathrm{s}$ SMT outcomes with independent Lean records for obligations left unresolved by SMT. 

\paragraph{Coverage and backend allocation.}
All-SMT verifies 59/81 certificates and All-Lean verifies 64/81. Table~\ref{tab:rq12_task} shows Full coverage of 81/81 with 43 Lean obligations and SMT-first coverage of 80/81 with 26 Lean obligations. Lean receives 14.5\% and 8.8\% of the active obligations, respectively. The SMT-first composition covers all ROA, forward-invariance, and safety certificates. Its remaining open obligation is radial unboundedness.

\begin{table}[!htbp]
\centering
\apptablestyle
\caption{Task-wise coverage and Lean allocation on the frozen certificates.}
\label{tab:rq12_task}
\begin{tabular*}{\apptablewidth}{@{\extracolsep{\fill}}lrrrrrr@{}}
\toprule
& \multicolumn{2}{c}{Single backend}
& \multicolumn{2}{c}{\textsc{Full}}
& \multicolumn{2}{c}{\textsc{SMT-first + Lean}} \\
\cmidrule(lr){2-3}\cmidrule(lr){4-5}\cmidrule(lr){6-7}
Task & All-SMT & All-Lean & Verified & Lean/active & Verified & Lean/active \\
\midrule
Stability          & 13/21 & 15/21 & 21/21 & 21/93  & 20/21 & 9/93 \\
ROA                & 6/8   & 6/8   & 8/8   & 3/41   & 8/8   & 3/41 \\
Forward-invariance & 4/7   & 6/7   & 7/7   & 3/24   & 7/7   & 3/24 \\
Safety             & 36/45 & 37/45 & 45/45 & 16/138 & 45/45 & 11/138 \\
\midrule
Total              & 59/81 & 64/81 & \textbf{81/81} & 43/296 & 80/81 & 26/296 \\
\bottomrule
\end{tabular*}
\apptabnote{All-SMT uses a $900\,\mathrm{s}$ obligation budget. Full is observed; SMT-first + Lean is a composition using $30\,\mathrm{s}$ SMT outcomes and independent Lean proofs.}
\end{table}

\paragraph{Verification time.}
Table~\ref{tab:rq12_task_time} averages verification time over the certificates verified by each strategy. The All-SMT and All-Lean means are $3.68\,\mathrm{s}$ and $579.17\,\mathrm{s}$. Full records $119.31\,\mathrm{s}$, and SMT-first + Lean records a mean of $94.28\,\mathrm{s}$.

\begin{table}[!htbp]
\centering
\apptablestyle
\caption{Mean verification time per verified certificate, in seconds.}
\label{tab:rq12_task_time}
\begin{tabular*}{\apptablewidth}{@{\extracolsep{\fill}}lrrrr@{}}
\toprule
Task & All-SMT & All-Lean & Full & SMT-first + Lean \\
\midrule
Stability          & 2.47 & 590.31 & 141.96 & 100.33 \\
ROA                & 1.97 & 489.94 & 143.35 & 80.35 \\
Forward-invariance & 0.85 & 697.66 & 108.72 & 129.01 \\
Safety             & 4.71 & 569.91 & 106.11 & 88.66 \\
\midrule
Overall            & 3.68 & 579.17 & 119.31 & 94.28 \\
\bottomrule
\end{tabular*}
\apptabnote{All-Lean uses the selected complete certificate records; Full uses its observed final-verification interval.}
\end{table}

\paragraph{Recorded model usage.}
Table~\ref{tab:rq12_cost} normalizes the selected records' token sum by the common 81-certificate cohort, including successful and unresolved records. All-Lean uses 24.87k tokens per system, compared with 8.34k for Full and 9.15k for SMT-first + Lean, reductions of 66.5\% and 63.2\%. Their mean times over verified certificates are 79.4\% and 83.7\% below All-Lean. Selective proof generation combines broad coverage with lower recorded model usage in this evaluation.

\begin{table}[!htbp]
\centering
\apptablestyle
\caption{Verification coverage, recorded model usage, and mean time.}
\label{tab:rq12_cost}
\begin{tabular*}{\apptablewidth}{@{\extracolsep{\fill}}lrrrr@{}}
\toprule
Strategy & Verified & Tokens/sys.\ (k) & Time/sys.\ (s) & Lean obl. \\
\midrule
\textsc{All-SMT}          & 59/81 & 0.00  & 3.68   & 0 \\
\textsc{All-Lean}         & 64/81 & 24.87 & 579.17 & 296 \\
\textsc{Routed (Full)}    & \textbf{81/81} & 8.34 & 119.31 & 43 \\
\textsc{SMT-first + Lean} & 80/81 & 9.15  & 94.28  & 26 \\
\bottomrule
\end{tabular*}
\apptabnote{Tokens/sys.\ uses all 81 selected certificate records; Time/sys.\ averages verified certificates. Lean obl.\ counts distinct assigned obligations.}
\end{table}

\paragraph{SMT-first verification paths.}
SMT discharges 270/296 obligations (91.2\%) by \texttt{unsat} within $30\,\mathrm{s}$. The 26 Lean paths comprise 8 after $\delta$-\texttt{sat}, 15 after unresolved SMT checks at the time limit, and 3 radial-unboundedness obligations without an SMT encoding.

\paragraph{Condition-wise verification.}
Table~\ref{tab:condition_cost} groups the SMT-first composition by obligation type. Continuous decrease or exponential decay and discrete inductiveness account for 20/26 Lean obligations (76.9\%) and 74.8\% of the summed condition time. SMT verifies the equilibrium, positivity,  and control-admissibility checks, concentrating Lean proof work on dynamical conditions.

\begin{table}[!htbp]
\centering
\apptablestyle
\caption{SMT-first + Lean allocation and summed time by obligation type.}
\label{tab:condition_cost}
\begin{tabular*}{\apptablewidth}{@{\extracolsep{\fill}}lrrr@{}}
\toprule
Obligation & Active & Lean & Time (s) \\
\midrule
Equilibrium/value              & 58 & 0  & 101.86 \\
Positivity/quadratic bounds     & 35 & 0  & 63.06 \\
Set containment                & 58 & 0  & 68.32 \\
Unsafe separation              & 52 & 3  & 884.32 \\
Control admissibility          & 9  & 0  & 24.32 \\
Continuous decrease/exp.\ decay & 73 & 17 & 5423.25 \\
Discrete inductiveness         & 8  & 3  & 916.01 \\
Radial unboundedness           & 3  & 3  & 990.21 \\
\midrule
Total                          & 296 & 26 & 8471.35 \\
\bottomrule
\end{tabular*}
\apptabnote{Time sums reconstructed intervals over all 296 obligations. Parallel intervals may overlap; this total measures obligation effort rather than certificate wall-clock time.}
\end{table}

\FloatBarrier
\Needspace{10\baselineskip}
\subsection{Additional Analysis for RQ3}
\label{app:rq3_additional}

Table~\ref{tab:repair_tokens} adds model usage to Table~\ref{tab:repair}. All three feedback strategies are re-run from the same frozen failed first attempts on 26 obligations from 19 systems. Usage receipts are available for 53/95 Generic Retry calls, 43/71 Raw Diagnostics calls, and 45/66 Structured Repair calls. Token columns sum these receipts; visible output excludes reasoning tokens.

\begin{table}[!htbp]
\centering
\apptablestyle
\caption{Recorded model usage for RQ3 repair; token values are in thousands.}
\label{tab:repair_tokens}
\begin{tabular*}{\apptablewidth}{@{\extracolsep{\fill}}lrrrrr@{}}
\toprule
Strategy & Calls & Input & Reasoning & Visible output & Total/obl. \\
\midrule
\textsc{Generic Retry}     & 95 & 534.64 & 129.07 & 68.02 & 28.14 \\
\textsc{Raw Diagnostics}   & 71 & 445.62 & 58.75  & 74.74 & 22.27 \\
\textsc{Structured Repair} & 66 & 534.71 & 70.92  & 90.59 & 26.78 \\
\bottomrule
\end{tabular*}
\apptabnote{Calls are post-failure agent calls. Input, reasoning, and visible output are arm totals; Total/obl.\ divides their sum by 26.}
\end{table}

Structured Repair completes 21/26 obligations, compared with 17/26 for Raw Diagnostics and 1/26 for Generic Retry, using 66 calls against 71 and 95. Its recorded token total is 4.9\% below Generic Retry and 20.2\% above Raw Diagnostics, while its mean timeout-charged repair time is the lowest. The repair cohort comprises 15 dynamical decrease or inductiveness obligations, six positivity or quadratic bounds, two radial-unboundedness obligations, and one each for equilibrium, unsafe separation, and set containment. It overlaps the separately defined SMT-first Lean set on 11 obligations.

\FloatBarrier
\Needspace{23\baselineskip}
\subsection{Analysis of Unresolved Systems}
\label{app:unresolved}

The Full experiment records 24 unresolved systems: six produce no parsed candidate and 18 enter Lean verification. Table~\ref{tab:unresolved_stage} groups the last recorded events to identify the stages where additional candidate or proof work can extend coverage.

\begin{table}[!htbp]
\centering
\apptablestyle
\caption{Last recorded event for the 24 unresolved Full runs.}
\label{tab:unresolved_stage}
\begin{tabularx}{\apptablewidth}{@{}Xr@{}}
\toprule
Last recorded event & Systems \\
\midrule
Lean call timeout or shared-deadline exhaustion & 10 \\
No parsed candidate before the shared deadline  & 6 \\
Invalid Lean proof output                      & 3 \\
Lean helper-module timeout                     & 2 \\
Lean compiler timeout                          & 1 \\
API rate limit during a Lean call              & 1 \\
Lean agent returned \texttt{cannot\_complete}  & 1 \\
\midrule
Total                                          & 24 \\
\bottomrule
\end{tabularx}
\apptabnote{Events identify the final recorded execution stage. Three runs also contain a replay-confirmed violation on another obligation; these events do not define mutually exclusive mathematical causes.}
\end{table}

%% file: sections/benchmark_details.tex
\section{\benchmarkname: Benchmark Details}
\label{benchmark}

\subsection{Benchmark Collection and Curation}

\paragraph{Candidate benchmark records.}

We collected 233 candidate records from fifteen representative source papers and BarrierBench \citep{taheri2025barrierbench}. Specifically, 133 records were extracted from the fifteen papers and 100 records were obtained from BarrierBench. Each record captures a concrete experimental configuration that comprises the plant dynamics along with its assigned verification domain, controller, and solver settings. Because multiple records often instantiate the same underlying system under different test conditions, we systematically deduplicate and curate them into an independent core benchmark suite.

\paragraph{Curation rules.}

To build a non-redundant and structurally rigorous benchmark, we identify each instance by its intrinsic plant dynamics and control specifications. Implementation-level variations—such as solver hyperparameters, neural architectures, or scaled verification domains—are consolidated under their corresponding systems. Following the operational criteria summarized in Table~\ref{tab:curation-reasons}, we apply five mutually exclusive rules:
(1) \textit{Rule~1 (Exact duplicates)}: merge records with identical dynamics and specifications across papers, retaining only the canonical instance;
(2) \textit{Rule~2 (Multiple configurations)}: consolidate records sharing identical vector fields under varying domains or hyperparameters into a single representative benchmark;
(3) \textit{Rule~3 (Family variants)}: retain a canonical representative for parametric sweeps or closely related models within the same system family;
(4) \textit{Rule~4 (Structural redundancy and degradation)}: exclude systems where nonlinearities are canceled by feedback linearization, dimensions are artificially padded with passive/decoupled coordinates, or dynamics reduce to simple linear contractions; and
(5) \textit{Rule~5 (Out-of-scope models)}: remove hybrid-reset systems, theoretical counterexamples, and simulation-only cases lacking formal verification specifications.

\begin{table*}[h]
\centering
\tiny
\setlength{\tabcolsep}{2.2pt}
\renewcommand{\arraystretch}{1.12}
\begin{tabularx}{\textwidth}{
@{}r
>{\raggedright\arraybackslash}p{0.15\textwidth}
>{\raggedright\arraybackslash}X
>{\raggedright\arraybackslash}X
r r r@{}}
\toprule
No. &
Curation rule &
Operational criterion &
Treatment &
Source papers &
BarrierBench &
Total \\
\midrule

1 &
Exact duplicate benchmark instance &
Dynamics, task, sets, controller, and constraints are equivalent;
only provenance, method, certificate architecture, or result differs. &
Retain one canonical instance and merge redundant provenance records. &
24 & 0 & 24 \\

2 &
Multiple configurations of the same dynamics &
The dynamics are identical, but the domain, task specification,
controller, or solver setting differs. &
Retain one canonical configuration and discard redundant variants. &
22 & 1 & 23 \\

3 &
Closely related variants in the same system family &
The basic plant and interaction structure are unchanged; differences
mainly concern parameters, coefficients, or minor local modifications. &
Retain one representative system from the family. &
31 & 13 & 44 \\

4 &
Simplified or structurally redundant dynamics &
Feedback cancels nonlinear terms, or additional states are linear,
passive, decoupled, or already represented by a simpler system. &
Exclude from the independent core benchmark suite. &
0 & 54 & 54 \\

5 &
Outside the current core experimental scope &
The record is a theoretical illustration, counterexample,
hybrid/abstraction model, or simulation-only case. &
Exclude from the continuous formal verification track. &
8 & 0 & 8 \\

\midrule

\multicolumn{4}{@{}l}{\textbf{Not counted as independent core systems}} &
\textbf{85} & \textbf{68} & \textbf{153} \\

\multicolumn{4}{@{}l}{\textbf{Retained core systems}} &
\textbf{48} & \textbf{32} & \textbf{80} \\

\midrule

\multicolumn{4}{@{}l}{\textbf{Candidate records}} &
\textbf{133} & \textbf{100} & \textbf{233} \\

\bottomrule
\end{tabularx}
\caption{
Curation of the 233 candidate records into independent core benchmark systems. Each non-retained record is mapped to exactly one primary curation rule in the order shown.
}
\label{tab:curation-reasons}
\end{table*}

\paragraph{Curation results and high-dimensional scarcity.}

The curation process accounts for all 153 excluded records: 24 exact duplicates, 23 alternative configurations, 44 parametric family variants, 54 structurally simplified systems, and 8 out-of-scope entries. Consequently, the Literature-derived tier contains
\textbf{80 instances}: 47 from the source papers,
one from the textbook, and 32 from BarrierBench, as
summarized in Table~\ref{tab:source-system-distribution}.

\begin{table}[h]
\centering
\scriptsize
\setlength{\tabcolsep}{2.8pt}
\renewcommand{\arraystretch}{1.08}
\begin{tabularx}{\linewidth}{
    @{}r
    >{\raggedright\arraybackslash}X
    >{\raggedright\arraybackslash}p{0.19\linewidth}
    rrrr@{}
}
\toprule
No. &
Source &
Assigned task(s) &
\shortstack{Reported\\records} &
\shortstack{Literature-\\derived} &
\shortstack{Reported\\$n\geq5$} &
\shortstack{Retained\\$n\geq5$} \\
\midrule

1 &
\cite{abate2021fossil} &
P1 (8), P4 (3) &
17 & 11 & 2 & 1 \\

2 &
\cite{edwards2024fossil} &
P1 (3), P2 (1), P4 (2) &
21 & 6 & 1 & 0 \\

3 &
\cite{zhao2020synthesizing} &
P4 (1) &
5 & 1 & 0 & 0 \\

4 &
\cite{liu2024tool} &
P2 (2) &
13 & 2 & 9 & 1 \\

5 &
\cite{sogokon2018vector} &
P4 (2) &
2 & 2 & 0 & 0 \\

6 &
\cite{wu2024completeness} &
P4 (3) &
22 & 3 & 0 & 0 \\

7 &
\cite{zou2025analytical} &
P1 (8) &
12 & 8 & 5 & 4 \\

8 &
\cite{zhou2022neural} &
P2 (2) &
3 & 2 & 0 & 0 \\

9 &
\cite{liu2015abstraction} &
P4 (2) &
8 & 2 & 0 & 0 \\

10 &
\cite{liu2023towards} &
P2 (2) &
14 & 2 & 0 & 0 \\

11 &
\cite{meng2024physics} &
P2 (1) &
6 & 1 & 2 & 0 \\

12 &
\cite{samanipour2026relu} &
P3 (3) &
3 & 3 & 0 & 0 \\

13 &
\cite{vaisi2025extracting} &
P3 (2) &
2 & 2 & 0 & 0 \\

14 &
\cite{liu2026globally} &
P1 (1) &
1 & 1 & 0 & 0 \\

15 &
\cite{wu2023neural} &
P3 (1) &
4 & 1 & 1 & 1 \\

16 &
\cite{khalil2002nonlinear} (Book) &
P1 (1) &
-- & 1 & -- & 0 \\

\midrule
\multicolumn{3}{@{}l}{
    \textbf{Subtotal: 15 source papers and one book}
} &
\textbf{133} &
\textbf{48} &
\textbf{20} &
\textbf{7} \\

\midrule

17 &
BarrierBench \cite{taheri2025barrierbench} &
P4 (32) &
100 & 32 & 9 & 0 \\

\midrule
\multicolumn{3}{@{}l}{\textbf{Total}} &
\textbf{233} &
\textbf{80} &
\textbf{29} &
\textbf{7} \\

\bottomrule
\end{tabularx}
\caption{
    Source contributions to the final Literature-derived tier. Reported counts refer to the 233-record candidate pool. The retained counts and task assignments refer to the 80 Literature-derived instances. P1--P4 denote stability certification, region-of-attraction verification, forward-invariant-set synthesis, and safety verification, respectively.
}
\label{tab:source-system-distribution}
\end{table}

Crucially, this curation process highlights the severe scarcity of genuinely high-dimensional systems ($n \ge 5$) in the existing literature. While 29 candidate records nominally feature $n \ge 5$, only \textbf{7} survive after filtering: the 9 high-dimensional records from \citet{liu2024tool} collapse to a single canonical system under Rule~2, while all 9 high-dimensional instances in BarrierBench are excluded under Rule~4 due to feedback linearization, linear-chain decoupling, or passive state padding.

\paragraph{Motivation for dataset extension.}
This acute shortage of coupled high-dimensional benchmarks motivates the three-tier taxonomy of our dataset:
(1) \textit{Literature-derived}: the curated core of 80 non-redundant baseline systems;
(2) \textit{Extended}: containing 11 instances that extend source systems through changes to state dimension,
verification domains, or control constraints; and
(3) \textit{New}, containing 14 constructed systems with
coupled, non-polynomial dynamics in five to ten dimensions. The additional tiers broaden the structural and verification settings represented in the benchmark.

Counting systems with $n \geq 5$ as high-dimensional benchmark instances, the Extended and New tiers contribute four and fourteen additional instances, respectively. The total number increases from 7 to 25, while their share rises from 8.75\% of the Literature-derived tier to 23.81\% of the full benchmark, an increase of 15.06 percentage points.

\subsection{Experiments details for main text}

We report each method's native numerical semantics. FOSSIL~2.0 uses neural CEGIS and SMT to refute the strict certificate conditions; its released dReal configuration has precision $\delta_{\mathrm F}=10^{-5}$, while the paper's Lyapunov example uses a Torus domain excluding $B_{0.01}(0)$ (the optional barrier belt affects training only) \citep{edwards2024fossil}. The RL method sets the dReal $\epsilon_{\mathrm R}=e^{-3}$ and the precision/margin to $\delta_{\mathrm R}=e^{-12}$, and checks $V(x)>\delta_{\mathrm R}$ and $L_fV(x)<-\delta_{\mathrm R}$ only on $D\setminus B_{\epsilon_{\mathrm R}}(0)$ \citep{zou2025analytical}. SynNLF learns a polynomial neural Lyapunov candidate and applies a floating-point SOS/LMI feasibility check through MOSEK. Also, its theorem introduces a slack $\varepsilon>0$, while the released validator uses a derivative offset $\tau_{\mathrm S}=5\times10^{-4}$ and a default upper bound $\beta_{\mathrm S}=10^{6}$ \citep{zhao2025learning}. We treat dReal $\delta$-\texttt{sat} and non-definitive numerical SOS outcomes as inconclusive, and do not transfer any method-specific precision, domain exclusion, or tolerance to the SMT+Lean verification.

\subsection{Problem 1 : Stability certification}

\begin{systeminstance}{Exercise~4.30, \cite{khalil2002nonlinear}}
\label{sys:khalil-4-30-shifted}

The vector flow field is:
\[
\dot{\mathbf{x}}
=
\begin{pmatrix}
\dot{x}_1\\
\dot{x}_2
\end{pmatrix}
=
\begin{pmatrix}
-\dfrac{1}{2}\tan\left(\dfrac{\pi x_1}{2} + \dfrac{\pi}{4}\right) + x_2 + \dfrac{1}{2}\\[1.5ex]
x_1 - \dfrac{1}{2}\tan\left(\dfrac{\pi x_2}{2} + \dfrac{\pi}{4}\right) + \dfrac{1}{2}
\end{pmatrix}.
\]

\begin{itemize}[label=\textendash, leftmargin=1.55em,
                labelsep=0.45em, topsep=0pt, itemsep=0pt]
    \item $x^\star=(0,0)$.
    \item Original $\mathcal{N}=\{\mathbf{x}\in\mathbb{R}^{2}\mid0.01^2\leq x_1^2+x_2^2\leq0.25^2\}$.
    \item Original verified type: $\epsilon$-asymptotic stability.
    \item Verified method in the original paper: analytical Lyapunov analysis.
    \item Our $\mathcal{N}=\{\mathbf{x}\in\mathbb{R}^{2}\mid x_1^2+x_2^2\leq0.25^2\}$.
    \item Our verified type: local asymptotic stability.
\end{itemize}
\end{systeminstance}

\begin{systeminstance}{Table~1 row~1 (NonPoly0), \cite{abate2021fossil}}
\label{sys:catalogue-sys0016}

The vector flow field is:
\[
\dot{\mathbf{x}}
=
\begin{pmatrix}
\dot{x}_1\\
\dot{x}_2
\end{pmatrix}
=
\begin{pmatrix}
-x_1+x_1x_2\\
-x_2
\end{pmatrix}.
\]
\begin{itemize}[label=\textendash, leftmargin=1.55em,
                labelsep=0.45em, topsep=0pt, itemsep=0pt]
    \item $x^\star=(0,0)$.
    \item Original $\mathcal{N}=\{\mathbf{x}\in\mathbb{R}^{2}\mid0.01^2\leq x_1^2+x_2^2\leq1^2\}$.
    \item Original verified type: local $\epsilon$-asymptotic stability.
    \item Verified method in the original paper: dReal.
    \item Our $\mathcal{N}=\{\mathbf{x}\in\mathbb{R}^{2}\mid x_1^2+x_2^2\leq1^2\}$.
    \item Our verified type: local asymptotic stability.
\end{itemize}
\end{systeminstance}

\begin{systeminstance}{Table~1 row~2 (NonPoly1), \cite{abate2021fossil}}
\label{sys:catalogue-sys0017}

The vector flow field is:
\[
\dot{\mathbf{x}}
=
\begin{pmatrix}
\dot{x}_1\\
\dot{x}_2
\end{pmatrix}
=
\begin{pmatrix}
-x_1+2x_1^2x_2\\
-x_2
\end{pmatrix}.
\]
\begin{itemize}[label=\textendash, leftmargin=1.55em,
                labelsep=0.45em, topsep=0pt, itemsep=0pt]
    \item $x^\star=(0,0)$.
    \item Original $\mathcal{N}=\{\mathbf{x}\in\mathbb{R}^{2}\mid x_1,x_2\geq0,\;0<x_1^2+x_2^2\leq10^2\}$.
    \item Original verified type: local asymptotic stability.
    \item Verified method in the original paper: Z3.
    \item Our $\mathcal{N}=\{\mathbf{x}\in\mathbb{R}^{2}\mid x_1,x_2\geq0,\;x_1^2+x_2^2\leq10^2\}$.
    \item Our verified type: local asymptotic stability.
\end{itemize}
\end{systeminstance}

\begin{systeminstance}{Table~1 row~3 (NonPoly2), \cite{abate2021fossil}}
\label{sys:catalogue-sys0018}

The vector flow field is:
\[
\dot{\mathbf{x}}
=
\begin{pmatrix}
\dot{x}_1\\
\dot{x}_2\\
\dot{x}_3
\end{pmatrix}
=
\begin{pmatrix}
-x_1\\
-2x_2+0.1x_1x_2^2+x_3\\
-x_3-1.5x_2
\end{pmatrix}.
\]
\begin{itemize}[label=\textendash, leftmargin=1.55em,
                labelsep=0.45em, topsep=0pt, itemsep=0pt]
    \item $x^\star=(0,0,0)$.
    \item Original $\mathcal{N}=\{\mathbf{x}\in\mathbb{R}^{3}\mid x_1,x_2,x_3\geq0,\;0<x_1^2+x_2^2+x_3^2\leq10^2\}$.
    \item Original verified type: local asymptotic stability.
    \item Verified method in the original paper: Z3.
    \item Our $\mathcal{N}=\{\mathbf{x}\in\mathbb{R}^{3}\mid x_1,x_2,x_3\geq0,\;x_1^2+x_2^2+x_3^2\leq10^2\}$.
    \item Our verified type: local asymptotic stability.
\end{itemize}
\end{systeminstance}

\begin{systeminstance}{Table~1 row~4 (NonPoly3), \cite{abate2021fossil}}
\label{sys:catalogue-sys0019}

The vector flow field is:
\[
\dot{\mathbf{x}}
=
\begin{pmatrix}
\dot{x}_1\\
\dot{x}_2\\
\dot{x}_3
\end{pmatrix}
=
\begin{pmatrix}
-3x_1-0.1x_1x_2^3\\
-x_2+x_3\\
-x_3
\end{pmatrix}.
\]
\begin{itemize}[label=\textendash, leftmargin=1.55em,
                labelsep=0.45em, topsep=0pt, itemsep=0pt]
    \item $x^\star=(0,0,0)$.
    \item Original $\mathcal{N}=\{\mathbf{x}\in\mathbb{R}^{3}\mid x_1,x_2,x_3\geq0,\;0<x_1^2+x_2^2+x_3^2\leq10^2\}$.
    \item Original verified type: local asymptotic stability.
    \item Verified method in the original paper: Z3.
    \item Our $\mathcal{N}=\{\mathbf{x}\in\mathbb{R}^{3}\mid x_1,x_2,x_3\geq0,\;x_1^2+x_2^2+x_3^2\leq10^2\}$.
    \item Our verified type: local asymptotic stability.
\end{itemize}
\end{systeminstance}

\begin{systeminstance}{Table~1 row~5 (Poly1), \cite{abate2021fossil}}
\label{sys:catalogue-sys0020}

The vector flow field is:
\[
\dot{\mathbf{x}}
=
\begin{pmatrix}
\dot{x}_1\\
\dot{x}_2\\
\dot{x}_3
\end{pmatrix}
=
\begin{pmatrix}
-x_1^3-x_1x_3^2\\
-x_2-x_1^2x_2\\
-x_3+3x_1^2x_3-3x_3
\end{pmatrix}.
\]
\begin{itemize}[label=\textendash, leftmargin=1.55em,
                labelsep=0.45em, topsep=0pt, itemsep=0pt]
    \item $x^\star=(0,0,0)$.
    \item Original $\mathcal{N}=\{\mathbf{x}\in\mathbb{R}^{3}\mid0.1^2\leq x_1^2+x_2^2+x_3^2\leq10^2\}$.
    \item Original verified type: local $\epsilon$-asymptotic stability.
    \item Verified method in the original paper: dReal.
    \item Our $\mathcal{N}=\{\mathbf{x}\in\mathbb{R}^{3}\mid x_1^2+x_2^2+x_3^2\leq10^2\}$.
    \item Our verified type: local asymptotic stability.
\end{itemize}
\end{systeminstance}

\begin{systeminstance}{Table~1 row~6 (Poly2), \cite{abate2021fossil}}
\label{sys:catalogue-sys0021}

The vector flow field is:
\[
\dot{\mathbf{x}}
=
\begin{pmatrix}
\dot{x}_1\\
\dot{x}_2
\end{pmatrix}
=
\begin{pmatrix}
-x_1^3+x_2\\
-x_1-x_2
\end{pmatrix}.
\]
\begin{itemize}[label=\textendash, leftmargin=1.55em,
                labelsep=0.45em, topsep=0pt, itemsep=0pt]
    \item $x^\star=(0,0)$.
    \item Original $\mathcal{N}=\{\mathbf{x}\in\mathbb{R}^{2}\mid0.01^2\leq x_1^2+x_2^2\leq10^2\}$.
    \item Original verified type: local $\epsilon$-asymptotic stability.
    \item Verified method in the original paper: Z3.
    \item Our $\mathcal{N}=\{\mathbf{x}\in\mathbb{R}^{2}\mid x_1^2+x_2^2\leq10^2\}$.
    \item Our verified type: local asymptotic stability.
\end{itemize}
\end{systeminstance}

\begin{systeminstance}{Table~1 row~7 (Poly3), \cite{abate2021fossil}}
\label{sys:catalogue-sys0022}

The vector flow field is:
\[
\dot{\mathbf{x}}
=
\begin{pmatrix}
\dot{x}_1\\
\dot{x}_2
\end{pmatrix}
=
\begin{pmatrix}
-x_1^3-x_2^2\\
x_1x_2-x_2^3
\end{pmatrix}.
\]
\begin{itemize}[label=\textendash, leftmargin=1.55em,
                labelsep=0.45em, topsep=0pt, itemsep=0pt]
    \item $x^\star=(0,0)$.
    \item Original $\mathcal{N}=\{\mathbf{x}\in\mathbb{R}^{2}\mid0.1^2\leq x_1^2+x_2^2\leq10^2\}$.
    \item Original verified type: local $\epsilon$-asymptotic stability.
    \item Verified method in the original paper: dReal.
    \item Our $\mathcal{N}=\{\mathbf{x}\in\mathbb{R}^{2}\mid x_1^2+x_2^2\leq10^2\}$.
    \item Our verified type: local asymptotic stability.
\end{itemize}
\end{systeminstance}

\begin{systeminstance}{Table~1 row~8 (Poly4), \cite{abate2021fossil}}
\label{sys:catalogue-sys0023}

The vector flow field is:
\[
\dot{\mathbf{x}}
=
\begin{pmatrix}
\dot{x}_1\\
\dot{x}_2
\end{pmatrix}
=
\begin{pmatrix}
-x_1-1.5x_1^2x_2^3\\
-x_2^3+0.5x_1^3x_2^2
\end{pmatrix}.
\]
\begin{itemize}[label=\textendash, leftmargin=1.55em,
                labelsep=0.45em, topsep=0pt, itemsep=0pt]
    \item $x^\star=(0,0)$.
    \item Original $\mathcal{N}=\{\mathbf{x}\in\mathbb{R}^{2}\mid0.1^2\leq x_1^2+x_2^2\leq10^2\}$.
    \item Original verified type: local $\epsilon$-asymptotic stability.
    \item Verified method in the original paper: dReal.
    \item Our $\mathcal{N}=\{\mathbf{x}\in\mathbb{R}^{2}\mid x_1^2+x_2^2\leq10^2\}$.
    \item Our verified type: local asymptotic stability.
\end{itemize}
\end{systeminstance}

\begin{systeminstance}{Table~4 \#2 (SineModel), \cite{edwards2024fossil}}
\label{sys:fossil2-f2-02-sinemodel}

The vector flow field is:
\[
\dot{\mathbf{x}}
=
\begin{pmatrix}
\dot{x}_1\\
\dot{x}_2
\end{pmatrix}
=
\begin{pmatrix}
x_2+u_1\\
u_2+\dfrac{mgL\sin(x_1)-bx_2}{mL^2}
\end{pmatrix},
\]
where $m=0.15$, $g=9.81$, $L=0.5$, and $b=0.1$.
\begin{itemize}[label=\textendash, leftmargin=1.55em,
                labelsep=0.45em, topsep=0pt, itemsep=0pt]
    \item $x^\star=(0,0)$.
    \item Original $\mathcal{N}=\{\mathbf{x}\in\mathbb{R}^{2}\mid0.1^2\leq x_1^2+x_2^2\leq1^2\}$.
    \item Original verified type: local $\epsilon$-asymptotic stability.
    \item Verified method in the original paper: dReal.
    \item Our $\mathcal{N}=\{\mathbf{x}\in\mathbb{R}^{2}\mid x_1^2+x_2^2\leq1^2\}$.
    \item Our verified type: local asymptotic stability.
\end{itemize}
\end{systeminstance}

\begin{systeminstance}{Table~4 \#14 (DTAhmadi2), \cite{edwards2024fossil}}
\label{sys:fossil2-f2-14-dtahmadi2}

The discrete-time transition map is:
\[
\mathbf{x}^{+}
=
F(\mathbf{x})
=
\begin{pmatrix}
0.687x_1+0.558x_2-0.0001x_1x_2\\
-0.292x_1+0.773x_2
\end{pmatrix}.
\]
\begin{itemize}[label=\textendash, leftmargin=1.55em,
                labelsep=0.45em, topsep=0pt, itemsep=0pt]
    \item $x^\star=(0,0)$.
    \item Original $\mathcal{N}=\{\mathbf{x}\in\mathbb{R}^{2}\mid0.1^2\leq x_1^2+x_2^2\leq1^2\}$.
    \item Original verified type: local  $\epsilon$-asymptotic stability.
    \item Verified method in the original paper: dReal.
    \item Our $\mathcal{N}=\{\mathbf{x}\in\mathbb{R}^{2}\mid x_1^2+x_2^2\leq1^2\}$.
    \item Our verified type: local asymptotic stability.
\end{itemize}
\end{systeminstance}

\begin{systeminstance}{Table~4 \#15 (DTAhmadi), \cite{edwards2024fossil}}
\label{sys:fossil2-f2-15-dtahmadi}

The discrete-time transition map is:
\[
\mathbf{x}^{+}
=
F(\mathbf{x},\mathbf{u})
=
\begin{pmatrix}
0.3x_1+u_1\\
-x_1+0.5x_2+\dfrac{7}{18}x_1^2+u_2
\end{pmatrix}.
\]
\begin{itemize}[label=\textendash, leftmargin=1.55em,
                labelsep=0.45em, topsep=0pt, itemsep=0pt]
    \item $x^\star=(0,0)$.
    \item Original $\mathcal{N}=\{\mathbf{x}\in\mathbb{R}^{2}\mid0.1^2\leq x_1^2+x_2^2\leq1^2\}$.
    \item Original verified type: local  $\epsilon$-asymptotic stability.
    \item Verified method in the original paper: dReal.
    \item Our $\mathcal{N}=\{\mathbf{x}\in\mathbb{R}^{2}\mid x_1^2+x_2^2\leq1^2\}$.
    \item Our verified type: local asymptotic stability.
\end{itemize}
\end{systeminstance}

\begin{systeminstance}{Table~1 row~4, \cite{zou2025analytical}}
\label{sys:catalogue-sys0227}

The vector flow field is:
\[
\dot{\mathbf{x}}
=
\begin{pmatrix}
\dot{x}_1\\
\dot{x}_2\\
\dot{x}_3
\end{pmatrix}
=
\begin{pmatrix}
-3x_1^3+3x_1x_3-9x_1\\
-x_1^3-5x_2+5x_3^3\\
-9x_3^3
\end{pmatrix}.
\]
\begin{itemize}[label=\textendash, leftmargin=1.55em,
                labelsep=0.45em, topsep=0pt, itemsep=0pt]
    \item $x^\star=(0,0,0)$.
    \item Original $\mathcal{N}=\{\mathbf{x}\in\mathbb{R}^{3}\mid e^{-6}\leq x_1^2+x_2^2+x_3^2,\;|x_i|\leq1,\ i=1,2,3\}$.
    \item Original verified type: local  $\epsilon$-asymptotic stability.
    \item Verified method in the original paper: dReal.
    \item Our $\mathcal{N}=\{\mathbf{x}\in\mathbb{R}^{3}\mid |x_i|\leq1,\ i=1,2,3\}$.
    \item Our verified type: local asymptotic stability.
\end{itemize}
\end{systeminstance}

\begin{systeminstance}{Table~1 row~5, \cite{zou2025analytical}}
\label{sys:catalogue-sys0228}

The vector flow field is:
\[
\dot{\mathbf{x}}
=
\begin{pmatrix}
\dot{x}_1\\
\dot{x}_2\\
\dot{x}_3
\end{pmatrix}
=
\begin{pmatrix}
x_2\\
-\sin(x_1)\cos(x_1)-x_2-\sin(x_3)\cos(x_3)\\
x_2-x_3
\end{pmatrix}.
\]
\begin{itemize}[label=\textendash, leftmargin=1.55em,
                labelsep=0.45em, topsep=0pt, itemsep=0pt]
    \item $x^\star=(0,0,0)$.
    \item Original $\mathcal{N}=\{\mathbf{x}\in\mathbb{R}^{3}\mid e^{-6}\leq x_1^2+x_2^2+x_3^2,\;|x_i|\leq1.5,\ i=1,2,3\}$.
    \item Original verified type: local  $\epsilon$-asymptotic stability.
    \item Verified method in the original paper: dReal.
    \item Our $\mathcal{N}=\{\mathbf{x}\in\mathbb{R}^{3}\mid |x_i|\leq1.5,\ i=1,2,3\}$.
    \item Our verified type: local asymptotic stability.
\end{itemize}
\end{systeminstance}

\begin{systeminstance}{Table~1 row~6, \cite{zou2025analytical}}
\label{sys:catalogue-sys0229}

The source system is expressed in the center-of-inertia coordinates
$(\delta_1,\delta_2,\omega_1,\omega_2)$, which satisfy $\delta_1+\delta_2=0$. We therefore remove the redundant common-angle coordinate and define $q:=\delta_1-\delta_2, \delta_1=\frac{q}{2}, \delta_2=-\frac{q}{2}$. Let $\mathbf{x}=(q,\omega_1,\omega_2)\in\mathbb{R}^3$. The reduced vector field is
\[
\dot{\mathbf{x}}
=
\begin{pmatrix}
\dot q\\
\dot\omega_1\\
\dot\omega_2
\end{pmatrix}
=
\begin{pmatrix}
\omega_1-\omega_2\\[0.4ex]
\frac12\bigl(-1-2\omega_1-\sin(q)+\cos(q)\bigr)\\[0.4ex]
\frac12\bigl(-1-2\omega_2+\sin(q)+\cos(q)\bigr)
\end{pmatrix}.
\]

\begin{itemize}[
    label=\textendash,
    leftmargin=1.55em,
    labelsep=0.45em,
    topsep=0pt,
    itemsep=0pt
]
    \item $x^\star=(0,0,0)$.

    \item Original $
    \mathcal{N}
    =
    \left\{
    (q,\omega_1,\omega_2)\in\mathbb{R}^3
    \;\middle|\;
    e^{-6}
    \le
    \frac12 q^2+\omega_1^2+\omega_2^2,\;
    |q|\le1.5,\;
    |\omega_i|\le2,\ i=1,2
    \right\}.
$
    \item Original verified type:
    local $\epsilon$-asymptotic stability.

    \item Verified method in the original paper: dReal.

    \item Our $
    \mathcal{N}
    =
    \left\{
    (q,\omega_1,\omega_2)\in\mathbb{R}^3
    \;\middle|\;
    |q|\le1.5,\;
    |\omega_i|\le2,\ i=1,2
    \right\}.
    $

    \item Our target:
    local asymptotic stability of the origin.
\end{itemize}
\end{systeminstance}

\begin{systeminstance}{Table~1 row~7, \cite{zou2025analytical}}
\label{sys:catalogue-sys0230}

The vector flow field is:
\[
\dot{\mathbf{x}}
=
\begin{pmatrix}
\dot{x}_1\\ \dot{x}_2\\ \dot{x}_3\\ \dot{x}_4\\ \dot{x}_5\\ \dot{x}_6
\end{pmatrix}
=
\begin{pmatrix}
-x_1+0.5x_2-0.1x_5^2\\
-0.5x_1-x_2\\
-x_3+0.5x_4-0.1x_1^2\\
-0.5x_3-x_4\\
-x_5+0.5x_6\\
-0.5x_5-x_6+0.1x_2^2
\end{pmatrix}.
\]
\begin{itemize}[label=\textendash, leftmargin=1.55em,
                labelsep=0.45em, topsep=0pt, itemsep=0pt]
    \item $x^\star=(0,0,0,0,0,0)$.
    \item Original $\mathcal{N}=\{\mathbf{x}\in\mathbb{R}^{6}\mid e^{-6}\leq\sum_{i=1}^{6}x_i^2,\;|x_i|\leq1,\ i=1,\ldots,6\}$.
    \item Original verified type: local  $\epsilon$-asymptotic stability.
    \item Verified method in the original paper: dReal.
    \item Our $\mathcal{N}=\{\mathbf{x}\in\mathbb{R}^{6}\mid |x_i|\leq1,\ i=1,\ldots,6\}$.
    \item Our verified type: local asymptotic stability.
\end{itemize}
\end{systeminstance}

\begin{systeminstance}{Table~1 row~8, \cite{zou2025analytical}}
\label{sys:catalogue-sys0231}

The vector flow field is:
\[
\dot{\mathbf{x}}
=
\begin{pmatrix}
\dot{x}_1\\ \dot{x}_2\\ \dot{x}_3\\ \dot{x}_4\\ \dot{x}_5\\ \dot{x}_6
\end{pmatrix}
=
\begin{pmatrix}
x_2\\
-\tfrac32x_6x_4-2x_1-\tfrac52x_2-\tfrac12x_4\sin(x_2)\cos(x_4)\\
x_4\\
\tfrac32x_6x_2-2x_3-10x_4+\tfrac12x_2\sin(x_2)\cos(x_4)\\
x_6\\
-5x_5-\tfrac45x_6
\end{pmatrix}.
\]
\begin{itemize}[label=\textendash, leftmargin=1.55em,
                labelsep=0.45em, topsep=0pt, itemsep=0pt]
    \item $x^\star=(0,0,0,0,0,0)$.
    \item Original $\mathcal{N}=\{\mathbf{x}\in\mathbb{R}^{6}\mid e^{-6}\leq\sum_{i=1}^{6}x_i^2,\;|x_i|\leq3,\ i=1,\ldots,6\}$.
    \item Original verified type: local  $\epsilon$-asymptotic stability.
    \item Verified method in the original paper: dReal and LaSalle's invariance principle.
    \item Our $\mathcal{N}=\{\mathbf{x}\in\mathbb{R}^{6}\mid |x_i|\leq3,\ i=1,\ldots,6\}$.
    \item Our verified type: local asymptotic stability.
\end{itemize}
\end{systeminstance}

\begin{systeminstance}{Table~1 row~9, \cite{zou2025analytical}}
\label{sys:catalogue-sys0232}

The source system is expressed in the center-of-inertia coordinates
$(\delta_1,\delta_2,\delta_3,\omega_1,\omega_2,\omega_3)$, which satisfy
$\delta_1+\delta_2+\delta_3=0$. We therefore remove the redundant
common-angle coordinate and define
$\delta_3:=-(\delta_1+\delta_2)$ and
$\bar{\omega}:=\frac13(\omega_1+\omega_2+\omega_3)$.
Let
$\mathbf{x}=(\delta_1,\delta_2,\omega_1,\omega_2,\omega_3)
\in\mathbb{R}^5$.
The reduced vector field is
\[
\dot{\mathbf{x}}
=
\begin{pmatrix}
\dot{\delta}_1\\
\dot{\delta}_2\\
\dot{\omega}_1\\
\dot{\omega}_2\\
\dot{\omega}_3
\end{pmatrix}
=
\begin{pmatrix}
\omega_1-\bar{\omega}\\[0.4ex]
\omega_2-\bar{\omega}\\[0.4ex]
-\omega_1-\frac12\sum_{j=1}^{3}\sin(\delta_1-\delta_j)\\[0.4ex]
-\omega_2-\frac12\sum_{j=1}^{3}\sin(\delta_2-\delta_j)\\[0.4ex]
-\omega_3-\frac12\sum_{j=1}^{3}\sin(\delta_3-\delta_j)
\end{pmatrix}.
\]

\begin{itemize}[
label=\textendash,
leftmargin=1.55em,
labelsep=0.45em,
topsep=0pt,
itemsep=0pt
]
\item $x^\star=(0,0,0,0,0)$.

\item Original
$\mathcal{N}
=
\{(\delta_1,\delta_2,\omega_1,\omega_2,\omega_3)\in\mathbb{R}^5
\mid
e^{-6}\le
\delta_1^2+\delta_2^2+(\delta_1+\delta_2)^2
+\sum_{i=1}^{3}\omega_i^2,
\allowbreak\;
|\delta_1|\le0.75,\;
|\delta_2|\le0.75,\;
|\delta_1+\delta_2|\le0.75,
\allowbreak\;
|\omega_i|\le1.2,\ i=1,2,3\}$.
\item Original verified type:
local $\epsilon$-asymptotic stability.

\item Verified method in the original paper:
dReal and LaSalle's invariance principle.

\item Our
$\mathcal{N}
=
\{(\delta_1,\delta_2,\omega_1,\omega_2,\omega_3)\in\mathbb{R}^5
\mid
|\delta_1|\leq0.75,\;
\allowbreak
|\delta_2|\leq0.75,\;
\allowbreak
|\delta_1+\delta_2|\leq0.75,\;
\allowbreak
|\omega_i|\leq1.2,\ i=1,2,3
\}$.

\item Our target:
local asymptotic stability of the origin.

\end{itemize}
\end{systeminstance}

\begin{remark}
The original system is stated in coordinates with $\delta_1+\delta_2+\delta_3=0$. We substitute $\delta_3 := -(\delta_1+\delta_2)$ and impose $|\delta_1+\delta_2|\le0.75$ to eliminate the equilibrium line $\{(c,c,c,0,0,0)\}$ present in the six-dimensional formulation. This reduction makes the single-point stability problem well-defined.
\end{remark}

\begin{systeminstance}{Table~1 row~10, \cite{zou2025analytical}}
\label{sys:catalogue-sys0233}

The vector flow field is:
\[
\dot{\mathbf{x}}
=
\begin{pmatrix}
\dot{x}_1\\ \dot{x}_2\\ \dot{x}_3\\ \dot{x}_4\\ \dot{x}_5\\
\dot{x}_6\\ \dot{x}_7\\ \dot{x}_8\\ \dot{x}_9
\end{pmatrix}
=
\begin{pmatrix}
-x_1+0.5x_2-0.1x_5^2\\
-0.5x_1-x_2+0.1x_8\\
-x_3+0.5x_4-0.1x_1^2\\
-0.5x_3-x_4\\
-x_5+0.5x_6\\
-0.5x_5-x_6+0.1x_2^2\\
x_8\\
-\sin(x_7)\cos(x_7)-x_8-\sin(x_9)\cos(x_9)-0.1x_2\\
x_8-x_9
\end{pmatrix}.
\]
\begin{itemize}[label=\textendash, leftmargin=1.55em,
                labelsep=0.45em, topsep=0pt, itemsep=0pt]
    \item $x^\star=(0,0,0,0,0,0,0,0,0)$.
    \item Original $\mathcal{N}=\{\mathbf{x}\in\mathbb{R}^{9}\mid e^{-6}\leq\sum_{i=1}^{9}x_i^2,\;|x_i|\leq1.5,\ i=1,\ldots,9\}$.
    \item Original verified type: local  $\epsilon$-asymptotic stability.
    \item Verified method in the original paper: dReal.
    \item Our $\mathcal{N}=\{\mathbf{x}\in\mathbb{R}^{9}\mid |x_i|\leq1.5,\ i=1,\ldots,9\}$.
    \item Our verified type: local asymptotic stability.
\end{itemize}
\end{systeminstance}

\begin{systeminstance}{Appendix Table~6, \cite{zou2025analytical}}
\label{sys:catalogue-sys0235}

The vector flow field is:
\[
\dot{\mathbf{x}}
=
\begin{pmatrix}
\dot{x}_1\\
\dot{x}_2\\
\dot{x}_3
\end{pmatrix}
=
\begin{pmatrix}
-8x_1x_2^2-10x_2^4\\
-8x_2^3+3x_2^2-8x_2\\
-x_3
\end{pmatrix}.
\]
\begin{itemize}[label=\textendash, leftmargin=1.55em,
                labelsep=0.45em, topsep=0pt, itemsep=0pt]
    \item Original $\mathcal{N}=\{\mathbf{x}\in\mathbb{R}^{3}\mid e^{-6}\leq x_1^2+x_2^2+x_3^2,\;|x_i|\leq1,\ i=1,2,3\}$.
    \item Original verified type: local  $\epsilon$-stability
    \item Verified method in the original paper: dReal.
    \item Our $\mathcal{N}=\{\mathbf{x}\in\mathbb{R}^{3}\mid |x_i|\leq1,\ i=1,2,3\}$.
    \item Our verified type: local stability.
\end{itemize}
\end{systeminstance}

\begin{systeminstance}{Main theorem, \cite{liu2026globally}}
\label{sys:liu2026-globally}

The vector flow field is:
\[
 \dot{\mathbf{x}}=
 \begin{pmatrix}\dot{x}\\\dot{y}\end{pmatrix}
 =
 \begin{pmatrix}4x^3-x^2y-6xy^2-y^3\\x^3+4x^2y+xy^2-6y^3\end{pmatrix}.
\]
\begin{itemize}[label=\textendash, leftmargin=1.55em,
                labelsep=0.45em, topsep=0pt, itemsep=0pt]
    \item $x^\star=(0,0)$.
    \item Original $\mathcal{N}=\mathbb{R}^{2}$.
    \item Original verified type: global asymptotic stability.
    \item Verified method in the original paper: human verified by product.
    \item Our $\mathcal{N}=\mathbb{R}^{2}$.
    \item Our verified type: global asymptotic stability.
\end{itemize}
\end{systeminstance}

\subsection{Problem 2 : Region-of-attraction verification}

\begin{systeminstance}{Section~5.3 and Appendix~C.3.3, \cite{zhou2022neural}}
\label{sys:catalogue-sys0033}

The closed-loop vector flow field is:
\[
\dot{\mathbf{x}}
=
\begin{pmatrix}
\dot{\theta}\\
\dot{\omega}
\end{pmatrix}
=
\begin{pmatrix}
\omega\\
\dfrac{mg\ell\sin(\theta)+u-0.1\omega}{m\ell^2}
\end{pmatrix}.
\]
\[
u=20\tanh(-23.286329\theta-5.27055\omega),
\]
where $m=0.15,\ell=0.5$ and $g=9.81$.
\begin{itemize}[label=\textendash, leftmargin=1.55em,
                labelsep=0.45em, topsep=0pt, itemsep=0pt]
    \item $x^\star=(0,0)$.
    \item Original $\mathcal{N}=\{\mathbf{x}\in\mathbb{R}^{2}\mid0.4^2\leq x_1^2+x_2^2\leq4^2\}$.
    \item Original verified type: $\epsilon$-ROA.
    \item Verified method in the original paper: dReal.
    \item Our $\mathcal{N}=\{\mathbf{x}\in\mathbb{R}^{2}\mid x_1^2+x_2^2\leq4^2\}$.
    \item Our verified type: ROA verification.
\end{itemize}
\end{systeminstance}

\begin{systeminstance}{Section~5.1 and Appendix~C.3.1, \cite{zhou2022neural}}
\label{sys:catalogue-sys0034}

The vector flow field is:
\[
\dot{\mathbf{x}}
=
\begin{pmatrix}
\dot{x}_1\\
\dot{x}_2
\end{pmatrix}
=
\begin{pmatrix}
-x_2\\
x_1+(x_1^2-1)x_2
\end{pmatrix}.
\]
\begin{itemize}[label=\textendash, leftmargin=1.55em,
                labelsep=0.45em, topsep=0pt, itemsep=0pt]
    \item $x^\star=(0,0)$.
    \item Original $\mathcal{N}=\{\mathbf{x}\in\mathbb{R}^{2}\mid0.2^2\leq x_1^2+x_2^2\leq1.2^2\}$.
    \item Original verified type: $\epsilon$-ROA.
    \item Verified method in the original paper: dReal.
    \item Our $\mathcal{N}=\{\mathbf{x}\in\mathbb{R}^{2}\mid x_1^2+x_2^2\leq1.2^2\}$.
    \item Our verified type: ROA verification.
\end{itemize}
\end{systeminstance}

\begin{systeminstance}{Example~3.1, \cite{liu2023towards}}
\label{sys:catalogue-sys0036}

The vector flow field is:
\[
\dot{\mathbf{x}}
=
\begin{pmatrix}
\dot{x}
\end{pmatrix}
=
\begin{pmatrix}
-x+x^3
\end{pmatrix}.
\]
\begin{itemize}[label=\textendash, leftmargin=1.55em,
                labelsep=0.45em, topsep=0pt, itemsep=0pt]
    \item $x^\star=0$.
    \item Original $\mathcal{N}=(-1,1)$.
    \item Original verified type: analytical ROA construction.
    \item Verified method in the original paper: analytical Zubov construction.
    \item Our $\mathcal{N}=(-1,1)$.
    \item Our verified type: ROA verification.
\end{itemize}
\end{systeminstance}

\begin{systeminstance}{Example~6.2 and Table~II, \cite{liu2023towards}}
\label{sys:catalogue-sys0047}

The vector flow field is:
\[
\dot{\mathbf{x}}
=
\begin{pmatrix}
\dot{x}_1\\
\dot{x}_2
\end{pmatrix}
=
\begin{pmatrix}
x_2\\
-2x_1+\tfrac13x_1^3-x_2
\end{pmatrix}.
\]
\begin{itemize}[label=\textendash, leftmargin=1.55em,
                labelsep=0.45em, topsep=0pt, itemsep=0pt]
    \item $x^\star=(0,0)$.
    \item Original $\mathcal{N}=[-6,6]^2$.
    \item Original verified type: ROA verification.
    \item Verified method in the original paper: dReal with Jacobian-based local verification.
    \item Our $\mathcal{N}=[-6,6]^2$.
    \item Our verified type: ROA verification.
\end{itemize}
\end{systeminstance}

\begin{systeminstance}{Table~4 \#3 (LorenzSystem), \cite{edwards2024fossil}}
\label{sys:fossil2-f2-03-lorenz-roa}

The vector flow field is:
\[
\dot{\mathbf{x}}
=
\begin{pmatrix}
\dot{x}_1\\
\dot{x}_2\\
\dot{x}_3
\end{pmatrix}
=
\begin{pmatrix}
-10(x_1-x_2)+u_1\\
28x_1-x_2-x_1x_3+u_2\\
x_1x_2-\dfrac83x_3+u_3
\end{pmatrix}.
\]
\begin{itemize}[label=\textendash, leftmargin=1.55em,
                labelsep=0.45em, topsep=0pt, itemsep=0pt]
    \item $x^\star=(0,0,0)$.
    \item Original $\mathcal{N}=\{\mathbf{x}\in\mathbb{R}^{3}\mid0.01^2\leq x_1^2+x_2^2+x_3^2\leq2^2\}$.
    \item $\mathcal{X}_I=\{\mathbf{x}\in\mathbb{R}^{3}\mid x_1^2+x_2^2+x_3^2\leq0.3^2\}$.
    \item $u_1,u_2,u_3 \in \mathbb{R}$\
    \item Original verified type: $\epsilon$-ROA.
    \item Verified method in the original paper: dReal.
    \item Our $\mathcal{N}=\{\mathbf{x}\in\mathbb{R}^{3}\mid x_1^2+x_2^2+x_3^2\leq2^2\}$.
    \item Our verified type: ROA verification.
\end{itemize}
\end{systeminstance}

\begin{systeminstance}{Example~5.2 and Table~2, \cite{liu2024tool}}
\label{sys:catalogue-sys0152}
The vector flow field is:
\[
\dot{\mathbf{x}}
=
\begin{pmatrix}
\dot{x}_1\\
\dot{x}_2
\end{pmatrix}
=
\begin{pmatrix}
x_2\\
-0.5x_2-\sin\left(x_1+\dfrac{\pi}{3}\right)
+\sin\left(\dfrac{\pi}{3}\right)
\end{pmatrix}.
\]
\begin{itemize}[label=\textendash, leftmargin=1.55em,
                labelsep=0.45em, topsep=0pt, itemsep=0pt]
    \item $x^\star=(0,0)$.
    \item Original $\mathcal{N}=[-2,3]\times[-3,1.5]$.
    \item Original verified type: ROA verification.
    \item Verified method in the original paper: dReal with Jacobian-based local verification.
    \item Our $\mathcal{N}=[-2,3]\times[-3,1.5]$.
    \item Our verified type: ROA verification.
\end{itemize}
\end{systeminstance}

\begin{systeminstance}{Example~5.4 and Table~3, \cite{liu2024tool}}
\label{sys:catalogue-sys0154}

The vector flow field is:
\[
\dot{\mathbf{x}}
=
\begin{pmatrix}
\dot{x}_1\\ \dot{x}_2\\ \dot{x}_3\\ \dot{x}_4\\ \dot{x}_5\\
\dot{x}_6\\ \dot{x}_7\\ \dot{x}_8\\ \dot{x}_9\\ \dot{x}_{10}
\end{pmatrix}
=
\begin{pmatrix}
-x_1+0.5x_2-0.1x_9^2\\
-0.5x_1-x_2\\
-x_3+0.5x_4-0.1x_1^2\\
-0.5x_3-x_4\\
-x_5+0.5x_6+0.1x_7^2\\
-0.5x_5-x_6\\
-x_7+0.5x_8\\
-0.5x_7-x_8\\
-x_9+0.5x_{10}\\
-0.5x_9-x_{10}+0.1x_2^2
\end{pmatrix}.
\]
\begin{itemize}[label=\textendash, leftmargin=1.55em,
                labelsep=0.45em, topsep=0pt, itemsep=0pt]
    \item $x^\star=\mathbf{0}$.
    \item Original $\mathcal{N}=[-1,1]^{10}$.
    \item Original verified type: ROA verification.
    \item Verified method in the original paper: dReal with Jacobian-based local verification.
    \item Our $\mathcal{N}=[-1,1]^{10}$.
    \item Our verified type: ROA verification.
\end{itemize}
\end{systeminstance}

\begin{systeminstance}{Section~5.1, with verification details in Appendix~E.2
(\(n=1\)), \cite{meng2024physics}}
\label{sys:catalogue-sys0254}

The closed-loop scalar dynamics are:
\[
\dot{x}=-x-x^3,
\]

\begin{itemize}[label=\textendash, leftmargin=1.55em,
                labelsep=0.45em, topsep=0pt, itemsep=0pt]
    \item Equilibrium: $x^\star=0$.
    \item Original $
    \mathcal{N}
    =\{x\in\mathbb{R}\mid 0.1\le |x|\le1\}.
    $
    \item Original verified type: $\epsilon$-ROA
    \item Verified method in the original paper: dReal.
    \item Our $
    \mathcal{N}
    =\{x\in\mathbb{R}\mid |x|\leq1\}.
    $
    \item Our verified type: ROA verification.
\end{itemize}
\end{systeminstance}

\subsection{Problem 3 :Forward-invariant set synthesis}

\begin{systeminstance}{Appendix~A.1 (PVTOL), \cite{wu2023neural}}
\label{sys:dt-pvtol6-invariance}

Let the state be ordered as
\[
\mathbf{x}
=
(p_x,p_z,\phi,v_x,v_z,\omega)^\top \in \mathbb{R}^{6},
\]
where $(p_x,p_z)$ are the planar position, $\phi$ is the roll
angle, $(v_x,v_z)$ are the body-frame velocities, and $\omega$
is the angular rate. Using the sampling period $\Delta t=0.05$,
the discrete-time transition map is:
\[
\mathbf{x}^{+}
=
F(\mathbf{x},\mathbf{u})
=
\begin{pmatrix}
p_x + 0.05\left(v_x\cos\phi-v_z\sin\phi\right)
\\[1ex]
p_z + 0.05\left(v_x\sin\phi+v_z\cos\phi\right)
\\[1ex]
\phi + 0.05\,\omega
\\[1ex]
v_x + 0.05\left(v_z\omega-9.8\sin\phi\right)
\\[1ex]
v_z + 0.05\left(
-v_x\omega-9.8\cos\phi+\dfrac{u_1+u_2}{4}
\right)
\\[1ex]
\omega + 0.05\left(
\dfrac{0.25}{0.0475}(u_1-u_2)
\right)
\end{pmatrix},
\]
where $\mathbf{u}=(u_1,u_2)^\top\in[0,39.2]^2.$

\begin{itemize}[label=\textendash, leftmargin=1.55em,
                labelsep=0.45em, topsep=0pt, itemsep=0pt]
    \item $
    \mathcal{X}
    =
    \left\{
    \mathbf{x}\in\mathbb{R}^{6}
    \mid
    \lVert\mathbf{x}\rVert_\infty\leq 1
    \right\}.
    $

    \item Unsafe region:
    $
    \mathcal{X}_U
    =
    \left\{
    \mathbf{x}\in\mathbb{R}^{6}
    \mid
    \lVert\mathbf{x}\rVert_\infty>1
    \right\}.
    $
    \item Verification target: forward-invariance verification.
\end{itemize}

\end{systeminstance}

\begin{systeminstance}{Example~1, \cite{samanipour2026relu}}
\label{sys:inverted-pendulum-saturation}

The vector flow field is:
\[
\dot{\mathbf{x}} =
\begin{pmatrix}
\dot{x}_1\\
\dot{x}_2
\end{pmatrix}
=
\begin{pmatrix}
x_2\\[1ex]
\sin(x_1) + u
\end{pmatrix},
\]
where $u = -3x_1 - 3x_2$.

\begin{itemize}[label=\textendash, leftmargin=1.55em,
                labelsep=0.45em, topsep=0pt, itemsep=0pt]
    \item $\mathcal{X} = \{x\in\mathbb{R}^{2}\mid \lVert x\rVert_\infty \le \pi\}$.
    \item $\mathcal{X}_U = \{x\in\mathbb{R}^{2}\mid \lVert x\rVert_\infty > \pi\}$.
    \item Verified type: forward-invariance verification.
    \item Verified method in the original paper: linear programming and facet-wise verification.
\end{itemize}

\end{systeminstance}

\begin{systeminstance}{Example~2, \cite{samanipour2026relu}}
\label{sys:2d-quadratic-system}

The vector flow field is:
\[
\dot{\mathbf{x}} = 
\begin{pmatrix}
\dot{x}_{1}\\
\dot{x}_{2}
\end{pmatrix}
=
\begin{pmatrix}
x_{1}^2 + x_{1}x_{2} + x_{1}\\[1ex]
x_{1}x_{2} + x_{2}^2 + x_{2}
\end{pmatrix}.
\]

\begin{itemize}[label=\textendash, leftmargin=1.55em,
                labelsep=0.45em, topsep=0pt, itemsep=0pt]
    \item $\mathcal{X} = \{x\in\mathbb{R}^{2}\mid \lVert x\rVert_\infty \le 2\}$.
    \item $\mathcal{X}_U = \{x\in\mathbb{R}^{2}\mid -1 \le x_1, x_2 \le 0\}$.
    \item Verified type: forward-invariance verification.
    \item Verified method in the original paper: linear programming and facet-wise verification.
\end{itemize}
 
\end{systeminstance}

\begin{systeminstance}{Example~3, \cite{samanipour2026relu}}
\label{sys:cartpole-saturation}

The vector flow field is:
\[
\dot{\mathbf{x}} =
\begin{pmatrix}
\dot{x}_1\\
\dot{x}_2\\
\dot{x}_3\\
\dot{x}_4
\end{pmatrix}
=
\begin{pmatrix}
x_2\\[1ex]
\dfrac{u + 0.1\,x_4^2 \sin x_3 - 0.981 \sin x_3 \cos x_3}{1 + 0.1\sin^2 x_3}\\[2ex]
x_4\\[1ex]
\dfrac{10.791 \sin x_3 - u \cos x_3 - 0.1\,x_4^2 \sin x_3 \cos x_3}{1 + 0.1\sin^2 x_3}
\end{pmatrix},
\]
where $u = x_1 + 2.4109\,x_2 + 34.3620\,x_3 + 10.7009\,x_4$.

\begin{itemize}[label=\textendash, leftmargin=1.55em,
                labelsep=0.45em, topsep=0pt, itemsep=0pt]
    \item $\mathcal{X} = \{x\in\mathbb{R}^{4}\mid \lVert x\rVert_\infty \le 1\}$.
    \item $\mathcal{X}_U = \{x\in\mathbb{R}^{4}\mid \lVert x\rVert_\infty > 1\}$.
    \item Verified type: forward-invariance verification.
    \item Verified method in the original paper: linear programming and facet-wise verification.
\end{itemize}

\end{systeminstance}

\begin{systeminstance}{Section~6.1.1, \cite{vaisi2025extracting}}
\label{sys:dt-inverted-pendulum}

The discrete-time transition map is:
\[
\mathbf{x}^{+}
=
F(\mathbf{x})
=
\begin{pmatrix}
x_1 + 0.01\,x_2\\[1ex]
x_2 + 0.01(9.8\sin x_1 + u)
\end{pmatrix},
\]
where $u \in [-10, 10]$.

\begin{itemize}[label=\textendash, leftmargin=1.55em,
                labelsep=0.45em, topsep=0pt, itemsep=0pt]
    \item $\mathcal{X} = \{x\in\mathbb{R}^{2}\mid \lVert x\rVert_\infty \le \pi/4\}$.
    \item $\mathcal{X}_U = \{x\in\mathbb{R}^{2}\mid \lVert x\rVert_\infty > \pi/4\}$.
    \item Verified type: forward-invariance verification.
    \item Verified method in the original paper: neural-network reachability and hyperplane-region enumeration.
\end{itemize}

\end{systeminstance}

\begin{systeminstance}{Section~6.1.2, \cite{vaisi2025extracting}}
\label{sys:dt-steerable-bicycle}

The discrete-time transition map is:
\[
\mathbf{x}^{+}
=
F(\mathbf{x})
=
\begin{pmatrix}
x_2\\[1ex]
29.4 \sin x_1 + 300 \cos x_1 \tan x_3 + 15 u \cos x_1 \cos^2 x_3\\[1ex]
u
\end{pmatrix},
\]
where $u \in [-10, 10]$.

\begin{itemize}[label=\textendash, leftmargin=1.55em,
                labelsep=0.45em, topsep=0pt, itemsep=0pt]
    \item $\mathcal{X} = \{x\in\mathbb{R}^{3}\mid \lVert x\rVert_\infty \le 2.2 \setminus x_3 = \pm\frac{\pi}{2}\}$.
    \item $\mathcal{X}_U = \{x\in\mathbb{R}^{3}\mid \lVert x\rVert_\infty > 2.2\}$.
    \item Verified type: forward-invariance verification.
    \item Verified method in the original paper: neural-network reachability and hyperplane-region enumeration.
\end{itemize}

\end{systeminstance}

\subsection{Problem 4 :Safety verification}

\begin{systeminstance}{Theorem~4, \cite{sogokon2018vector}}
\label{sys:catalogue-sys0009}

The vector flow field is:
\[
\dot{\mathbf{x}}
=
\begin{pmatrix}
\dot{x}_1\\
\dot{x}_2
\end{pmatrix}
=
\begin{pmatrix}
x_2\\
x_1
\end{pmatrix}.
\]
\begin{itemize}[label=\textendash, leftmargin=1.55em,
                labelsep=0.45em, topsep=0pt, itemsep=0pt]
    \item $\mathcal{X}=\mathbb{R}^{2}$.
    \item $\mathcal{X}_I= \{(x_1,x_2)\mid x_1\leq0\wedge x_2\leq0\}$.
    \item $\mathcal{X}_U=\{(x_1,x_2)\mid x_1>0\}$.
    \item Verified type: barrier/safety verification.
    \item Verified method in the original paper: vector barrier certificates.
\end{itemize}
\end{systeminstance}

\begin{systeminstance}{Example~1, \cite{sogokon2018vector}}
\label{sys:catalogue-sys0010}

The vector flow field is:
\[
\dot{\mathbf{x}}
=
\begin{pmatrix}
\dot{x}_1\\
\dot{x}_2\\
\dot{x}_3
\end{pmatrix}
=
\begin{pmatrix}
2x_1+x_2+2x_3-x_1^2+x_1x_3-x_3^2\\
-2+x_1-x_2-x_2^2\\
-2-x_2-x_3+x_1^2-x_1x_3
\end{pmatrix}.
\]
\begin{itemize}[label=\textendash, leftmargin=1.55em,
                labelsep=0.45em, topsep=0pt, itemsep=0pt]
    \item $\mathcal{X}=\mathbb{R}^{3}$.
    \item $\mathcal{X}_I= \{\mathbf{x}\in\mathbb{R}^{3}\mid-x_i\geq0,\ i=1,2,3\}$.
    \item $\mathcal{X}_U= \{\mathbf{x}\in\mathbb{R}^{3}\mid x_1+x_2+x_3\geq1\}$.
    \item Verified type: barrier/safety verification.
    \item Verified method in the original paper: vector barrier certificates and SOS programming.
\end{itemize}
\end{systeminstance}

\begin{systeminstance}{Example~3.1, \cite{zhao2020synthesizing}}
\label{sys:catalogue-sys0011}

The vector flow field is:
\[
\dot{\mathbf{x}}
=
\begin{pmatrix}
\dot{x}_1\\
\dot{x}_2
\end{pmatrix}
=
\begin{pmatrix}
x_2\\
-x_1+\tfrac13x_1^3-x_2
\end{pmatrix}.
\]
\begin{itemize}[label=\textendash, leftmargin=1.55em,
                labelsep=0.45em, topsep=0pt, itemsep=0pt]
    \item $\mathcal{X}= \{\mathbf{x}\in\mathbb{R}^{2}\mid-3.5\leq x_1\leq2,\ -2\leq x_2\leq1\}$.
    \item $\mathcal{X}_I= \{\mathbf{x}\in\mathbb{R}^{2}\mid(x_1-1.5)^2+x_2^2\leq0.25\}$.
    \item $\mathcal{X}_U= \{\mathbf{x}\in\mathbb{R}^{2}\mid(x_1+1)^2+(x_2+1)^2\leq0.16\}$.
    \item Verified type: barrier/safety verification.
    \item Verified method in the original paper: iSAT3.
\end{itemize}
\end{systeminstance}

\begin{systeminstance}{Table~1 row~9 (Barr1), \cite{abate2021fossil}}
\label{sys:catalogue-sys0024}

The vector flow field is:
\[
\dot{\mathbf{x}}
=
\begin{pmatrix}
\dot{x}_1\\
\dot{x}_2
\end{pmatrix}
=
\begin{pmatrix}
x_2+2x_1x_2\\
-x_1-x_2^2+2x_1^2
\end{pmatrix}.
\]
\begin{itemize}[label=\textendash, leftmargin=1.55em,
                labelsep=0.45em, topsep=0pt, itemsep=0pt]
    \item $\mathcal{X}=[-2,2]^2$.
    \item $\mathcal{X}_I=[0,1]\times[1,2]$.
    \item $\mathcal{X}_U= \{(x_1,x_2)\in\mathbb{R}^{2}\mid x_1+x_2^2\leq0\}$.
    \item Verified type: barrier/safety verification.
    \item Verified method in the original paper: dReal.
\end{itemize}
\end{systeminstance}

\begin{systeminstance}{Table~1 row~12 (Barr4), \cite{abate2021fossil}}
\label{sys:catalogue-sys0027}

The vector flow field is:
\[
\dot{\mathbf{x}}
=
\begin{pmatrix}
\dot{x}_1\\
\dot{x}_2\\
\dot{\phi}
\end{pmatrix}
=
\begin{pmatrix}
\sin(\phi)\\
\cos(\phi)\\
-\sin(\phi)+
\dfrac{3(x_1\sin(\phi)+x_2\cos(\phi))}
{0.5+x_1^2+x_2^2}
\end{pmatrix}.
\]
\begin{itemize}[label=\textendash, leftmargin=1.55em,
                labelsep=0.45em, topsep=0pt, itemsep=0pt]
    \item $\mathcal{X}=[-2,2]\times[-2,2]\times[-1.57,1.57]$.
    \item $\mathcal{X}_I=[-0.1,0.1]\times[-2,-1.8]\times[-0.52,0.52]$.
    \item $\mathcal{X}_U= \{(x_1,x_2,\phi)\in\mathbb{R}^{3}\mid x_1^2+x_2^2\leq0.04\}$.
    \item Verified type: barrier/safety verification.
    \item Verified method in the original paper: dReal.
\end{itemize}
\end{systeminstance}

\begin{systeminstance}{Table~1 row~17 (Hi-Ord8), \cite{abate2021fossil}}
\label{sys:catalogue-sys0032}

The vector flow field is:
\[
\dot{\mathbf{x}}
=
\begin{pmatrix}
\dot{x}_0\\ \dot{x}_1\\ \dot{x}_2\\ \dot{x}_3\\
\dot{x}_4\\ \dot{x}_5\\ \dot{x}_6\\ \dot{x}_7
\end{pmatrix}
=
\begin{pmatrix}
x_1\\
x_2\\
x_3\\
x_4\\
x_5\\
x_6\\
x_7\\
-\left(
\begin{aligned}
&20x_7+170x_6+800x_5+2273x_4\\[-0.15ex]
&{}+3980x_3+4180x_2+2400x_1+576x_0
\end{aligned}
\right)
\end{pmatrix}.
\]
\begin{itemize}[label=\textendash, leftmargin=1.55em,
                labelsep=0.45em, topsep=0pt, itemsep=0pt]
    \item $\mathcal{X}=[-2.2,2.2]^8$.
    \item $\mathcal{X}_I=[0.9,1.1]^8$.
    \item $\mathcal{X}_U=[-2.2,-1.8]^8$.
    \item Verified type: barrier/safety verification.
    \item Verified method in the original paper: dReal.
\end{itemize}
\end{systeminstance}

\begin{systeminstance}{Table~1 (arch2[36], instance~1), \cite{wu2024completeness}}
\label{sys:catalogue-sys0060}

The vector flow field is:
\[
\dot{\mathbf{x}}
=
\begin{pmatrix}
\dot{x}_1\\
\dot{x}_2
\end{pmatrix}
=
\begin{pmatrix}
x_1^2+x_2^2-1\\
5(x_1x_2-1)
\end{pmatrix}.
\]
\begin{itemize}[label=\textendash, leftmargin=1.55em,
                labelsep=0.45em, topsep=0pt, itemsep=0pt]
    \item $\mathcal{X}=\mathbb{R}^{2}$.
    \item $\mathcal{X}_I= \{\mathbf{x}\in\mathbb{R}^{2}\mid 2.25-x_1^2-(x_2-2)^2\geq0\}$.
    \item $\mathcal{X}_U= \{\mathbf{x}\in\mathbb{R}^{2}\mid -1-x_1\geq0\wedge x_1+3\geq0\wedge -2-x_2\geq0\wedge x_2+3\geq0\}$.
    \item Verified type: unbounded-domain barrier verification.
    \item Verified method in the original paper: SDP-based barrier synthesis.
\end{itemize}
\end{systeminstance}

\begin{systeminstance}{Table~1 (arch3[36], instance~1), \cite{wu2024completeness}}
\label{sys:catalogue-sys0062}

The vector flow field is:
\[
\dot{\mathbf{x}}
=
\begin{pmatrix}
\dot{x}_1\\
\dot{x}_2
\end{pmatrix}
=
\begin{pmatrix}
x_1-x_1^3+x_2-x_1x_2^2\\
-x_1+x_2-x_1^2x_2-x_2^3
\end{pmatrix}.
\]
\begin{itemize}[label=\textendash, leftmargin=1.55em,
                labelsep=0.45em, topsep=0pt, itemsep=0pt]
    \item $\mathcal{X}=\mathbb{R}^{2}$.
    \item $\mathcal{X}_I= \{\mathbf{x}\in\mathbb{R}^{2}\mid0.04-x_1^2-x_2^2\geq0\}$.
    \item $\mathcal{X}_U= \{\mathbf{x}\in\mathbb{R}^{2}\mid 0.25-(x_1-2.5)^2-(x_2-2.5)^2\geq0\}$.
    \item Verified type: unbounded-domain barrier verification.
    \item Verified method in the original paper: SDP-based barrier synthesis.
\end{itemize}
\end{systeminstance}

\begin{systeminstance}{Example~5 and Table~1, \cite{liu2015abstraction}}
\label{sys:catalogue-sys0005}

The vector flow field is:
\[
\dot{\mathbf{x}}
=
\begin{pmatrix}
\dot{x}_1\\
\dot{x}_2
\end{pmatrix}
=
\begin{pmatrix}
e^{-x_1}+x_2-1\\
-\sin^2(x_1)
\end{pmatrix}.
\]
\begin{itemize}[label=\textendash, leftmargin=1.55em,
                labelsep=0.45em, topsep=0pt, itemsep=0pt]
    \item $\mathcal{X}=[-2,2]^2$.
    \item $\mathcal{X}_I= \{(x_1,x_2)\mid(x_1+0.5)^2+(x_2-0.5)^2\leq0.16\}$.
    \item $\mathcal{X}_U= \{(x_1,x_2)\mid(x_1-0.7)^2+(x_2+0.7)^2\leq0.09\}$.
    \item Verified type: safety verification
    \item Verified method in the original paper: polynomial abstraction and SOS programming.
\end{itemize}
\end{systeminstance}

\begin{systeminstance}{Example~6 and Table~1, \cite{liu2015abstraction}}
\label{sys:catalogue-sys0006}

The vector flow field is:
\[
\dot{\mathbf{u}}
=
\begin{pmatrix}
\dot{u}_1\\
\dot{u}_2\\
\dot{u}_3
\end{pmatrix}
=
\begin{pmatrix}
-\dfrac{\beta c u_1u_2}{u_1+u_2+u_3}-\mu u_1\\
\dfrac{\beta c u_1u_2}{u_1+u_2+u_3}-(\mu+\nu)u_2\\
\nu u_2-\alpha u_3
\end{pmatrix},
\]
where $\beta=0.2$, $c=10$, $\mu=0.008$, $\alpha=0.95$, and $\nu=0.1$.
\begin{itemize}[label=\textendash, leftmargin=1.55em,
                labelsep=0.45em, topsep=0pt, itemsep=0pt]
    \item $\mathcal{X}=\{\mathbf{u}\in\mathbb{R}^{3}\mid u_i\geq0,\ u_1+u_2+u_3\leq10.013\}$ with not equal to 0 at the same time.
    \item $\mathcal{X}_I= [9.985,9.995]\times[0.005,0.015]\times[0,0.003]$.
    \item $\mathcal{X}_U=\{\mathbf{u}\mid u_3>1\}$.
    \item Verified type: safety verification.
    \item Verified method in the original paper: polynomial abstraction and SOS programming.
\end{itemize}
\end{systeminstance}

\begin{systeminstance}{Table~1 (lotka[14], instance~1), \cite{wu2024completeness}}
\label{sys:catalogue-sys0070}

The vector flow field is:
\[
\dot{\mathbf{x}}
=
\begin{pmatrix}
\dot{x}_1\\
\dot{x}_2\\
\dot{x}_3
\end{pmatrix}
=
\begin{pmatrix}
x_1(1-x_3)\\
x_2(1-2x_3)\\
x_3(x_1+x_2-1)
\end{pmatrix}.
\]
\begin{itemize}[label=\textendash, leftmargin=1.55em,
                labelsep=0.45em, topsep=0pt, itemsep=0pt]
    \item $\mathcal{X}=\mathbb{R}^{3}$.
    \item $\mathcal{X}_I= \{\mathbf{x}\in\mathbb{R}^{3}\mid 0.09-(x_1-1)^2-(x_2-1)^2-(x_3-1.5)^2\geq0\}$.
    \item $\mathcal{X}_U= \{\mathbf{x}\in\mathbb{R}^{3}\mid 0.09-(x_1+1)^2-(x_2+1)^2-(x_3+1.5)^2\geq0\}$.
    \item Verified type: unbounded-domain barrier verification.
    \item Verified method in the original paper: SDP-based barrier synthesis.
\end{itemize}
\end{systeminstance}

\begin{systeminstance}{Table~4 \#5 (ctrlbarr obstacle), \cite{edwards2024fossil}}
\label{sys:catalogue-sys0084}

The vector flow field is:
\[
\dot{\mathbf{x}}
=
\begin{pmatrix}
\dot{x}_1\\
\dot{x}_2\\
\dot{\phi}
\end{pmatrix}
=
\begin{pmatrix}
\sin(\phi)\\
\cos(\phi)\\
-\sin(\phi)+u_1
\end{pmatrix}.
\]
\begin{itemize}[label=\textendash, leftmargin=1.55em,
                labelsep=0.45em, topsep=0pt, itemsep=0pt]
    \item $\mathcal{X}=[-10,10]\times[-10,10]\times[-\pi,\pi]$.
    \item $\mathcal{X}_I=[4,6]\times[4,6]\times [-\tfrac{\pi}{2},\tfrac{\pi}{2}]$.
    \item $\mathcal{X}_U=[-9,-7]\times[-9,-6]\times [-\tfrac{\pi}{2},\tfrac{\pi}{2}]$.
    \item Verified type: barrier/safety verification.
    \item Verified method in the original paper: dReal.
\end{itemize}
\end{systeminstance}

\begin{systeminstance}{Table~4 \#17 (ctrlbarr 2roomtemp DT), \cite{edwards2024fossil}}
\label{sys:catalogue-sys0096}

The discrete-time transition map is:
\[
\mathbf{x}^{+}
=
F(\mathbf{x},\mathbf{u})
=
\begin{pmatrix}
0.725x_1+0.25x_2+0.375+0.018(55-x_1)u_1\\
0.25x_1+0.71x_2+0.6+0.018(55-x_2)u_2
\end{pmatrix}.
\]
\begin{itemize}[label=\textendash, leftmargin=1.55em,
                labelsep=0.45em, topsep=0pt, itemsep=0pt]
    \item $\mathcal{X}=[17,30]^2$.
    \item $\mathcal{X}_I=[17,18]^2$.
    \item $\mathcal{X}_U=[28,30]^2$.
    \item Verified type: barrier/safety verification.
    \item Verified method in the original paper: dReal.
\end{itemize}
\end{systeminstance}

\begin{systeminstance}{~\texttt{S001}, \cite{taheri2025barrierbench}}
\label{sys:bb-s001}

The vector flow field is:
\[
\dot{\mathbf{x}} =
\begin{pmatrix}
\dot{x}_{1}\\
\dot{x}_{2}\\
\dot{x}_{3}
\end{pmatrix}
=
\begin{pmatrix}
x_{2} + u_{0}\\
-0.8\,x_{1} - 0.15\,x_{2} + 0.2\,x_{1}^{2} + u_{1}\\
-0.1\,x_{3} + 0.1\,x_{1} + u_{2}
\end{pmatrix}.
\]

\begin{itemize}[label=\textendash, leftmargin=1.55em,
                labelsep=0.45em, topsep=0pt, itemsep=0pt]
    \item $\mathcal{X} = \mathbb{R}^{3}$.
    \item $\mathcal{X}_0 = \{x\in\mathbb{R}^{3}\mid \lVert x\rVert_2\le 0.4\}$.
    \item $\mathcal{X}_U = \mathbb{R}^{3}\setminus \{x\in\mathbb{R}^{3}\mid \lVert x\rVert_2\le 3\}$.
    \item Verified type: barrier/safety verification.
    \item Verified method in the original paper: Z3/Yices.
\end{itemize}

\end{systeminstance}

\begin{systeminstance}{~\texttt{S002}, \cite{taheri2025barrierbench}} 
\label{sys:bb-s002}

The vector flow field is:
\[
\dot{\mathbf{x}} =
\begin{pmatrix}
\dot{x}_{1}\\
\dot{x}_{2}\\
\dot{x}_{3}\\
\dot{x}_{4}
\end{pmatrix}
=
\begin{pmatrix}
x_{2} + u_{0}\\
-0.95\,\sin(x_{1}) - 0.02\,x_{2} + 0.01\,x_{3} + u_{1}\\
-0.1\,x_{3} + 0.02\,x_{1} + u_{2}\\
-0.03\,x_{4} + 0.01\,x_{2} + u_{3}
\end{pmatrix}.
\]

\begin{itemize}[label=\textendash, leftmargin=1.55em,
                labelsep=0.45em, topsep=0pt, itemsep=0pt]
    \item $\mathcal{X} = \mathbb{R}^{4}$.
    \item $\mathcal{X}_0 = \{x\in\mathbb{R}^{4}\mid \lVert x\rVert_2\le 0.3\}$.
    \item $\mathcal{X}_U = \mathbb{R}^{4}\setminus \{x\in\mathbb{R}^{4}\mid \lVert x\rVert_2\le 2.5\}$.
    \item Verified type: barrier/safety verification.
    \item Verified method in the original paper: Z3/Yices with Taylor approximations.
\end{itemize}
\end{systeminstance}

\begin{systeminstance}{~\texttt{S004}, \cite{taheri2025barrierbench}}
\label{sys:bb-s004}

The vector flow field is:
\[
\dot{\mathbf{x}} =
\begin{pmatrix}
\dot{x}_{1}\\
\dot{x}_{2}\\
\dot{x}_{3}\\
\dot{x}_{4}
\end{pmatrix}
=
\begin{pmatrix}
10\,(x_{2} - x_{1}) + u_{0}\\
28\,x_{1} - x_{2} - x_{1}\,x_{3} + u_{1}\\
x_{1}\,x_{2} - \frac{8}{3}\,x_{3} + u_{2}\\
-x_{2}\,x_{3} + 2.7\,x_{4} + u_{3}
\end{pmatrix}.
\]

\begin{itemize}[label=\textendash, leftmargin=1.55em,
                labelsep=0.45em, topsep=0pt, itemsep=0pt]
    \item $\mathcal{X} = \mathbb{R}^{4}$.
    \item $\mathcal{X}_0 = \{x\in\mathbb{R}^{4}\mid \lVert x\rVert_2\le 0.1\}$.
    \item $\mathcal{X}_U = \mathbb{R}^{4}\setminus \{x\in\mathbb{R}^{4}\mid \lVert x\rVert_2\le 25\}$.
    \item Verified type: barrier/safety verification.
    \item Verified method in the original paper: Z3/Yices.
\end{itemize}

\end{systeminstance}

\begin{systeminstance}{~\texttt{S008}, \cite{taheri2025barrierbench}}
\label{sys:bb-s008}

The vector flow field is:
\[
\dot{\mathbf{x}} =
\begin{pmatrix}
\dot{x}_{1}\\
\dot{x}_{2}\\
\dot{x}_{3}
\end{pmatrix}
=
\begin{pmatrix}
-2\,x_{1} + 6.7\,x_{2} - x_{2}\,x_{3} + u_{0}\\
x_{1} + u_{1}\\
-x_{3} + x_{2}^{2} + u_{2}
\end{pmatrix}.
\]

\begin{itemize}[label=\textendash, leftmargin=1.55em,
                labelsep=0.45em, topsep=0pt, itemsep=0pt]
    \item $\mathcal{X} = \mathbb{R}^{3}$.
    \item $\mathcal{X}_0 = \{x\in\mathbb{R}^{3}\mid \lVert x\rVert_2\le 0.2\}$.
    \item $\mathcal{X}_U = \mathbb{R}^{3}\setminus \{x\in\mathbb{R}^{3}\mid \lVert x\rVert_2\le 5\}$.
    \item Verified type: barrier/safety verification.
    \item Verified method in the original paper: Z3/Yices.
\end{itemize}

\end{systeminstance}

\begin{systeminstance}{~\texttt{S011}, \cite{taheri2025barrierbench}}
\label{sys:bb-s011}

The vector flow field is:
\[
\dot{\mathbf{x}} =
\begin{pmatrix}
\dot{x}_{1}\\
\dot{x}_{2}
\end{pmatrix}
=
\begin{pmatrix}
x_{2} + u_{0}\\
-0.3\,x_{1} - 0.2\,(1 + \cos(x_{1}))\,x_{2} - 0.1\,x_{1}^{3} + u_{1}
\end{pmatrix}.
\]

\begin{itemize}[label=\textendash, leftmargin=1.55em,
                labelsep=0.45em, topsep=0pt, itemsep=0pt]
    \item $\mathcal{X} = \mathbb{R}^{2}$.
    \item $\mathcal{X}_0 = \{x\in\mathbb{R}^{2}\mid \lVert x\rVert_2\le 0.6\}$.
    \item $\mathcal{X}_U = \{x\in\mathbb{R}^{2}\mid \lVert x-(2,2)^{\mathsf T}\rVert_2\le 1\}$.
    \item Verified type: barrier/safety verification.
    \item Verified method in the original paper: Z3/Yices with Taylor approximations.
\end{itemize}
\end{systeminstance}

\begin{systeminstance}{~\texttt{S019}, \cite{taheri2025barrierbench}}
\label{sys:bb-s019}

The vector flow field is:
\[
\dot{\mathbf{x}} =
\begin{pmatrix}
\dot{x}_{1}\\
\dot{x}_{2}\\
\dot{x}_{3}
\end{pmatrix}
=
\begin{pmatrix}
-x_{1} + x_{2}\,x_{3}\\
-x_{2} + x_{1}\,x_{3}\\
-x_{3} - x_{1}\,x_{2}
\end{pmatrix}.
\]

\begin{itemize}[label=\textendash, leftmargin=1.55em,
                labelsep=0.45em, topsep=0pt, itemsep=0pt]
    \item $\mathcal{X} = \mathbb{R}^{3}$.
    \item $\mathcal{X}_0 = \{x\in\mathbb{R}^{3}\mid \lVert x\rVert_2\le 0.2\}$.
    \item $\mathcal{X}_U = \mathbb{R}^{3}\setminus \{x\in\mathbb{R}^{3}\mid \lVert x\rVert_2\le 1\}$.
    \item Verified type: barrier/safety verification.
    \item Verified method in the original paper: Z3/Yices.
\end{itemize}

\end{systeminstance}

\begin{systeminstance}{~\texttt{S024}, \cite{taheri2025barrierbench}}
\label{sys:bb-s024}

The vector flow field is:
\[
\dot{\mathbf{x}} =
\begin{pmatrix}
\dot{x}_{1}\\
\dot{x}_{2}\\
\dot{x}_{3}\\
\dot{x}_{4}
\end{pmatrix}
=
\begin{pmatrix}
x_{2} + u_{0}\\
-x_{1} - 0.4\,x_{2} + 0.15\,\sin(x_{3}) + u_{1}\\
x_{4} + u_{2}\\
-0.6\,x_{3} - 0.3\,x_{4} + u_{3}
\end{pmatrix}.
\]

\begin{itemize}[label=\textendash, leftmargin=1.55em,
                labelsep=0.45em, topsep=0pt, itemsep=0pt]
    \item $\mathcal{X} = \mathbb{R}^{4}$.
    \item $\mathcal{X}_0 = [0.5,0.8]\times [0.3,0.6]\times [0.5,0.8]\times [0.3,0.6]$.
    \item $\mathcal{X}_U = [2.5,3.5]\times [2,3]\times [2.5,3.5]\times [2,3]\cup [-3.5,-2.5]\times [-3,-2]\times [-3.5,-2.5]\times [-3,-2]$.
    \item Verified type: barrier/safety verification.
    \item Verified method in the original paper: Z3/Yices with Taylor approximations.
\end{itemize}
\end{systeminstance}

\begin{systeminstance}{~\texttt{S028}, \cite{taheri2025barrierbench}}
\label{sys:bb-s028}

The vector flow field is:
\[
\dot{\mathbf{x}} =
\begin{pmatrix}
\dot{x}_{1}\\
\dot{x}_{2}\\
\dot{x}_{3}
\end{pmatrix}
=
\begin{pmatrix}
-0.4\,x_{1} + 0.25\,x_{2}\,x_{3} - 0.15\,x_{1}^{3}\\
-0.5\,x_{2} + 0.3\,x_{1}\,x_{3} - 0.2\,x_{2}^{3}\\
-0.6\,x_{3} + 0.2\,x_{1}\,x_{2} - 0.18\,x_{1}^{2}\,x_{2}
\end{pmatrix}.
\]

\begin{itemize}[label=\textendash, leftmargin=1.55em,
                labelsep=0.45em, topsep=0pt, itemsep=0pt]
    \item $\mathcal{X} = \mathbb{R}^{3}$.
    \item $\mathcal{X}_0 = \{x\in\mathbb{R}^{3}\mid \lVert x\rVert_2\le 0.5\}$.
    \item $\mathcal{X}_U = \mathbb{R}^{3}\setminus \{x\in\mathbb{R}^{3}\mid \lVert x\rVert_2\le 3.5\}$.
    \item Verified type: barrier/safety verification.
    \item Verified method in the original paper: Z3/Yices.
\end{itemize}

\end{systeminstance}

\begin{systeminstance}{~\texttt{S031}, \cite{taheri2025barrierbench}}
\label{sys:bb-s031}

The vector flow field is:
\[
\dot{\mathbf{x}} =
\begin{pmatrix}
\dot{x}_{1}\\
\dot{x}_{2}\\
\dot{x}_{3}\\
\dot{x}_{4}
\end{pmatrix}
=
\begin{pmatrix}
-0.3\,x_{1} + 0.2\,x_{2}\,x_{4} - 0.15\,x_{3}^{2}\\
-0.4\,x_{2} + 0.25\,x_{1}\,x_{3} - 0.18\,x_{2}\,x_{4}\\
-0.5\,x_{3} + 0.3\,x_{1}\,x_{4} + 0.12\,x_{2}^{2}\\
-0.6\,x_{4} + 0.2\,x_{2}\,x_{3} - 0.1\,x_{1}^{2}\,x_{4}
\end{pmatrix}.
\]

\begin{itemize}[label=\textendash, leftmargin=1.55em,
                labelsep=0.45em, topsep=0pt, itemsep=0pt]
    \item $\mathcal{X} = \mathbb{R}^{4}$.
    \item $\mathcal{X}_0 = \{x\in\mathbb{R}^{4}\mid \lVert x\rVert_2\le 0.4\}$.
    \item $\mathcal{X}_U = \mathbb{R}^{4}\setminus \{x\in\mathbb{R}^{4}\mid \lVert x\rVert_2\le 3\}$.
    \item Verified type: barrier/safety verification.
    \item Verified method in the original paper: Z3/Yices.
\end{itemize}

\end{systeminstance}

\begin{systeminstance}{~\texttt{S036}, \cite{taheri2025barrierbench}}
\label{sys:bb-s036}

The update map is:
\[
\mathbf{x}_{k+1} =
\begin{pmatrix}
x_{1,k+1}\\
x_{2,k+1}
\end{pmatrix}
=
\begin{pmatrix}
0.85\,x_{1,k} + u_{0,k}\\
0.8\,x_{2,k} + 0.15\,x_{1,k} + u_{1,k}
\end{pmatrix}.
\]

\begin{itemize}[label=\textendash, leftmargin=1.55em,
                labelsep=0.45em, topsep=0pt, itemsep=0pt]
    \item $\mathcal{X} = \mathbb{R}^{2}$.
    \item $\mathcal{X}_0 = [1.5,2]\times [1.2,1.8]$.
    \item $\mathcal{X}_U = [4.5,5.5]\times [4,5]\cup [-5.5,-4.5]\times [-5,-4]$.
    \item Verified type: barrier/safety verification.
\end{itemize}

\end{systeminstance}

\begin{systeminstance}{~\texttt{S037}, \cite{taheri2025barrierbench}}
\label{sys:bb-s037}

The vector flow field is:
\[
\dot{\mathbf{x}} =
\begin{pmatrix}
\dot{x}_{1}\\
\dot{x}_{2}
\end{pmatrix}
=
\begin{pmatrix}
x_{2} + u_{0}\\
-0.2\,x_{1} - 0.1\,(x_{1}^{2} - 1)\,x_{2} + u_{1}
\end{pmatrix}.
\]

\begin{itemize}[label=\textendash, leftmargin=1.55em,
                labelsep=0.45em, topsep=0pt, itemsep=0pt]
    \item $\mathcal{X} = \mathbb{R}^{2}$.
    \item $\mathcal{X}_0 = \{x\in\mathbb{R}^{2}\mid \lVert x\rVert_2\le 0.5\}$.
    \item $\mathcal{X}_U = \mathbb{R}^{2}\setminus \{x\in\mathbb{R}^{2}\mid \lVert x\rVert_2\le 2.8\}$.
    \item Verified type: barrier/safety verification.
    \item Verified method in the original paper: Z3/Yices.
\end{itemize}

\end{systeminstance}

\begin{systeminstance}{~\texttt{S042}, \cite{taheri2025barrierbench}}
\label{sys:bb-s042}

The vector flow field is:
\[
\dot{\mathbf{x}} =
\begin{pmatrix}
\dot{x}_{1}\\
\dot{x}_{2}\\
\dot{x}_{3}\\
\dot{x}_{4}
\end{pmatrix}
=
\begin{pmatrix}
x_{2} + u_{0}\\
-0.6\,x_{1} - 0.1\,x_{2} + 0.05\,x_{3} + u_{1}\\
x_{4} + u_{2}\\
-0.8\,x_{3} - 0.15\,x_{4} + 0.05\,x_{1} + u_{3}
\end{pmatrix}.
\]

\begin{itemize}[label=\textendash, leftmargin=1.55em,
                labelsep=0.45em, topsep=0pt, itemsep=0pt]
    \item $\mathcal{X} = \mathbb{R}^{4}$.
    \item $\mathcal{X}_0 = \{x\in\mathbb{R}^{4}\mid \lVert x\rVert_2\le 0.6\}$.
    \item $\mathcal{X}_U = \mathbb{R}^{4}\setminus \{x\in\mathbb{R}^{4}\mid \lVert x\rVert_2\le 3.5\}$.
    \item Verified type: barrier/safety verification.
    \item Verified method in the original paper: Z3/Yices.
\end{itemize}

\end{systeminstance}

\begin{systeminstance}{~\texttt{S048}, \cite{taheri2025barrierbench}}
\label{sys:bb-s048}

The vector flow field is:
\[
\dot{\mathbf{x}} =
\begin{pmatrix}
\dot{x}_{1}\\
\dot{x}_{2}\\
\dot{x}_{3}
\end{pmatrix}
=
\begin{pmatrix}
-0.5\,x_{1} + 0.3\,x_{2}\,\sin(x_{3}) - 0.1\,x_{1}^{2}\,x_{2}\\
-0.6\,x_{2} + 0.5\,x_{1}\,\cos(x_{3}) + 0.15\,x_{3}^{2}\,x_{1}\\
-0.7\,x_{3} - 0.1\,x_{3}^{3}
\end{pmatrix}.
\]

\begin{itemize}[label=\textendash, leftmargin=1.55em,
                labelsep=0.45em, topsep=0pt, itemsep=0pt]
    \item $\mathcal{X} = \mathbb{R}^{3}$.
    \item $\mathcal{X}_0 = \{x\in\mathbb{R}^{3}\mid \lVert x\rVert_2\le 0.4\}$.
    \item $\mathcal{X}_U = \mathbb{R}^{3}\setminus \{x\in\mathbb{R}^{3}\mid \lVert x\rVert_2\le 3.5\}$.
    \item Verified type: barrier/safety verification.
    \item Verified method in the original paper: Z3/Yices with Taylor approximations.
\end{itemize}
\end{systeminstance}

\begin{systeminstance}{~\texttt{S052}, \cite{taheri2025barrierbench}}
\label{sys:bb-s052}

The vector flow field is:
\[
\dot{\mathbf{x}} =
\begin{pmatrix}
\dot{x}_{1}\\
\dot{x}_{2}\\
\dot{x}_{3}
\end{pmatrix}
=
\begin{pmatrix}
x_{2} + u_{0}\\
x_{3} + u_{1}\\
-6\,x_{3} - 2.92\,x_{2} - 1.2\,x_{1} + x_{1}^{2} + u_{2}
\end{pmatrix}.
\]

\begin{itemize}[label=\textendash, leftmargin=1.55em,
                labelsep=0.45em, topsep=0pt, itemsep=0pt]
    \item $\mathcal{X} = \mathbb{R}^{3}$.
    \item $\mathcal{X}_0 = \{x\in\mathbb{R}^{3}\mid \lVert x\rVert_2\le 0.1\}$.
    \item $\mathcal{X}_U = \mathbb{R}^{3}\setminus \{x\in\mathbb{R}^{3}\mid \lVert x\rVert_2\le 2\}$.
    \item Verified type: barrier/safety verification.
    \item Verified method in the original paper: Z3/Yices.
\end{itemize}

\end{systeminstance}

\begin{systeminstance}{~\texttt{S057}, \cite{taheri2025barrierbench}}
\label{sys:bb-s057}

The vector flow field is:
\[
\dot{\mathbf{x}} =
\begin{pmatrix}
\dot{x}_{1}\\
\dot{x}_{2}
\end{pmatrix}
=
\begin{pmatrix}
-x_{1} + 0.08\,x_{2} + x_{1}^{2}\,x_{2} + u_{0}\\
0.6 - 0.08\,x_{2} - x_{1}^{2}\,x_{2} + u_{1}
\end{pmatrix}.
\]

\begin{itemize}[label=\textendash, leftmargin=1.55em,
                labelsep=0.45em, topsep=0pt, itemsep=0pt]
    \item $\mathcal{X} = \mathbb{R}^{2}$.
    \item $\mathcal{X}_0 = [0.1,0.5]\times [1,2]$.
    \item $\mathcal{X}_U = \mathbb{R}^{2}\setminus \{x\in\mathbb{R}^{2}\mid \lVert x\rVert_2\le 5\}$.
    \item Verified type: barrier/safety verification.
    \item Verified method in the original paper: Z3/Yices.
\end{itemize}

\end{systeminstance}

\begin{systeminstance}{~\texttt{S058}, \cite{taheri2025barrierbench}}
\label{sys:bb-s058}

The vector flow field is:
\[
\dot{\mathbf{x}} =
\begin{pmatrix}
\dot{x}_{1}\\
\dot{x}_{2}\\
\dot{x}_{3}
\end{pmatrix}
=
\begin{pmatrix}
x_{2} + u_{0}\\
-0.1\,x_{1} - 0.05\,x_{2} + 0.02\,x_{1}\,x_{3} + u_{1}\\
-0.2\,x_{3} + 0.1\,x_{1} + u_{2}
\end{pmatrix}.
\]

\begin{itemize}[label=\textendash, leftmargin=1.55em,
                labelsep=0.45em, topsep=0pt, itemsep=0pt]
    \item $\mathcal{X} = \mathbb{R}^{3}$.
    \item $\mathcal{X}_0 = \{x\in\mathbb{R}^{3}\mid \lVert x\rVert_2\le 0.5\}$.
    \item $\mathcal{X}_U = \mathbb{R}^{3}\setminus \{x\in\mathbb{R}^{3}\mid \lVert x\rVert_2\le 3\}$.
    \item Verified type: barrier/safety verification.
    \item Verified method in the original paper: Z3/Yices.
\end{itemize}

\end{systeminstance}

\begin{systeminstance}{~\texttt{S059}, \cite{taheri2025barrierbench}}
\label{sys:bb-s059}

The vector flow field is:
\[
\dot{\mathbf{x}} =
\begin{pmatrix}
\dot{x}_{1}\\
\dot{x}_{2}\\
\dot{x}_{3}
\end{pmatrix}
=
\begin{pmatrix}
-0.25\,x_{1} - x_{2}^{2} - x_{3}^{2} + 0.75 + u_{0}\\
-x_{2} + x_{1}\,x_{3} + u_{1}\\
-x_{3} + x_{1}\,x_{2} - 0.1\,x_{1} + u_{2}
\end{pmatrix}.
\]

\begin{itemize}[label=\textendash, leftmargin=1.55em,
                labelsep=0.45em, topsep=0pt, itemsep=0pt]
    \item $\mathcal{X} = \mathbb{R}^{3}$.
    \item $\mathcal{X}_0 = \{x\in\mathbb{R}^{3}\mid \lVert x\rVert_2\le 0.2\}$.
    \item $\mathcal{X}_U = \mathbb{R}^{3}\setminus \{x\in\mathbb{R}^{3}\mid \lVert x\rVert_2\le 3\}$.
    \item Verified type: barrier/safety verification.
    \item Verified method in the original paper: Z3/Yices.
\end{itemize}

\end{systeminstance}

\begin{systeminstance}{~\texttt{S060}, \cite{taheri2025barrierbench}}
\label{sys:bb-s060}

The vector flow field is:
\[
\dot{\mathbf{x}} =
\begin{pmatrix}
\dot{x}_{1}\\
\dot{x}_{2}
\end{pmatrix}
=
\begin{pmatrix}
x_{1}\,(x_{1} - 0.1)\,(1 - x_{1}) - 0.3\,x_{1}\,x_{2} + u_{0}\\
-0.2\,x_{2} + 0.1\,x_{1}\,x_{2} + u_{1}
\end{pmatrix}.
\]

\begin{itemize}[label=\textendash, leftmargin=1.55em,
                labelsep=0.45em, topsep=0pt, itemsep=0pt]
    \item $\mathcal{X} = \mathbb{R}^{2}$.
    \item $\mathcal{X}_0 = [0.2,0.8]\times [0.1,0.6]$.
    \item $\mathcal{X}_U = [1.5,2]\times [1,1.5]$.
    \item Verified type: barrier/safety verification.
    \item Verified method in the original paper: Z3/Yices.
\end{itemize}

\end{systeminstance}

\begin{systeminstance}{~\texttt{S062}, \cite{taheri2025barrierbench}}
\label{sys:bb-s062}

The vector flow field is:
\[
\dot{\mathbf{x}} =
\begin{pmatrix}
\dot{x}_{1}\\
\dot{x}_{2}\\
\dot{x}_{3}
\end{pmatrix}
=
\begin{pmatrix}
x_{2} + u_{0}\\
-x_{1} + x_{2}\,x_{3} + u_{1}\\
1 - x_{2}^{2} + u_{2}
\end{pmatrix}.
\]

\begin{itemize}[label=\textendash, leftmargin=1.55em,
                labelsep=0.45em, topsep=0pt, itemsep=0pt]
    \item $\mathcal{X} = \mathbb{R}^{3}$.
    \item $\mathcal{X}_0 = \{x\in\mathbb{R}^{3}\mid \lVert x\rVert_2\le 0.2\}$.
    \item $\mathcal{X}_U = \mathbb{R}^{3}\setminus \{x\in\mathbb{R}^{3}\mid \lVert x\rVert_2\le 2.5\}$.
    \item Verified type: barrier/safety verification.
    \item Verified method in the original paper: Z3/Yices.
\end{itemize}

\end{systeminstance}

\begin{systeminstance}{~\texttt{S076}, \cite{taheri2025barrierbench}}
\label{sys:bb-s076}

The vector flow field is:
\[
\dot{\mathbf{x}} =
\begin{pmatrix}
\dot{x}_{1}\\
\dot{x}_{2}\\
\dot{x}_{3}
\end{pmatrix}
=
\begin{pmatrix}
x_{3} + x_{2}\,x_{1} + u_{0}\\
-x_{2} - x_{1}^{2} + u_{1}\\
-x_{1} - 0.5\,x_{3} + u_{2}
\end{pmatrix}.
\]

\begin{itemize}[label=\textendash, leftmargin=1.55em,
                labelsep=0.45em, topsep=0pt, itemsep=0pt]
    \item $\mathcal{X} = \mathbb{R}^{3}$.
    \item $\mathcal{X}_0 = \{x\in\mathbb{R}^{3}\mid \lVert x\rVert_2\le 0.3\}$.
    \item $\mathcal{X}_U = \mathbb{R}^{3}\setminus \{x\in\mathbb{R}^{3}\mid \lVert x\rVert_2\le 3.5\}$.
    \item Verified type: barrier/safety verification.
    \item Verified method in the original paper: Z3/Yices.
\end{itemize}

\end{systeminstance}

\begin{systeminstance}{~\texttt{S077}, \cite{taheri2025barrierbench}}
\label{sys:bb-s077}

The vector flow field is:
\[
\dot{\mathbf{x}} =
\begin{pmatrix}
\dot{x}_{1}\\
\dot{x}_{2}\\
\dot{x}_{3}\\
\dot{x}_{4}
\end{pmatrix}
=
\begin{pmatrix}
x_{2} + u_{0}\\
-0.8\,x_{1} - 0.3\,x_{2}\,(1 + x_{1}^{2}) + 0.1\,x_{3} + u_{1}\\
x_{4} + u_{2}\\
-0.9\,x_{3} - 0.25\,x_{4}\,(1 + x_{3}^{2}) + 0.1\,x_{1} + u_{3}
\end{pmatrix}.
\]

\begin{itemize}[label=\textendash, leftmargin=1.55em,
                labelsep=0.45em, topsep=0pt, itemsep=0pt]
    \item $\mathcal{X} = \mathbb{R}^{4}$.
    \item $\mathcal{X}_0 = \{x\in\mathbb{R}^{4}\mid \lVert x\rVert_2\le 0.35\}$.
    \item $\mathcal{X}_U = \mathbb{R}^{4}\setminus \{x\in\mathbb{R}^{4}\mid \lVert x\rVert_2\le 2.8\}$.
    \item Verified type: barrier/safety verification.
    \item Verified method in the original paper: Z3/Yices.
\end{itemize}

\end{systeminstance}

\begin{systeminstance}{~\texttt{S087}, \cite{taheri2025barrierbench}}
\label{sys:bb-s087}

The update map is:
\[
\mathbf{x}_{k+1} =
\begin{pmatrix}
x_{1,k+1}\\
x_{2,k+1}
\end{pmatrix}
=
\begin{pmatrix}
x_{1,k} + 0.01\,(-100\,x_{1,k} - 1\,x_{2,k})\\
x_{2,k} + 0.01\,(1\,x_{1,k} - 100\,x_{2,k})
\end{pmatrix}.
\]

\begin{itemize}[label=\textendash, leftmargin=1.55em,
                labelsep=0.45em, topsep=0pt, itemsep=0pt]
    \item $\mathcal{X} = \mathbb{R}^{2}$.
    \item $\mathcal{X}_0 = [0.1,0.4]\times [0.1,0.55]$.
    \item $\mathcal{X}_U = [0.45,0.5]\times [0.6,1]$.
    \item Verified type: barrier/safety verification.
\end{itemize}

\end{systeminstance}

\begin{systeminstance}{~\texttt{U000}, \cite{taheri2025barrierbench}}
\label{sys:bb-u000}

The vector flow field is:
\[
\dot{\mathbf{x}} =
\begin{pmatrix}
\dot{x}_{1}\\
\dot{x}_{2}\\
\dot{x}_{3}\\
\dot{x}_{4}
\end{pmatrix}
=
\begin{pmatrix}
-0.5\,x_{1} + 0.3\,x_{2}\\
-0.4\,x_{2} + 0.15\,\cos(x_{1})\,\sin(x_{3})\\
-0.6\,x_{3} + 0.1\,\sin(x_{2})\,\cos(x_{4})\\
-0.5\,x_{4}
\end{pmatrix}.
\]

\begin{itemize}[label=\textendash, leftmargin=1.55em,
                labelsep=0.45em, topsep=0pt, itemsep=0pt]
    \item $\mathcal{X} = \mathbb{R}^{4}$.
    \item $\mathcal{X}_0 = \{x\in\mathbb{R}^{4}\mid \lVert x\rVert_2\le 0.3\}$.
    \item $\mathcal{X}_U = \mathbb{R}^{4}\setminus \{x\in\mathbb{R}^{4}\mid \lVert x\rVert_2\le 2.5\}$.
    \item Verified type: barrier/safety verification.
\end{itemize}

\end{systeminstance}

\begin{systeminstance}{~\texttt{U001}, \cite{taheri2025barrierbench}}
\label{sys:bb-u001}

The vector flow field is:
\[
\dot{\mathbf{x}} =
\begin{pmatrix}
\dot{x}_{1}\\
\dot{x}_{2}\\
\dot{x}_{3}
\end{pmatrix}
=
\begin{pmatrix}
-0.3\,x_{1} + 0.5\,\sin(x_{2})\\
-0.4\,x_{2} - 0.6\,\sin(x_{1}) + 0.1\,\sin(x_{3})\\
-0.5\,x_{3} + 0.2\,\sin(x_{1})\,\sin(x_{2})
\end{pmatrix}.
\]

\begin{itemize}[label=\textendash, leftmargin=1.55em,
                labelsep=0.45em, topsep=0pt, itemsep=0pt]
    \item $\mathcal{X} = \mathbb{R}^{3}$.
    \item $\mathcal{X}_0 = \{x\in\mathbb{R}^{3}\mid \lVert x\rVert_2\le 0.4\}$.
    \item $\mathcal{X}_U = \mathbb{R}^{3}\setminus \{x\in\mathbb{R}^{3}\mid \lVert x\rVert_2\le 3\}$.
    \item Verified type: barrier/safety verification.
\end{itemize}

\end{systeminstance}

\begin{systeminstance}{~\texttt{U002}, \cite{taheri2025barrierbench}}
\label{sys:bb-u002}

The vector flow field is:
\[
\dot{\mathbf{x}} =
\begin{pmatrix}
\dot{x}_{1}\\
\dot{x}_{2}\\
\dot{x}_{3}
\end{pmatrix}
=
\begin{pmatrix}
-0.3\,x_{1} + 0.2\,\sin(x_{2})\,\exp(-x_{3}^{2})\\
-0.4\,x_{2} + 0.15\,\cos(x_{1})\,\exp(-x_{1}^{2})\\
-0.5\,x_{3} + 0.1\,\sin(x_{1} + x_{2})
\end{pmatrix}.
\]

\begin{itemize}[label=\textendash, leftmargin=1.55em,
                labelsep=0.45em, topsep=0pt, itemsep=0pt]
    \item $\mathcal{X} = \mathbb{R}^{3}$.
    \item $\mathcal{X}_0 = \{x\in\mathbb{R}^{3}\mid \lVert x\rVert_2\le 0.45\}$.
    \item $\mathcal{X}_U = \mathbb{R}^{3}\setminus \{x\in\mathbb{R}^{3}\mid \lVert x\rVert_2\le 3.2\}$.
    \item Verified type: barrier/safety verification.
\end{itemize}

\end{systeminstance}

\begin{systeminstance}{~\texttt{U003}, \cite{taheri2025barrierbench}}
\label{sys:bb-u003}

The update map is:
\[
\mathbf{x}_{k+1} =
\begin{pmatrix}
x_{1,k+1}\\
x_{2,k+1}\\
x_{3,k+1}
\end{pmatrix}
=
\begin{pmatrix}
0.8\,x_{1,k} + 0.1\,\sin(x_{2,k})\\
0.85\,x_{2,k} + 0.08\,x_{1,k}\,x_{3,k}\\
0.9\,x_{3,k} + 0.08\,\cos(x_{1,k}\,x_{2,k})
\end{pmatrix}.
\]

\begin{itemize}[label=\textendash, leftmargin=1.55em,
                labelsep=0.45em, topsep=0pt, itemsep=0pt]
    \item $\mathcal{X} = \mathbb{R}^{3}$.
    \item $\mathcal{X}_0 = [0.1,0.4]\times [0.1,0.3]\times [0.1,0.4]$.
    \item $\mathcal{X}_U = [2,2.5]\times [1.8,2.2]\times [1.8,2.3]$.
    \item Verified type: barrier/safety verification.
\end{itemize}

\end{systeminstance}

\begin{systeminstance}{~\texttt{U004}, \cite{taheri2025barrierbench}}
\label{sys:bb-u004}

The vector flow field is:
\[
\dot{\mathbf{x}} =
\begin{pmatrix}
\dot{x}_{1}\\
\dot{x}_{2}\\
\dot{x}_{3}\\
\dot{x}_{4}
\end{pmatrix}
=
\begin{pmatrix}
-0.65\,x_{1} + 0.1\,x_{2}^{2}\,\cos(x_{3})\\
-0.7\,x_{2} + 0.12\,\sin(x_{1})\,\tanh(x_{4})\\
-0.75\,x_{3} + 0.08\,x_{1}\,x_{2}\\
-0.6\,x_{4} + 0.06\,(x_{1} + x_{2})
\end{pmatrix}.
\]

\begin{itemize}[label=\textendash, leftmargin=1.55em,
                labelsep=0.45em, topsep=0pt, itemsep=0pt]
    \item $\mathcal{X} = \mathbb{R}^{4}$.
    \item $\mathcal{X}_0 = [0.2,0.5]\times [-0.3,0]\times [0.1,0.4]\times [-0.2,0.1]$.
    \item $\mathcal{X}_U = [-2.2,-1.7]\times [1.6,2.1]\times [-1.9,-1.4]\times [1.3,1.8]$.
    \item Verified type: barrier/safety verification.
\end{itemize}

\end{systeminstance}

\begin{systeminstance}{~\texttt{U005}, \cite{taheri2025barrierbench}}
\label{sys:bb-u005}

The vector flow field is:
\[
\dot{\mathbf{x}} =
\begin{pmatrix}
\dot{x}_{1}\\
\dot{x}_{2}\\
\dot{x}_{3}
\end{pmatrix}
=
\begin{pmatrix}
-0.58\,x_{1} + 0.12\,\sin(x_{2})\,\cos(x_{3})\\
-0.62\,x_{2} + 0.1\,\exp(-x_{1}^{2})\,\tanh(x_{3})\\
-0.66\,x_{3} + 0.08\,\sin(x_{1}x_{2})\,\exp(-x_{3}^{2})
\end{pmatrix}.
\]

\begin{itemize}[label=\textendash, leftmargin=1.55em,
                labelsep=0.45em, topsep=0pt, itemsep=0pt]
    \item $\mathcal{X} = \mathbb{R}^{3}$.
    \item $\mathcal{X}_0 = [-0.3,0.2]\times [0.1,0.4]\times [-0.2,0.3]$.
    \item $\mathcal{X}_U = [-2.3,-1.8]\times [1.7,2.2]\times [-2,-1.5]$.
    \item Verified type: barrier/safety verification.
\end{itemize}

\end{systeminstance}

\begin{systeminstance}{~\texttt{U006}, \cite{taheri2025barrierbench}}
\label{sys:bb-u006}
The vector flow field is:
\[
\dot{\mathbf{x}} =
\begin{pmatrix}
\dot{x}_{1}\\
\dot{x}_{2}\\
\dot{x}_{3}
\end{pmatrix}
=
\begin{pmatrix}
u_{0}\cos(x_{3})\\
u_{0}\sin(x_{3})\\
u_{1}
\end{pmatrix}.
\]

\begin{itemize}[label=\textendash, leftmargin=1.55em,
                labelsep=0.45em, topsep=0pt, itemsep=0pt]
    \item $\mathcal{X} = \mathbb{R}^{3}$.
    \item $\mathcal{X}_0 = [-1,1]\times [-1,1]\times [-0.5,0.5]$.
    \item $\mathcal{X}_U
=
\mathbb{R}^{3}\setminus
\bigl([-5,5]\times[-5,5]\times[-3.14,3.14]\bigr)$.
    \item Verified type: barrier/safety verification.
\end{itemize}

\end{systeminstance}

\begin{systeminstance}{~\texttt{U007}, \cite{taheri2025barrierbench}}
\label{sys:bb-u007}

The update map is:
\[
\mathbf{x}_{k+1} =
\begin{pmatrix}
x_{1,k+1}\\
x_{2,k+1}\\
x_{3,k+1}
\end{pmatrix}
=
\begin{pmatrix}
0.75\,x_{1,k} + 0.1\,\tanh(x_{2,k})\,\tanh(x_{3,k})\\
0.82\,x_{2,k} + 0.09\,\sin(x_{1,k})\,\exp(-x_{3,k}^{2})\\
0.79\,x_{3,k} + 0.08\,\cos(x_{1,k})\,\cos(x_{2,k})
\end{pmatrix}.
\]

\begin{itemize}[label=\textendash, leftmargin=1.55em,
                labelsep=0.45em, topsep=0pt, itemsep=0pt]
    \item $\mathcal{X} = \mathbb{R}^{3}$.
    \item $\mathcal{X}_0 = \{x\in\mathbb{R}^{3}\mid \lVert x-(1,-0.5,0.8)^{\mathsf T}\rVert_2\le 0.5\}$.
    \item $\mathcal{X}_U = \{x\in\mathbb{R}^{3}\mid \lVert x-(-2.5,2,-2.2)^{\mathsf T}\rVert_2\le 0.6\}$.
    \item Verified type: barrier/safety verification.
\end{itemize}

\end{systeminstance}

\begin{systeminstance}{~\texttt{U008}, \cite{taheri2025barrierbench}}
\label{sys:bb-u008}

The update map is:
\[
\mathbf{x}_{k+1} =
\begin{pmatrix}
x_{1,k+1}\\
x_{2,k+1}\\
x_{3,k+1}
\end{pmatrix}
=
\begin{pmatrix}
0.73\,x_{1,k} + 0.13\,\sin(x_{2,k})\,\tanh(x_{3,k})\\
0.89\,x_{2,k} + 0.17\,\cos(x_{1,k}^{2})\,\exp(-x_{3,k}^{2})\\
0.64\,x_{3,k} + 0.22\,\tanh(x_{1,k}\,x_{2,k})
\end{pmatrix}.
\]

\begin{itemize}[label=\textendash, leftmargin=1.55em,
                labelsep=0.45em, topsep=0pt, itemsep=0pt]
    \item $\mathcal{X} = \mathbb{R}^{3}$.
    \item $\mathcal{X}_0 = \{x\in\mathbb{R}^{3}\mid \lVert x-(0.9,-0.7,1.2)^{\mathsf T}\rVert_2\le 0.55\}$.
    \item $\mathcal{X}_U = \{x\in\mathbb{R}^{3}\mid \lVert x-(-3.1,2.8,-2.6)^{\mathsf T}\rVert_2\le 0.75\}$.
    \item Verified type: barrier/safety verification.
\end{itemize}

\end{systeminstance}

\begin{systeminstance}{~\texttt{U009}, \cite{taheri2025barrierbench}}
\label{sys:bb-u009}

The vector flow field is:
\[
\dot{\mathbf{x}} =
\begin{pmatrix}
\dot{x}_{1}\\
\dot{x}_{2}\\
\dot{x}_{3}
\end{pmatrix}
=
\begin{pmatrix}
-0.62\,x_{1} + 0.1\,\sin(x_{2} + x_{3})\\
-0.67\,x_{2} + 0.08\,\cos(x_{1})\,\exp(-x_{3}^{2})\\
-0.64\,x_{3} + 0.09\,\tanh(x_{1})\,\sin(x_{2})
\end{pmatrix}.
\]

\begin{itemize}[label=\textendash, leftmargin=1.55em,
                labelsep=0.45em, topsep=0pt, itemsep=0pt]
    \item $\mathcal{X} = \mathbb{R}^{3}$.
    \item $\mathcal{X}_0 = [-0.2,0.3]\times [0.1,0.5]\times [-0.3,0.2]$.
    \item $\mathcal{X}_U = [2,2.6]\times [-2.4,-1.8]\times [2.1,2.7]$.
    \item Verified type: barrier/safety verification.
\end{itemize}

\end{systeminstance}

\subsection{Extended Tier: Scaled and Perturbed Variants of Retained Systems}

We further extend some of the systems described above in several ways, including strengthening the verification objective from asymptotic stability to exponential stability, incorporating control constraints, scaling existing systems to higher dimensions, and introducing parameter variations.

\textbf{Problem.~1: }

\begin{systeminstance}{Extend-Table~1 row~1 (NonPoly0), \cite{abate2021fossil}}
\label{sys:catalogue-sys0016-ext}

The vector flow field is:
\[
\dot{\mathbf{x}}
=
\begin{pmatrix}
\dot{x}\\
\dot{y}
\end{pmatrix}
=
\begin{pmatrix}
-x+xy\\
-y
\end{pmatrix}.
\]
\begin{itemize}[label=\textendash, leftmargin=1.55em,
                labelsep=0.45em, topsep=0pt, itemsep=0pt]
    \item $x^\star=(0,0)$.
    \item $\mathcal{N}=\mathbb{R}^{2}$.
    \item Verified type: Global asymptotic stability 
\end{itemize}
\end{systeminstance}
We modify the local asymptotic stability to the global asymptotic stability.

\begin{systeminstance}{Extend-Table~1 row~6 (Poly2), \cite{abate2021fossil}}
\label{sys:catalogue-sys0021-ext}

The vector flow field is:
\[
\dot{\mathbf{x}}
=
\begin{pmatrix}
\dot{x}_1\\
\dot{x}_2
\end{pmatrix}
=
\begin{pmatrix}
-x_1^3+x_2\\
-x_1-x_2
\end{pmatrix}.
\]
\begin{itemize}[label=\textendash, leftmargin=1.55em,
                labelsep=0.45em, topsep=0pt, itemsep=0pt]
    \item $x^\star=(0,0)$.
    \item $\mathcal{N}=\mathbb{R}^{2}$.
    \item Verified type: Global exponential stability 
\end{itemize}
\end{systeminstance}
We modify the local asymptotic stability to the global exponential stability.

\begin{systeminstance}{Extend-Table~4 \#2 (SineModel), \cite{edwards2024fossil}}
\label{sys:fossil2-f2-02-sinemodel-ext}

The vector flow field is:
\[
\dot{\mathbf{x}}
=
\begin{pmatrix}
\dot{x}_1\\
\dot{x}_2
\end{pmatrix}
=
\begin{pmatrix}
x_2+u_1\\
u_2+\dfrac{mgL\sin(x_1)-bx_2}{mL^2}
\end{pmatrix},
\]
where $m=0.15$, $g=9.81$, $L=0.5$, and $b=0.1$.
\begin{itemize}[label=\textendash, leftmargin=1.55em,
                labelsep=0.45em, topsep=0pt, itemsep=0pt]
    \item $x^\star=(0,0)$.
    \item $\mathcal{N}=\{\mathbf{x}\in\mathbb{R}^{2}\mid x_1^2+x_2^2\leq1^2\}$.
    \item $u_1,u_2 \in [-5,5]$ when $\mathbf{x}\in \mathcal{N}$ and $u_1(0)=u_2(0)=0$.
    \item Verified type: Exponentially stable on $\mathcal{N}$.
    \item Verified method in the original paper: dReal.
\end{itemize}
\end{systeminstance}

We add control constraints $u_1,u_2 \in [-5,5], u_1(0)=u_2(0)=0$ and change the local asymptotic stability to local exponential stability.

\textbf{Problem.~2: }

\begin{systeminstance}{Extend-Section~5.2 and Appendix~C.3.2, \cite{zhou2022neural}}
\label{sys:catalogue-sys0035}

The closed-loop vector flow field is:
\[
\dot{\mathbf{x}}
=
\begin{pmatrix}
\dot{x}_1\\
\dot{x}_2
\end{pmatrix}
=
\begin{pmatrix}
\sin(x_2)\\
u-\dfrac{\cos(x_2)}{1-x_1}
\end{pmatrix}
\]
\begin{itemize}[label=\textendash, leftmargin=1.55em,
                labelsep=0.45em, topsep=0pt, itemsep=0pt]
    \item $x^\star=(0,0)$.
    \item Original $\mathcal{N}=\{\mathbf{x}\in\mathbb{R}^{2}\mid0.1^2\leq x_1^2+x_2^2\leq0.8^2\}$. 
    \item Original verified type: $\epsilon$-ROA.
    \item Verified method in the original paper: dReal.
    \item Our $\mathcal{N}=\{\mathbf{x}\in\mathbb{R}^{2}\mid x_1^2+x_2^2\leq0.8^2\}$. 
    \item Our verified type: ROA verification.
    \item $u \in [-5,5]$ when $\mathbf{x}\in \mathcal{N}$ and $u(0,0)=1$.
\end{itemize}
\end{systeminstance}
In the original paper, the control policy was $u=5\tanh(q_1x_1+q_2x_2+b)$, under which $(0,0)^\top$ is not an equilibrium point. We therefore change the goal to finding a policy $u\in[-5,5]$ such that $u\bigl((0,0)^\top\bigr)=1$.

\textbf{Problem.~3: }

\begin{systeminstance}{Extend--2D quadrotor, \cite{meng2024physics}}
\label{sys:extend-meng-2d-quadrotor}

The vector flow field is:
\[
\dot{\mathbf{x}} =
\begin{pmatrix}
\dot{x}_1\\
\dot{x}_2\\
\dot{x}_3\\
\dot{x}_4\\
\dot{x}_5\\
\dot{x}_6
\end{pmatrix}
=
\begin{pmatrix}
x_4\\[1ex]
x_5\\[1ex]
x_6\\[1ex]
9.81\sin(x_3)+\sin(x_3)u_1\\[1ex]
-9.81+9.81\cos(x_3)+\cos(x_3)u_1\\[1ex]
u_2
\end{pmatrix},
\]
where $\mathbf{u}=(u_1,u_2)^\top$ is the control input to be synthesized.

\begin{itemize}[label=\textendash, leftmargin=1.55em,
                labelsep=0.45em, topsep=0pt, itemsep=0pt]
    \item $\mathcal{X} = \{x\in\mathbb{R}^{6}\mid \lVert x\rVert_\infty \le 0.2\}$.
    \item $\mathcal{X}_U = \{x\in\mathbb{R}^{6}\mid \lVert x\rVert_\infty > 0.2\}$.
    \item $u_1,u_2 \in [-5,5]$ when $\mathbf{x}\in \mathcal{X}$. 
    \item Verified type: controlled forward-invariant set synthesis.
    \item Verified method in the original paper: no formal invariance certificate was reported for this instance.
\end{itemize}
\end{systeminstance}
We add control constraints $u_1,u_2 \in [-5,5]$.

\begin{systeminstance}{Extend--3D quadrotor, \cite{meng2024physics}}
\label{sys:extend-meng-3d-quadrotor}

The vector flow field is:
\[
\dot{\mathbf{x}} =
\begin{pmatrix}
\dot{x}_1\\
\dot{x}_2\\
\dot{x}_3\\
\dot{x}_4\\
\dot{x}_5\\
\dot{x}_6\\
\dot{x}_7\\
\dot{x}_8\\
\dot{x}_9
\end{pmatrix}
=
\begin{pmatrix}
x_4\\[1ex]
x_5\\[1ex]
x_6\\[1ex]
-9.81\sin(x_8)-\sin(x_8)u_1\\[1ex]
9.81\cos(x_8)\sin(x_7)
+\cos(x_8)\sin(x_7)u_1\\[1ex]
9.81\cos(x_8)\cos(x_7)-9.81
+\cos(x_8)\cos(x_7)u_1\\[1ex]
u_2\\[1ex]
u_3\\[1ex]
u_4
\end{pmatrix},
\]
where $\mathbf{u}=(u_1,u_2,u_3,u_4)^\top$ is the control input to be synthesized.

\begin{itemize}[label=\textendash, leftmargin=1.55em,
                labelsep=0.45em, topsep=0pt, itemsep=0pt]
    \item $\mathcal{X} = \{x\in\mathbb{R}^{9}\mid \lVert x\rVert_\infty \le 0.2\}$.
    \item $\mathcal{X}_U = \{x\in\mathbb{R}^{9}\mid \lVert x\rVert_\infty > 0.2\}$.
    \item $u_i \in [-5,5]$ when $\mathbf{x}\in \mathcal{X}\quad i=1,2,3,4$. %
    \item Verified type: controlled forward-invariant set synthesis.
    \item Verified method in the original paper: no formal invariance certificate was reported for this instance.
\end{itemize}

\end{systeminstance}

We add the control constraints $u_i \in [-5,5]$ when $\mathbf{x}\in \mathcal{X}\quad i=1,2,3,4$.

\textbf{Problem.~4: }

\begin{systeminstance}{Extend-\texttt{S002}, \cite{taheri2025barrierbench}} 
\label{sys:bb-s002-ext}

The vector flow field is:
\[
\dot{\mathbf{x}} =
\begin{pmatrix}
\dot{x}_{1}\\
\dot{x}_{2}\\
\dot{x}_{3}\\
\dot{x}_{4}
\end{pmatrix}
=
\begin{pmatrix}
x_{2} + u_{0}\\
-0.95\,\sin(x_{1}) - 0.02\,x_{2} + 0.01\,x_{3} + u_{1}\\
-0.1\,x_{3} + 0.02\,x_{1} + u_{2}\\
-0.03\,x_{4} + 0.01\,x_{2} + u_{3}
\end{pmatrix}.
\]

\begin{itemize}[label=\textendash, leftmargin=1.55em,
                labelsep=0.45em, topsep=0pt, itemsep=0pt]
    \item $\mathcal{X} = \mathbb{R}^{4}$.
    \item $\mathcal{X}_0 = \{x\in\mathbb{R}^{4}\mid \lVert x\rVert_2\le 0.3\}$.
    \item $\mathcal{X}_U = \mathbb{R}^{4}\setminus \{x\in\mathbb{R}^{4}\mid \lVert x\rVert_2\le 2.5\}$.
    \item $u_i\in [-5,5](i=0,1,2,3)$ when $\mathbf{x}\in \mathcal{X}$
    \item Verified type: safety verification.
    \item Verified method in the original paper: Z3/Yices with Taylor approximations.
\end{itemize}
\end{systeminstance}

We add the control constraints $u_i\in [-5,5]; i=0,1,2,3$.

\begin{systeminstance}{Extend-\texttt{S011}, \cite{taheri2025barrierbench}}
\label{sys:bb-s011-ext}

The vector flow field is:
\[
\dot{\mathbf{x}} =
\begin{pmatrix}
\dot{x}_{1}\\
\dot{x}_{2}
\end{pmatrix}
=
\begin{pmatrix}
x_{2} + u_{0}\\
-0.3\,x_{1} - 0.2\,(1 + \cos(x_{1}))\,x_{2} - 0.1\,x_{1}^{3} + u_{1}
\end{pmatrix}.
\]

\begin{itemize}[label=\textendash, leftmargin=1.55em,
                labelsep=0.45em, topsep=0pt, itemsep=0pt]
    \item $\mathcal{X} = \mathbb{R}^{2}$.
    \item $\mathcal{X}_0 = \{x\in\mathbb{R}^{2}\mid \lVert x\rVert_2\le 0.6\}$.
    \item $\mathcal{X}_U = \{x\in\mathbb{R}^{2}\mid \lVert x-(2,2)^{\mathsf T}\rVert_2\le 1\}$.
    \item $u_{0,1}\in [-1,1]$ when $x \in \mathcal{X}$.
    \item Verified type: safety verification.
    \item Verified method in the original paper: Z3/Yices with Taylor approximations.
\end{itemize}
\end{systeminstance}
We add the control constraints $u_{0,1}\in [-1,1]^2$.

\begin{systeminstance}{Extend-\texttt{S024}, \cite{taheri2025barrierbench}}
\label{sys:bb-s024-ext}

The vector flow field is:
\[
\dot{\mathbf{x}} =
\begin{pmatrix}
\dot{x}_{1}\\
\dot{x}_{2}\\
\dot{x}_{3}\\
\dot{x}_{4}
\end{pmatrix}
=
\begin{pmatrix}
x_{2} + u_{0}\\
-x_{1} - 0.4\,x_{2} + 0.15\,\sin(x_{3}) + u_{1}\\
x_{4} + u_{2}\\
-0.6\,x_{3} - 0.3\,x_{4} + u_{3}
\end{pmatrix}.
\]

\begin{itemize}[label=\textendash, leftmargin=1.55em,
                labelsep=0.45em, topsep=0pt, itemsep=0pt]
    \item $\mathcal{X} = \mathbb{R}^{4}$.
    \item $\mathcal{X}_0 = [0.5,0.8]\times [0.3,0.6]\times [0.5,0.8]\times [0.3,0.6]$.
    \item $\mathcal{X}_U = [2.5,3.5]\times [2,3]\times [2.5,3.5]\times [2,3]\cup [-3.5,-2.5]\times [-3,-2]\times [-3.5,-2.5]\times [-3,-2]$.
    \item $u_{i}\in [-1,1], i=0,1,2,3$ when $x \in \mathcal{X}$.
    \item Verified type: safety verification.
    \item Verified method in the original paper: Z3/Yices with Taylor approximations.
\end{itemize}
\end{systeminstance}
We add the control constraints $u_{i}\in [-1,1], i=0,1,2,3$.

\begin{systeminstance}{Extend--Hi-Ord8 with bounded sinusoidal disturbance, \cite{abate2021fossil}}
\label{sys:extend-hiord8-safety}

The vector flow field is:
\[
\dot{\mathbf{x}}
=
\begin{pmatrix}
\dot{x}_0\\ \dot{x}_1\\ \dot{x}_2\\ \dot{x}_3\\
\dot{x}_4\\ \dot{x}_5\\ \dot{x}_6\\ \dot{x}_7
\end{pmatrix}
=
\begin{pmatrix}
x_1\\
x_2\\
x_3\\
x_4\\
x_5\\
x_6\\
x_7\\
-\begin{aligned}[t]
&20x_7+170x_6+800x_5+2273x_4\\[-0.15ex]
&{}+3980x_3+4180x_2+2400x_1+576x_0\\[-0.15ex]
&{}+\sin(\omega x_0)
\end{aligned}
\end{pmatrix}.
\]

\begin{itemize}[label=\textendash, leftmargin=1.55em,
                labelsep=0.45em, topsep=0pt, itemsep=0pt]
    \item $\mathcal{X}=[-2.2,2.2]^8$.
    \item $\mathcal{X}_I=[0.9,1.1]^8$.
    \item $\mathcal{X}_U=[-2.2,-1.8]^8$.
    \item $\omega\in[0.5,2]$.
    \item Verified type: safety verification 
\end{itemize}
\end{systeminstance}

The safety property is verified uniformly for all admissible values of $\omega\in[0.5,2]$ which is a time-invariant parameter. 

\begin{remark}
    We require a single barrier function that depends only on $x$ and not on the uncertain parameter $\omega$, i.e., the barrier conditions must hold for every $\omega \in [0.5,2]$.
\end{remark}

\begin{systeminstance}{Safety extension of Example~5.4, \cite{liu2024tool}}
\label{sys:extend-liu2024-example54-safety}

The vector flow field is:
\[
\dot{\mathbf{x}}
=
\begin{pmatrix}
\dot{x}_1\\
\dot{x}_2\\
\dot{x}_3\\
\dot{x}_4\\
\dot{x}_5\\
\dot{x}_6\\
\dot{x}_7\\
\dot{x}_8\\
\dot{x}_9\\
\dot{x}_{10}
\end{pmatrix}
=
\begin{pmatrix}
-x_1+0.5x_2-0.1x_9^2+\sin(x_2)\\
-0.5x_1-x_2\\
-x_3+0.5x_4-0.1x_1^2+e^{-x_3}-1\\
-0.5x_3-x_4\\
-x_5+0.5x_6+0.1x_7^2\\
-0.5x_5-x_6\\
-x_7+0.5x_8\\
-0.5x_7-x_8\\
-x_9+0.5x_{10}\\
-0.5x_9-x_{10}+0.1x_2^2
\end{pmatrix}.
\]

\begin{itemize}[label=\textendash, leftmargin=1.55em,
                labelsep=0.45em, topsep=0pt, itemsep=0pt]
    \item $\mathcal{X}=[-1,1]^{10}$.
    \item $\mathcal{X}_I=[-0.35,0.35]^{10}$.
    \item $\mathcal{X}_U= [0.45,0.8]\times[-0.2,0.2]\times[-0.4,0.4]^8$.
    \item Verified type: safety verification 
\end{itemize}
\end{systeminstance}
We augment the dynamics by introducing two perturbations, $\sin(x_2)$ to the vector field of $x_1$ and  $e^{-x_3} -1 $ to the vector field of $x_3$. 

\subsection{Newly Designed Challenge Instances}

Given the limited availability of non-polynomial and high-dimensional systems in the existing literature, we additionally introduce the following systems.

\textbf{Problem.~1 :}

\begin{systeminstance}{8D Rational--Cosine Ring}
The vector flow field is:
\[
\dot{\mathbf{x}}
=
\begin{pmatrix}
\dot{x}_1\\
\dot{x}_2\\
\dot{x}_3\\
\dot{x}_4\\
\dot{x}_5\\
\dot{x}_6\\
\dot{x}_7\\
\dot{x}_8
\end{pmatrix}
=
\begin{pmatrix}
-3 x_1 + x_2 \frac{\cos x_1}{1 + x_8^2}\\[6pt]
-3 x_2 + x_3 \frac{\cos x_2}{1 + x_1^2}\\[6pt]
-3 x_3 + x_4 \frac{\cos x_3}{1 + x_2^2}\\[6pt]
-3 x_4 + x_5 \frac{\cos x_4}{1 + x_3^2}\\[6pt]
-3 x_5 + x_6 \frac{\cos x_1}{1 + x_4^2}\\[6pt]
-3 x_6 + x_7 \frac{\cos x_2}{1 + x_5^2}\\[6pt]
-3 x_7 + x_8 \frac{\cos x_3}{1 + x_6^2}\\[6pt]
-3 x_8 + x_1 \frac{\cos x_4}{1 + x_7^2}
\end{pmatrix}.
\]
\begin{itemize}[label=\textendash, leftmargin=1.55em,
                labelsep=0.45em, topsep=0pt, itemsep=0pt]
    \item $x^\star=(0,0,0,0,0,0,0,0)$.
    \item $\mathcal{N}=\mathbb{R}^{8}$.
    \item Verified type: Global exponential stability
\end{itemize}
\end{systeminstance}

\begin{systeminstance}{7D Heterogeneous Trigonometric--Rational Network}

The vector flow field is:
\[
\dot{\mathbf{x}} =
\begin{pmatrix}
-5x_1 + x_2\dfrac{\cos(x_1)}{1+x_7^2}\\[4pt]
-\dfrac{9}{2}x_2
+\dfrac{6}{5}x_4\dfrac{\sin(2x_2)}{1+2x_1^2}\\[4pt]
-6x_3
+\dfrac{4}{5}x_6\dfrac{\cos(x_3/2)}
 {1+\frac{3}{2}x_2^2}\\[4pt]
-\dfrac{11}{2}x_4
+\dfrac{3}{2}x_7\dfrac{\sin(3x_4/2)}
 {1+3x_3^2}\\[4pt]
-5x_5
+\dfrac{9}{10}x_1\dfrac{\cos(2x_5)}
 {1+\frac{4}{3}x_4^2}\\[4pt]
-\dfrac{9}{2}x_6
+\dfrac{7}{5}x_5\dfrac{\sin(x_6/3)}
 {1+\frac{5}{2}x_5^2}\\[4pt]
-6x_7
+\dfrac{1}{2}x_3\dfrac{\cos(5x_7/2)}
 {1+\frac{1}{2}x_5^2}
\end{pmatrix}.
\]

\begin{itemize}[label=\textendash, leftmargin=1.55em,
labelsep=0.45em, topsep=0pt, itemsep=0pt]
\item $x^\star=(0,0,0,0,0,0,0)$.
\item $\mathcal{N}=\mathbb{R}^{7}$.
\item Verified type: Global exponential stability.
\end{itemize}

\end{systeminstance}

\textbf{Problem.~2 :}

\begin{systeminstance}{Our cyclic ten-dimensional rational system}

The vector flow field is
\[
\dot{x}_i=-x_i+\frac{1}{4}x_{i+1}+\frac{1}{20}\frac{x_{i+2}}{1+x_{i+2}^2},\qquad i=1,\ldots,10,
\]
where the indices are interpreted cyclically, with $x_{11}=x_1$ and $x_{12}=x_2$.

\begin{itemize}[label=\textendash, leftmargin=1.55em, labelsep=0.45em, topsep=0pt, itemsep=0pt]
\item $x^\star=\mathbf{0}\in\mathbb{R}^{10}$.
\item $\mathcal{N}=\mathbb{R}^{10}$.
\item Verified type: ROA.
\end{itemize}

\end{systeminstance}

\begin{systeminstance}{Seven-dimensional multi-machine swing system}
\label{sys:swing7-roa}

The vector flow field is:
\[
\dot{\mathbf{x}}
=
\begin{pmatrix}
\dot{x}_1\\
\dot{x}_2\\
\dot{x}_3\\
\dot{x}_4\\
\dot{x}_5\\
\dot{x}_6\\
\dot{x}_7
\end{pmatrix}
=
\begin{pmatrix}
x_2\\[1mm]
-\dfrac12x_2-\dfrac32\sin x_1-\sin(x_1-x_3)-\dfrac32\sin(x_1-x_7)\\[2mm]
x_4\\[1mm]
-\dfrac12x_4-\dfrac12\sin x_3-\sin(x_3-x_1)-\sin(x_3-x_5)\\[2mm]
x_6\\[1mm]
-\dfrac12x_6-\sin x_5-\sin(x_5-x_3)-\dfrac32\sin(x_5-x_7)\\[2mm]
-\dfrac15\sin x_7-\dfrac32\sin(x_7-x_1)-\dfrac32\sin(x_7-x_5)
\end{pmatrix}.
\]
\begin{itemize}[label=\textendash, leftmargin=1.55em,
                labelsep=0.45em, topsep=0pt, itemsep=0pt]
    \item $x^\star=(0,0,0,0,0,0,0)$.
    \item $\mathcal{N}=[-\pi,\pi]^7$.
    \item $\mathcal{X}_I=[-0.6,0.6]^7$.
    \item Our verified type: ROA verification.
\end{itemize}
\end{systeminstance}

\begin{systeminstance}{Seven-dimensional coupled saturating oscillator}
\label{sys:ours-roa-sat-7d}

The vector flow field is:
\[
\dot{\mathbf{x}}
=
\begin{pmatrix}
\dot{x}_1\\ \dot{x}_2\\ \dot{x}_3\\ \dot{x}_4\\ \dot{x}_5\\ \dot{x}_6\\ \dot{x}_7
\end{pmatrix}
=
\begin{pmatrix}
x_2\\[1mm]
-\dfrac{2x_1}{1+x_1^2}+\dfrac12x_3-0.3\,x_2+0.2\,x_5\\[2mm]
x_4\\[1mm]
\dfrac12x_1-\dfrac32x_3-\sin(x_3)-0.2\,x_4+0.1\,x_6\\[2mm]
-0.5\,x_5-0.2\,x_2+0.1\,x_6\\[1mm]
-0.4\,x_6-0.1\,x_5-0.1\,x_4+0.1\,x_7\\[1mm]
-0.3\,x_7-0.1\,x_6
\end{pmatrix}.
\]
\begin{itemize}[label=\textendash, leftmargin=1.55em,
                labelsep=0.45em, topsep=0pt, itemsep=0pt]
    \item $x^\star=(0,0,0,0,0,0,0)$.
    \item $\mathcal{N}=\{\mathbf{x}\in\mathbb{R}^{7}\mid -5\le x_i\le 5,\ i=1,\ldots,7\}$.
    \item $\mathcal{X}_I=\{\mathbf{x}\in\mathbb{R}^{7}\mid -0.6\le x_i\le 0.6,\ i=1,\ldots,7\}$.
    \item Our verified type: ROA verification.
\end{itemize}
\end{systeminstance}

\textbf{Problem.~3 :}

\begin{systeminstance}{New-6D Sin Cascade Invariance}
\label{sys:new-p3-cosine-cascade-bounded}

The vector flow field is:
\[
\dot{\mathbf{x}}
=
\begin{pmatrix}
\dot{x}_1\\
\dot{x}_2\\
\dot{x}_3\\
\dot{x}_4\\
\dot{x}_5\\
\dot{x}_6
\end{pmatrix}
=
\begin{pmatrix}
x_2+u_0\\[1mm]
-x_1-\dfrac1{10}\sin(x_1)-x_2+x_3\\[2mm]
-2x_3+x_4\\[1mm]
-\dfrac32x_4+x_5\\[2mm]
-2x_5+x_6\\[1mm]
-\dfrac32x_6+u_1
\end{pmatrix}.
\]

\begin{itemize}[label=\textendash, leftmargin=1.55em, labelsep=0.45em, topsep=0pt, itemsep=0pt]
\item $\mathcal X=[-1,1]^6$.
\item $\mathcal X_{\mathrm{seed}}=\left\{x\in\mathbb R^6\mid \lVert x\rVert_2\le\frac12\right\}$.
\item $\mathcal X_U=\left\{x\in\mathbb R^6\mid \left\lVert x-\left(0,\frac45,0,0,0,0\right)^{\mathsf T}\right\rVert_2\le\frac1{10}\right\}$.
\item $u_0,u_1\in[-1,1]$ for every $x\in\mathcal X$. %
\item The requested output is a nonempty controlled forward-invariant set $\mathcal S\subseteq\mathcal X\setminus\mathcal X_U$ containing $\mathcal X_{\mathrm{seed}}$.
\item Task type: controlled forward-invariant set synthesis
\end{itemize}

\end{systeminstance}

\begin{systeminstance}{New-6D Exponential Cascade Invariance}
\label{sys:new-p3-exponential-cascade-bounded}

The vector flow field is:
\[
\dot{\mathbf{x}}=
\begin{pmatrix}
\dot{x}_1\\
\dot{x}_2\\
\dot{x}_3\\
\dot{x}_4\\
\dot{x}_5\\
\dot{x}_6
\end{pmatrix}
=
\begin{pmatrix}
x_2+u\\[1mm]
-x_1-\dfrac1{20}\left(\exp(x_1)-\exp(-x_1)\right)-x_2+x_3\\[2mm]
-2x_3+x_4\\
-\dfrac32x_4+x_5\\
-2x_5+x_6\\
-\dfrac32x_6
\end{pmatrix}.
\]

\begin{itemize}[label=\textendash, leftmargin=1.55em, labelsep=0.45em, topsep=0pt, itemsep=0pt]
\item $\mathcal X=[-1,1]^6$.
\item $\mathcal X_I=\left\{x\in\mathbb R^6\mid \left(x_1+\frac14\right)^2+\left(x_2-\frac14\right)^2+x_3^2+x_4^2+x_5^2+x_6^2\le\frac1{64}\right\}$.
\item $\mathcal X_U=\left\{x\in\mathbb R^6\mid \left(x_1-\frac7{10}\right)^2+\left(x_2+\frac7{10}\right)^2+x_3^2+x_4^2+x_5^2+x_6^2\le\frac1{100}\right\}$.
\item $u\in[-1,1]$ for every $\mathbf{x}\in\mathcal X$.%
\item The requested output is a controlled forward-invariant set $\mathcal S\subseteq\mathcal X\setminus\mathcal X_U$ containing $\mathcal X_I$.
\item Task type: controlled forward-invariant set synthesis.
\end{itemize}
\end{systeminstance}

\begin{systeminstance}{New-5D Mixed Nonlinear Cascade Invariance}
\label{sys:new-p3-mixed-nonlinear-cascade-5d}

The vector flow field is:
\[
\dot{\mathbf{x}}
=
\begin{pmatrix}
\dot{x}_1\\
\dot{x}_2\\
\dot{x}_3\\
\dot{x}_4\\
\dot{x}_5
\end{pmatrix}
=
\begin{pmatrix}
x_2-\dfrac12x_1\\[2mm]
-x_1-\dfrac{x_1}{5(1+x_1^2)}-2x_2+\dfrac{x_3}{10(1+x_3^2)}\\[3mm]
x_4-\dfrac12x_3\\[2mm]
-x_3-\dfrac1{40}\left(\exp(x_3)-\exp(-x_3)\right)-2x_4+\dfrac15x_5\\[2mm]
-\dfrac32x_5-\dfrac1{10}x_5^3+u
\end{pmatrix}.
\]

\begin{itemize}[label=\textendash, leftmargin=1.55em, labelsep=0.45em, topsep=0pt, itemsep=0pt]
\item $\mathcal X=[-1,1]^5$.
\item $\mathcal X_U=\left\{x\in\mathbb R^5\mid \left(x_1-\frac7{10}\right)^2+\left(x_2+\frac7{10}\right)^2+x_3^2+x_4^2+x_5^2\le\frac1{100}\right\}$.
\item $u\in[-1,1]$ for every $x\in\mathcal X$.
\item Task type: controlled forward-invariant set synthesis.
\end{itemize}

\end{systeminstance}

\begin{systeminstance}{New-6D Opinion Dynamics}
\label{sys:new-6d-opinion-dynamics}

Consider an undirected network $\mathcal G=(\mathcal V,\mathcal E)$ with $N=6$ nodes and adjacency matrix
\[
A=
\begin{pmatrix}
0&1&1&0&0&0\\
1&0&1&1&0&0\\
1&1&0&0&1&0\\
0&1&0&0&1&0\\
0&0&1&1&0&1\\
0&0&0&0&1&0
\end{pmatrix},
\]
where $A_{ij}=1$ means that nodes $i$ and $j$ are connected. The discrete-time dynamics are
\[
x_i(k+1)=x_i(k)+\frac{1}{4}\sum_{j=1}^{6}A_{ij}\tanh\bigl(x_j(k)-x_i(k)\bigr),
\qquad i=1,\ldots,6.
\]

\begin{itemize}[label=\textendash, leftmargin=1.55em, labelsep=0.45em, topsep=0pt, itemsep=0pt]
\item $\mathcal X=[-1,1]^6$.
\item $\mathcal X_U=\{x\in\mathcal X\mid \exists i\in\{1,\ldots,6\},\ |x_i|\ge\frac45\}$.
\item Verified type: forward invariance verification.
\end{itemize}

\end{systeminstance}

\begin{systeminstance}{Invariance Verification of a Discrete-Time Controlled Hopfield Network}
\label{sys:catalogue-hopfield-invariance}

Consider the closed-loop $10$-dimensional Hopfield network
$u(k)=\pi(x(k))$. For every $i\in\{1,\dots,10\}$,
\[
x_i(k+1)=x_i(k)
+0.05\bigl(1-x_i(k)^2\bigr)
\left[
-\operatorname{arctanh}\bigl(x_i(k)\bigr)
+\sum_{j=1}^{10} W_{ij}x_j(k)
+\pi_{i}(x(k))
\right],
\] 
where the symmetric weight matrix is
\[
W=
\begin{pmatrix}
-\frac32 & \frac14 & 0 & 0 & 0 & 0 & 0 & 0 & 0 & 0\\
\frac14 & -\frac32 & \frac14 & 0 & 0 & 0 & 0 & 0 & 0 & 0\\
0 & \frac14 & -\frac32 & \frac14 & 0 & 0 & 0 & 0 & 0 & 0\\
0 & 0 & \frac14 & -\frac32 & \frac14 & 0 & 0 & 0 & 0 & 0\\
0 & 0 & 0 & \frac14 & -\frac32 & \frac14 & 0 & 0 & 0 & 0\\
0 & 0 & 0 & 0 & \frac14 & -\frac32 & \frac14 & 0 & 0 & 0\\
0 & 0 & 0 & 0 & 0 & \frac14 & -\frac32 & \frac14 & 0 & 0\\
0 & 0 & 0 & 0 & 0 & 0 & \frac14 & -\frac32 & \frac14 & 0\\
0 & 0 & 0 & 0 & 0 & 0 & 0 & \frac14 & -\frac32 & \frac14\\
0 & 0 & 0 & 0 & 0 & 0 & 0 & 0 & \frac14 & -\frac32
\end{pmatrix}.
\]

\begin{itemize}[label=\textendash, leftmargin=1.55em,
                labelsep=0.45em, topsep=0pt, itemsep=0pt]
    \item Domain set:
    $
    \mathcal{X}=[-0.9,0.9]^{10}.
    $
     \item $\mathcal{X}_U = \mathbb{R}^{10}\setminus \mathcal{X}$.
    \item Control constraint:
    $
    \pi(x)\in[-2,2].
    $
    \item Verified type: forward invariance verification
\end{itemize}
\end{systeminstance}

\textbf{Problem.~4 :}

\begin{systeminstance}{New-8D Chain Safety}
\label{sys:new-8d-logarithmic-chain-safety}

The discrete-time state update is:
\[
\mathbf{x}(k+1)=
\begin{pmatrix}
x_1(k)+\dfrac1{200}\left(
-\dfrac{x_1(k)}{1+x_1(k)^2}-4x_2(k)
\right)
\\[2mm]
x_2(k)+\dfrac1{200}\left(
\dfrac{4x_1(k)}{1+x_1(k)^2}
-x_2(k)-\dfrac{3x_3(k)}{1+x_3(k)^2}
\right)
\\[2mm]
x_3(k)+\dfrac1{200}\left(
3x_2(k)-\dfrac{x_3(k)}{1+x_3(k)^2}-5x_4(k)
\right)
\\[2mm]
x_4(k)+\dfrac1{200}\left(
\dfrac{5x_3(k)}{1+x_3(k)^2}
-x_4(k)-\dfrac{2x_5(k)}{1+x_5(k)^2}
\right)
\\[2mm]
x_5(k)+\dfrac1{200}\left(
2x_4(k)-\dfrac{x_5(k)}{1+x_5(k)^2}-4x_6(k)
\right)
\\[2mm]
x_6(k)+\dfrac1{200}\left(
\dfrac{4x_5(k)}{1+x_5(k)^2}
-x_6(k)-\dfrac{3x_7(k)}{1+x_7(k)^2}
\right)
\\[2mm]
x_7(k)+\dfrac1{200}\left(
3x_6(k)-\dfrac{x_7(k)}{1+x_7(k)^2}-6x_8(k)
\right)
\\[2mm]
x_8(k)+\dfrac1{200}\left(
\dfrac{6x_7(k)}{1+x_7(k)^2}-x_8(k)
\right)
\end{pmatrix}.
\]

\begin{itemize}[label=\textendash, leftmargin=1.55em,
labelsep=0.45em, topsep=0pt, itemsep=0pt]
\item $\mathcal X=[-1,1]^8$.
\item $\mathcal X_I=[-\frac3{20},\frac3{20}]^8$.
\item $\mathcal X_U=[\frac34,\frac9{10}]
\times[-\frac1{10},\frac1{10}]^7$.
\item Task type: safety verification.
\end{itemize}

\end{systeminstance}

\begin{systeminstance}{New-6D Arctan Chain Safety}
\label{sys:new-6d-arctangent-chain-safety}

The discrete-time state update is:
\[
\mathbf{x}(k+1)=
\begin{pmatrix}
x_1(k)+\dfrac1{200}\left(-\arctan(x_1(k))-8x_2(k)\right)\\[2mm]
x_2(k)+\dfrac1{200}\left(8\arctan(x_1(k))-x_2(k)-6\arctan(x_3(k))\right)\\[2mm]
x_3(k)+\dfrac1{200}\left(6x_2(k)-\arctan(x_3(k))-10x_4(k)\right)\\[2mm]
x_4(k)+\dfrac1{200}\left(10\arctan(x_3(k))-x_4(k)-4\arctan(x_5(k))\right)\\[2mm]
x_5(k)+\dfrac1{200}\left(4x_4(k)-\arctan(x_5(k))-8x_6(k)\right)\\[2mm]
x_6(k)+\dfrac1{200}\left(8\arctan(x_5(k))-x_6(k)\right)
\end{pmatrix}.
\]

\begin{itemize}[label=\textendash, leftmargin=1.55em,
                labelsep=0.45em, topsep=0pt, itemsep=0pt]
\item $\mathcal X=[-1,1]^6$.
\item $\mathcal X_I=[-\frac3{20},\frac3{20}]^6$.
\item $\mathcal X_U=[\frac34,\frac9{10}]\times[-\frac1{10},\frac1{10}]^5$.
\item Task type: safety verification.
\end{itemize}

\end{systeminstance}

\begin{systeminstance}{New-8D Parametric Nonlinear Safety}
\label{sys:new-8d-parametric-nonlinear-safety}

The vector flow field is:
\[
\dot{\mathbf{x}}=
\begin{pmatrix}
x_2 \\[1mm]
-x_1-0.1\sin(\omega x_1)-x_2+x_3 \\[2mm]
-2x_3+x_4 \\[1mm]
-1.5x_4+x_5 \\[2mm]
-2x_5+x_6 \\[1mm]
-1.5x_6+0.2x_7 \\[2mm]
-2x_7+0.2x_6+0.2x_8
-0.1x_7\sin(x_8) \\[1mm]
-1.5x_8+0.2x_7
-0.1x_8\cos(x_7)
\end{pmatrix}.
\]

\begin{itemize}[label=\textendash, leftmargin=1.55em,
labelsep=0.45em, topsep=0pt, itemsep=0pt]
\item $\mathcal X=[-1,1]^8$.
\item $\mathcal X_I=
\left\{x\in\mathbb R^8\mid \lVert x\rVert_2\le 0.5\right\}$.
\item $\mathcal X_U=
\left\{x\in\mathbb R^8\mid
\left\|x-(0,0.8,0,0,0,0,0,0)^{\mathsf T}\right\|_2\le 0.1
\right\}$.
\item $\omega\in[0.5,2]$ is a fixed uncertain parameter.
\item The barrier certificate is state-only and uniform for every fixed
$\omega\in[0.5,2]$.
\item Task type: robust safety verification.
\end{itemize}

\end{systeminstance}

\begin{systeminstance}{New-5D Arctangent Perturbation Safety}
\label{sys:new-5d-arctangent-perturbation-safety}

The vector flow field is:
\[
\dot{\mathbf{x}}=
\begin{pmatrix}
x_2+\dfrac12x_5\\[2mm]
-x_1-\omega x_2+2x_4\\[1mm]
x_4\\[1mm]
-2x_3-\dfrac18\arctan(x_3)-\dfrac32x_4-2x_2+x_5\\[2mm]
-\dfrac12x_1-x_4-x_5-\dfrac18\arctan(x_5)
\end{pmatrix}.
\]

\begin{itemize}[label=\textendash, leftmargin=1.55em, labelsep=0.45em, topsep=0pt, itemsep=0pt]
\item $\mathcal X=[-1,1]^5$.
\item $\mathcal X_I=\left\{x\in\mathbb R^5\mid\lVert x\rVert_2\le\frac12\right\}$.
\item $\mathcal X_U=\left\{x\in\mathbb R^5\mid\left\lVert x-\left(0,0,\frac45,0,0\right)^{\mathsf T}\right\rVert_2\le\frac1{10}\right\}$.
\item $\omega\in[1,2]$.
\item Task type: robust safety verification.
\end{itemize}

\end{systeminstance}